\documentclass[journal]{IEEEtran}
\usepackage{graphicx} 
\usepackage{subfigure}
\usepackage{amsmath}
\usepackage{amsbsy}
\usepackage{amssymb}
\usepackage{graphicx,xparse}
\usepackage{mathtools}
\usepackage{placeins}
\usepackage{pdfpages}
\usepackage{mathptmx}
\usepackage{float}
\usepackage[acronym]{glossaries}
\usepackage{enumitem}
\setlist[itemize]{noitemsep, topsep=0pt}
\setlist[enumerate]{noitemsep, topsep=0pt}
\usepackage{longtable}

\usepackage{xcolor}
\usepackage{cite}
\usepackage{comment}
\definecolor{ashgrey}{rgb}{0.7, 0.75, 0.71}
\definecolor{apricot}{rgb}{0.98, 0.81, 0.69}
\definecolor{anti-flashwhite}{rgb}{0.95, 0.95, 0.96}
\definecolor{almond}{rgb}{0.94, 0.87, 0.8}
\definecolor{airforceblue}{rgb}{0.36, 0.54, 0.66}
\definecolor{asparagus}{rgb}{0.53, 0.66, 0.42}
\definecolor{azure(web)(azuremist)}{rgb}{0.94, 1.0, 1.0}
\definecolor{babyblue}{rgb}{0.54, 0.81, 0.94}
\definecolor{beige}{rgb}{0.96, 0.96, 0.86}
\definecolor{bisque}{rgb}{1.0, 0.89, 0.77}
\definecolor{aliceblue}{rgb}{0.94, 0.97, 1.0}
\definecolor{chestnut}{rgb}{0.8, 0.36, 0.36}
\usepackage[hidelinks]{hyperref}
\usepackage{xcolor}
\hypersetup{
    colorlinks,
    linkcolor={black!50!black},
    citecolor={black!90!black},
    urlcolor={black!80!black}
}

\usepackage{tikz}
\usetikzlibrary{shapes.geometric, arrows}
\usepackage{placeins}
\usepackage[edges]{forest}
\usepackage[linewidth=1pt]{mdframed}
\usepackage{array,multirow,graphicx}
\makeatletter
\newcommand{\thickhline}{%
    \noalign {\ifnum 0=`}\fi \hrule height 1pt
    \futurelet \reserved@a \@xhline
}
\newcolumntype{"}{@{\hskip\tabcolsep\vrule width 1pt\hskip\tabcolsep}}
\makeatother

\tikzstyle{startstop} = [rectangle, rounded corners, 
minimum width=2cm, 
minimum height=1cm,
text width=8cm,
text centered, 
draw=black, 
fill=yellow!70]
\tikzstyle{arrow} = [thick,->,>=stealth]
\definecolor{myframecolor}{RGB}{173, 216, 230}
\definecolor{mybackgroundcolor}{RGB}{255, 250, 205}

\newmdenv[
  linecolor=myframecolor,
  backgroundcolor=mybackgroundcolor,
  roundcorner=10pt,
  linewidth=2pt,
  skipabove=10pt,
  skipbelow=10pt
]{customframe}

\newacronym{2d}{2D}{two dimensional}
\newacronym{3d}{3D}{three dimensional}
\newacronym{5g}{5G}{fifth generation}
\newacronym{6g}{6G}{sixth generation}
\newacronym{ai}{AI}{artificial intelligence}
\newacronym{aoa}{AOA}{angle-of-arrival}
\newacronym{aod}{AOD}{angle-of-departure}
\newacronym{bs}{BS}{base station}
\newacronym{cnn}{CNN}{convolutional neural network}
\newacronym{crf}{CRF}{conventional radio frequency}
\newacronym{crb}{CRB}{Cramér-Rao bound}
\newacronym{crlb}{CRLB}{Cramér-Rao lower bound}
\newacronym{cs}{CS}{compressive sensing}
\newacronym{csi}{CSI}{channel state information}
\newacronym{dfl}{DFL}{device-free localization}
\newacronym{dl}{DL}{downlink}
\newacronym{dnn}{DNN}{deep neural network}
\newacronym{dsp}{DSP}{digital signal processors}
\newacronym{fim}{FIM}{Fisher information matrix}
\newacronym{fpga}{FPGA}{field-programmable gate arrays}
\newacronym{gdop}{GDoP}{geometric dilution of precision}
\newacronym{gnss}{GNSS}{global navigation satellite systems}
\newacronym{iot}{IoT}{internet of things}
\newacronym{isac}{ISAC}{integrated sensing and communication}
\newacronym{jcal}{JCAL}{joint communication and localization}
\newacronym{los}{LoS}{line-of-sight}
\newacronym{mmw}{mmW}{millimeter wave}
\newacronym{mimo}{MIMO}{multiple-input, multiple-output}
\newacronym{miso}{MISO}{multiple-input, single-output}
\newacronym{mcrb}{MCRB}{misspecified Cramér-Rao bound}
\newacronym{ml}{ML}{machine learning}
\newacronym{mle}{MLE}{maximum likelihood estimation}
\newacronym{mse}{MSE}{mean square error}
\newacronym{music}{MUSIC}{multiple signal classification}
\newacronym{nlos}{NLoS}{non-line-of-sight}
\newacronym{peb}{PEB}{position error bound}
\newacronym{reb}{REB}{rotation error bound}
\newacronym{ris}{RIS}{reconfigurable intelligent surfaces}
\newacronym{rl}{RL}{reinforcement learning}
\newacronym{rss}{RSS}{received signal strength}
\newacronym{rtt}{RTT}{round-trip time}
\newacronym{siso}{SISO}{single-input, single-output}
\newacronym{simo}{SIMO}{single-input, multiple-output}
\newacronym{sim}{SIM}{stacked intelligent metasurfaces}
\newacronym{sl}{SL}{sidelink}
\newacronym{slac}{SLAC}{simultaneous localization and communication}
\newacronym{slam}{SLAM}{simultaneous localization and mapping}
\newacronym{star}{STAR}{Simultaneously transmitting and reflecting}
\newacronym{snr}{SNR}{signal-to-noise ratio}
\newacronym{oeb}{OEB}{orientation error bound}
\newacronym{ofdm}{OFDM}{orthogonal frequency division multiplexing}
\newacronym{thz}{THz}{terahertz band}
\newacronym{toa}{TOA}{time-of-arrival}
\newacronym{tof}{TOF}{time-of-flight}
\newacronym{tdoa}{TDOA}{time-difference-of-arrival}
\newacronym{ue}{UE}{user equipment}
\newacronym{ul}{UL}{uplink}
\newacronym{uwb}{UWB}{ultra wide band}
\newacronym{vlp}{VLP}{visible light positioning}
\newacronym{uav}{UAV}{unmanned ariel vehicle}
\newacronym{ofts}{OFTS}{orthogonal time frequency space}

\begin{document}
\title{Reconfigurable Intelligent Surfaces in 6G Radio Localization: A Survey of Recent Developments, Opportunities, and Challenges}
\author{Anum~Umer, Ivo~M\"{u}\"{u}rsepp, Muhammad~Mahtab~Alam,~\IEEEmembership{Senior~Member,~IEEE,} Henk~Wymeersch,~\IEEEmembership{Fellow,~IEEE}
\thanks{This work has received funding partly from the European Union’s Horizon 2020 Research and Innovation Program under Grant 101058505 ‘5G-TIMBER’ and by the project ``Increasing the knowledge intensity of Ida-Viru entrepreneurship'' co-funded by the European Union as well as by NATO-SPS G7699 PROTECT project. Henk Wymeersch was supported by the European Commission through the H2020 project RISE-6G (grant agreement no.  101017011) and the SNS JU project 6G-DISAC under the EU’s Horizon Europe research and innovation program under Grant Agreement No 101139130.
}
\thanks{A. Umer, I. M\"{u}\"{u}rsepp, and M. M. Alam are with Thomas Johann Seebeck Department of Electronics, Tallinn University of Technology, 12616 Tallinn, Estonia (e-mail: anum.umer@taltech.ee; ivo.muursepp@ttu.ee; muhammad.alam@taltech.ee).}
\thanks{H. Wymeersch is with Department of Electrical Engineering, Chalmers
University of Technology, 412 58 Gothenburg, Sweden (e-mail:  henkw@chalmers.se).}}
\maketitle

\begin{abstract}
  In this survey paper, we present an extensive review of the use of \acrfull{ris} in 6G radio localization, highlighting their pivotal role as a low-cost, energy-efficient technology that reshapes wireless communication and localization landscapes. Investigating the versatile capabilities of \acrshort{ris}, we explore their dynamic control over electromagnetic wave manipulation, including reflection, refraction, and transmission, which opens new horizons in diverse applications ranging from \acrfull{iot} connectivity to advanced mobile communication, and various innovative applications in Industry 4.0. Our comprehensive review provides an overview of \acrshort{ris} use in 6G radio localization, highlighting recent progress in \acrshort{ris} technology assisted localization. It focuses on key aspects, including network scenarios, transmission bands, deployment environments, and near-field operations. We discuss studies to examine the state-of-the-art \acrshort{ris}-assisted localization and optimization techniques and their performance evaluation matrices. In addition, we present a detailed taxonomy of \acrshort{ris}-assisted radio localization, emphasizing the rapid evolution and potential of \acrshort{ris} technology in non-line-of-sight scenarios as an alternative to traditional base stations. Based on the careful investigation of the reviewed studies, the survey also sheds light on future research directions, technical challenges, and limitations, offering a clear perspective on the integration and optimization of \acrshort{ris} in 6G networks for enhanced localization capabilities.
   % Reconfigurable intelligent surfaces (\acrshort{ris}) are seen as a key enabler low-cost and energy-efficient technology for 6G radio communication and localization. In this paper, we aim to provide a comprehensive overview of the current research progress on the \acrshort{ris} technology in radio localization for 6G. Particularly, we discuss the \acrshort{ris}-assisted radio localization taxonomy and review the studies of \acrshort{ris}-assisted radio localization for different network scenarios, bands of transmission, deployment environments, as well as near-field operations. Based on this review, we highlight the future research directions, associated technical challenges, real-world applications, and limitations of \acrshort{ris}-assisted radio localization.
\end{abstract}


\section{Introduction}
\label{sec1}
%\textbf{Significance}:
The \acrfull{ris} are advanced metasurfaces designed with the remarkable capability of being able to be reprogrammed to alter their electromagnetic properties and functionalities according to specific requirements \cite{9215972}. These intelligent surfaces enable dynamic control over the reflection, transmission, and absorption of electromagnetic waves, allowing for unprecedented flexibility and adaptability in manipulating wireless signals and optimizing wireless communication systems. Through their programmability, \acrshort{ris} empower researchers and engineers to explore a wide range of applications, including wireless communication networks, smart environments, \acrfull{iot} connectivity, radar systems, and more \cite{9330512}. By harnessing the potential of \acrshort{ris} technology, we can revolutionize the way we interact with and shape the electromagnetic world around us. They provide the ability to control and program the wireless communication channel, making it a highly versatile tool for wireless communication \cite{chepuri2022integrated}. This feature makes \acrshort{ris} favorable for radio communication and localization since we can control the illumination of \acrfull{nlos} areas where direct signaling from the anchor node is not possible. Moreover, radio localization typically requires more than one anchor to function, in contrast, \acrshort{ris} offer a cost-effective and energy-efficient solution to replace additional anchor nodes and relays \cite{9765815}. This is due to the simpler hardware implementation of \acrshort{ris}, which are easier to deploy and maintain. \acrshort{ris}, with their limited power requirements, can be installed on surfaces like walls, billboards, or even unmanned aerial vehicles for emergency services \cite{9665421}.
%SUMMARY OF RECENT LOCALIZATION AND \acrshort{ris} TECHNOLOGY ASSISTED LOCALIZATION SURVEYS AND TUTORIALS
\begin{scriptsize}
    \centering
\begin{table*}
\centering
\caption{topic wise comparison with recent localization and \acrshort{ris}-assisted localization surveys and tutorials }\label{tableofsurveys}
    \begin{tabular}{c|c|c|c|c|c|c|c|c|c|c}
    \thickhline
         & \multirow{2}{*}{\textbf{Article}} &   \multirow{2}{*}{\textbf{Type}} & \multicolumn{3}{c|}{\textbf{Transmission bands}}  & \multicolumn{2}{c|}{\textbf{Environment}} & \multicolumn{3}{c}{\textbf{Applications}}\\
        \cline{4-11}
         & & & \textbf{$<$30GHz} & \textbf{mmW}& \textbf{THz} & \textbf{Indoor} & \textbf{Outdoor}  & \textbf{SLAM} & \textbf{ISAC} &\textbf{JCAL}\\
        \thickhline
        
         \multirow{18}{*}{\rotatebox[origin=c]{90}{\parbox[c]{0.2cm}{\centering \textbf{Localization}}}}

          & \cite{bresson2017simultaneous} & Survey &&&&&\checkmark&\checkmark&&\\
         \cline{2-11}
         
          & \cite{7927385} & Survey & & & & \checkmark&&&\\
         \cline{2-11}

          & \cite{8306879} & Survey &&&&&\checkmark&& \\
         \cline{2-11}

          & \cite{8409950} & Survey&\checkmark&&&\checkmark&&\checkmark&\\
         \cline{2-11}

          & \cite{8226757} & Survey & \checkmark &&&&&&&\\
         \cline{2-11}

          & \cite{8692423} & Survey &&&&\checkmark&&& \\
         \cline{2-11}

          & \cite{8734111} & Survey &\checkmark&&&&&&\\
         \cline{2-11}

          & \cite{WEN201921} &Survey&&\checkmark &&&&&&\\
         \cline{2-11}

%          & \cite{Burghal2020ACS} &Survey & \checkmark & \checkmark &\checkmark &\checkmark & & &  & \\
 %        \cline{2-11}

          & \cite{9159543} &Survey&&&&\checkmark&&&&\\
         \cline{2-11}

          & \cite{9523553} &Survey &\checkmark&&&&&&&\\
         \cline{2-11}

          & \cite{9328120} &Survey&&&&\checkmark&&&&\\
         \cline{2-11}

          & \cite{9356512} & Survey & \checkmark &\checkmark &&&&&&\\
         \cline{2-11}
          & \cite{xiao2022overview} & Survey &&\checkmark&&&&&&\\
         \cline{2-11}

          & \cite{s22010247} & Survey&&&&&&&&\\
         \cline{2-11}

          & \cite{Liu2021ASO} & Survey &&&&&&&\checkmark&\\
         \cline{2-11}

          & \cite{Trevlakis2023LocalizationAA}& Tutorial&\checkmark&\checkmark&\checkmark&\checkmark&\checkmark&\checkmark&\checkmark&\checkmark\\

         \thickhline

        \multirow{5}{*}{\rotatebox[origin=c]{90}{\parbox[c]{1.5cm}{\centering \textbf{\acrshort{ris}- Assisted Localization}}}}

%         &2020 & \cite{9215972} & Tutorial &\checkmark&\checkmark&&&&&&&&\\
 %        \cline{2-14}
         
%        &2021 & \cite{9330512} & Unidentified&\checkmark&\checkmark&\checkmark&\checkmark&\checkmark&\checkmark&\checkmark&\checkmark\\
 %       \cline{2-12}
        
          &\cite{9495362} & Survey &&&&&&\checkmark&&\\
         \cline{2-11}

          & \cite{9782674} & Tutorial & &&\checkmark&&&&&\\
         \cline{2-11}

          & \cite{9874802} & Survey & &\checkmark&&\checkmark&&&&\\
         \cline{2-11}

          & \cite{9847080} & Survey & &\checkmark &&&&&&\\
         \cline{2-11}

         \cline{2-11}
         & this work & Survey &  \checkmark&  \checkmark&  \checkmark&  \checkmark&\checkmark & \checkmark & \checkmark&\checkmark\\
         \thickhline
         
    \end{tabular}
\end{table*}
\end{scriptsize}

Besides reconfigurability, there are other key features that promise to make \acrshort{ris} suitable for use in wireless
networks.  \acrshort{ris} can be fabricated using low-cost materials such as printed circuit boards, making it affordable for widespread deployment \cite{9124848}. It can significantly reduce the energy consumption of wireless communication systems. By reflecting and focusing the signal towards the intended receiver, \acrshort{ris} can reduce the need for high-power transmitters and increase the energy efficiency of the whole system \cite{8288263}. RIS can mitigate interference in wireless communication systems by reflecting and manipulating the signal, it can create nulls in the directions where interference is present, leading to improved signal quality and channel capacity \cite{9201413, 8741198}.

\acrshort{ris} are, thus, perceived as state-of-the-art technology for the localization of users in \acrfull{6g} mobile communication networks 
%that do not have a direct Line of Sight (LoS) with the BS, 
provided the location of the \acrshort{ris} is already known \cite{9215972}. The research community is actively working on modeling and optimization of various aspects of \acrshort{ris}-assisted radio localization to enable bigger-impact techniques and applications for \acrshort{6g} networks, such as \acrfull{slac}, \acrfull{slam}, and numerous inventive applications in the realm of Industry 4.0, as elaborated further in the paper. While research in this area of radio localization is advancing rapidly, it is essential to consolidate the progress made in this field and pinpoint both past accomplishments and future avenues for exploration.
\par
%\textbf{Novelty}

\subsection{Related Work and Motivation}
In this subsection, we discuss the related works published in the recent years and highlight the motivation behind current study. Several review studies exist on localization that share the localization basics in common but each of them is primarily focused on either the type of signals used for localization, the localization environment, or the techniques for localization. For instance, indoor localization is discussed in \cite{7927385, 8409950, 8692423, 9159543, 9328120}, outdoor localization in \cite{bresson2017simultaneous, 8306879, 8710259} and localization in ground-air-space networks in \cite{10565792}. As for the types of signals, radio signals are covered in \cite{8409950, 8226757, 9328120} and \acrfull{vlp} in \cite{8358018}. Localization techniques such as \acrshort{slam} \cite{bresson2017simultaneous, 8409950}, multi-dimensional scaling \cite{8734111}, \acrfull{ml} \cite{9159543, 9523553} have also been discussed in the dedicated surveys. Several surveys exist on the application of localization such as \acrfull{dfl} \cite{8710259}, autonomous driving \cite{bresson2017simultaneous, 8306879, s22010247}, pedestrian localization \cite{9523553}, emergency response \cite{7927385} and network localization \cite{8409950}. Recent studies focus on envisioned applications and use cases of localization in \acrshort{6g} \cite{Trevlakis2023LocalizationAA}, technological enablers for beyond \acrfull{5g} and \acrshort{6g} localization including \acrshort{ris} \cite{9215972, 9330512}, surveys on localization signal processing techniques and algorithms for \acrshort{6g} \cite{9721205, 9495362, 9847080},  as well as the convergent communication, localization, and sensing including \acrfull{isac} \cite{Liu2021ASO, 9330512}, and high-frequency localization \cite{9215972, 9330512, 9782674, 9721205}. While \cite{Trevlakis2023LocalizationAA} provides a holistic overview of localization in 6G networks, our study is focused only on RIS-assisted localization in 6G networks. Several studies explore the potential and applications of \acrshort{ris} in \acrshort{6g} systems. Research has been conducted on the potential of \acrshort{ris} in radio localization and mapping, which are detailed in \cite{9215972} and \cite{9330512}. Further studies and surveys discuss signal processing in \acrshort{ris}-assisted networks, as can be found in \cite{9721205, 9495362, 9847080}. A tutorial that gives an overview on radio localization with \acrshort{ris} at higher frequencies is presented in \cite{9782674}. 
%IoT positioning is another area of interest, with a comprehensive survey available in \cite{9874802}. 
The domain of \acrshort{iot} positioning has been explored in a contemporary survey presented in \cite{9874802}. However, the breadth of this study is relatively limited. This constrained scope suggests that it may not fully encapsulate the entire spectrum of research conducted in \acrshort{ris}-assisted radio localization in 6G networks. Furthermore, the survey's content is not up-to-date with the most recent advancements in the field. Additionally, a notable limitation of this survey is its lack of reproducibility, which could be a critical factor for those seeking to validate or build upon its findings.
The concept of \acrshort{isac} with \acrshort{ris} is explained in a tutorial overview in \cite{chepuri2022integrated}. Lastly, the study in \cite{10044963} examines the use of \acrshort{ris} in different network scenarios, such as \acrfull{siso}, \acrfull{mimo}, \acrfull{miso}, and \acrfull{simo}. The recent study provided an overview of the performance improvements offered by \acrshort{ris} over traditional network designs in smart wireless environments. It specifically focused on the network architecture, deployment scenarios, bandwidth, and area of influence enabled by \acrshort{ris} \cite{9815617}. Topic wise comparison of recent articles on localization and \acrshort{ris}-assisted radio localization with our work is provided in Table \ref{tableofsurveys}.
%A recent study gave an overview of the situations in which \acrshort{ris} will provide significant performance improvements over traditional network designs with smart wireless environments enabled by \acrshort{ris} from the perspective of network architecture for deployment scenarios, bandwidth, and area of influence \cite{9815617}. .
%, encompassing only 19 distinct studies

The research community is actively exploring various aspects of \acrshort{ris} in radio localization. These include modeling, analysis, and optimization of various localization scenarios with \acrfull{bs}, \acrfull{ue}, and \acrshort{ris}, determining the number and type of \acrshort{ris} elements, as well as designing phase control and coefficient values for enhanced localization \cite{8264743, 9048973, 8815412, 9148744, 9129075, 9124848, 9500437, 9513781, 9410435, 9238887, 9193909, 9456027, 9201330, 9500663, 9508872, 9593200, 9593241, 10017173, 10044963, 9847080, 10054103, 9473890, guo2022dynamic, Keykhosravi2020SISORJ, Fascista2020RISAidedJL, Gong2022AsynchronousRL, 9538860, 9779402, pan2022risaided, 9777939, 9729782, isprs, 9902991}. Other factors being studied primarily include the placement of \acrshort{ris} in indoor and outdoor scenarios with variations in the number of antennas on \acrshort{bs} and \acrshort{ue}, and \acrshort{ris} operation at multiple frequency bands, i.e., frequency range 1 (FR1) (450 MHz to 6 GHz), frequency range 2 (FR2) (between 24.25 GHz and 52.6 GHz), \acrfull{mmw} (30-300 GHz) and \acrfull{thz} (0.1-10 THz), as well as the near-field and far-field operation of \acrshort{ris}-assisted localization \cite{9782674}, which are the focus of this article. Overall, there is significant interest and effort being dedicated to advancing the use of \acrshort{ris} for localization. To develop the understanding of basics of RIS-assisted radio localization and to determine the unexplored research avenues within this field, it is deemed necessary to carry out a comprehensive survey. We briefly shed light on role of RIS in 6G radio localization and compile the recent developments in terms of state-of-the-art areas of investigation as well as the techniques. %This will facilitate the identification of specific areas for the research community. 
In contrast to the existing surveys and tutorials, in this survey, we first highlight the role of RIS in radio localization in terms of its potential applications and by the introduction of RIS-assisted localization taxonomy. Then we provide a comprehensive overview of the recent studies on the role of \acrshort{ris} in radio localization in 6G networks. Our work consolidates and builds upon existing knowledge to promote the advancement of \acrshort{ris}-assisted localization in 6G networks.  %Such identification holds significance as it paves the way for the effective integration of RIS in radio localization, thereby enabling highly precise positioning at the centimeter level accuracy, along with optimal coverage in the context of 6G technology. 
In this study, we seek answers to the following questions:

\begin{enumerate}
    \item \label{q1}How has \acrshort{ris} been used for localization in 6G networks?
    \item \label{q2}What are the current trends and developments in the use of \acrshort{ris} technology for 6G radio localization?
    \item \label{q3}What are the future directions of \acrshort{ris}-assisted radio localization in 6G and what are the associated technical challenges, and limitations?
\end{enumerate}

The comparison of this survey with recent localization surveys, tutorials and review studies on the basis of aforementioned questions is given in Table \ref{qbased}.

\begin{scriptsize}
    \centering
\begin{table}
\caption{ Research questions based comparison o with existing surveys, tutorials and review studies with $\checkmark$ for fully addressed, $\partial$ for partially addressed and $\times$ of not addressed questions. }
\label{qbased}
\begin{tabular}{|c|p{1cm}|p{1cm}|p{1cm}|p{2cm}|}
\hline
\textbf{Reference} & \textbf{Q1} & \textbf{Q2} & \textbf{Q3} & \textbf{Type of study} \\ \hline
\cite{9495362} & $\times$ & $\times$ & $\times$ & Survey \\ \hline
\cite{9782674} & $\partial$ & $\partial$ & $\partial$ & Tutorial \\ \hline
\cite{10565792} & $\times$ & $\times$ & $\times$ & Tutorial \\ \hline
\cite{9874802} & $\partial$ & $\partial$ & $\partial$ & Review  \\ \hline
\cite{Trevlakis2023LocalizationAA} & $\partial$ & $\partial$ & $\partial$ & Review/ Tutorial \\ \hline
\cite{9847080} & $\partial$ & $\partial$ & $\times$ & Review/ Tutorial \\ \hline
\cite{10536135} & $\times$ & $\times$ & $\times$ & Review  \\ \hline
\cite{10188352} & $\times$ & $\times$ & $\times$ & Review  \\ \hline
\cite{10541024} & $\times$ & $\times$ & $\times$ & Review \\ \hline
This work & $\checkmark$ & $\checkmark$ & $\checkmark$ & Survey \\ \hline
\end{tabular}
\end{table}
\end{scriptsize}

\subsection{Review Method}\label{reviewmethod}
After having identified the research questions, we defined our search string and identified the appropriate databases to find the most relevant literature based on the inclusion and exclusion criteria. For the search string, we selected the keywords from our research questions, based on which two search strings were defined. The first string limits the investigation to the \acrshort{ris} and its synonyms. Similarly, the second string limits the literature to localization and its synonyms. Both search strings were combined using logical operators before being applied to the literature databases. The final search string is:  (\textit{``intelligent reflecting surface" OR ``reconfigurable intelligent surface" OR ``\acrshort{ris}" OR ``IRS" OR ``LIS" OR ``large intelligent surface") AND (``localization" OR ``positioning"}) AND (``6G"). The search string was used on digital databases, i.e., IEEE Xplore, to identify the most pertinent papers. We searched for journal, workshop, and conference papers in the databases. % until 30 June 2023.
As various studies use diverse terms to describe \acrshort{ris}-assisted radio localization, it is possible that our search string might not capture all relevant works. Therefore, we also performed comprehensive backward referencing. To ensure we did not miss pertinent articles, we included full texts in our analysis, even if they did not have our search terms in their titles or abstracts. In short, deliberate effort has been put to include all the works relevant to the topic of our survey. 
%These databases were chosen because they encompass the majority of publications in the fields of radio communication and localization.

After having collected all the relevant studies, we passed it through a rigorous inclusion and exclusion criterion to narrow down the database of relevant studies.  We excluded all the studies that fulfilled the following exclusion criteria: if multiple versions of the same work are collected, such as a peer-reviewed published version and its pre-print archive version, only the published version shall be kept.
%The study cannot be accessed digitally, i.e., cannot be accessed on web freely, or the study is a duplicate, i.e., multiple versions of same work available on web. %, and the paper consists of fewer than four pages. 
The remaining set of studies was evaluated for the following inclusion criteria: the study is relevant to the topic of \acrshort{ris} in radio localization in \acrshort{6g}, and it showcases or illustrates a method, approach, or technique for \acrshort{ris}-assisted radio localization. 
%\textbf{Problem statement}:
%\textcolor{red}{In this survey, our focus is to identify how is \acrshort{ris} used for localization in 6G, and how does it impact key performance indicators such as accuracy, coverage, latency, update rate, stability, scalability, and mobility of radio localization. }

\begin{table}[t!]
    \caption{Abbreviations}
    \centering
    \begin{tabular}{ll}
       2D & two dimensional\\
       3D & three dimensional\\
       5G & fifth generation\\
       6G & sixth generation \\
       AI & artificial intelligence\\
       AOA & angle-of-arrival\\
       AOD & angle-of-departure\\
       BS & base station\\
       CNN & convolutional neural network \\
       CRF & conventional radio frequency \\
       CRB & Cramér-Rao bound \\
       CRLB & Cramér-Rao lower bound \\
       CS & compressive sensing\\
       CSI & channel state information\\
       DFL & device-free localization\\
       DL & downlink\\
       DNN & deep neural network \\ 
%       DSP & digital signal processors\\
       FIM & Fisher information matrix\\
%       FPGA &  field-programmable gate arrays\\
       GDoP & geometric dilution of precision \\
       GNSS & global navigation satellite systems\\
       IoT & internet of things\\
       ISAC & integrated sensing and communication\\
       JCAL & joint communication and localization\\
       LoS & line-of-sight\\
       mmW   & millimeter wave \\
       MIMO & multiple-input, multiple-output\\
       MISO & multiple-input, single-output\\
       MCRB & misspecified Cramér-Rao bound\\
       ML & machine learning\\
       MLE & maximum likelihood estimation\\
       MSE & mean square error \\
       MUSIC & multiple signal classification\\
       NLoS & non-line-of-sight\\
       OEB & orientation error bound \\
       OFDM & orthogonal frequency division multiplexing\\
       OFTS & orthogonal time frequency space\\
       PEB & position error bound\\
       REB & rotation error bound\\
       RIS  & reconfigurable intelligent surfaces\\
       RL & reinforcement learning\\
       RSS & received signal strength\\
       RTT & round-trip time\\
       SISO & single-input, single-output\\
       SIMO & single-input, multiple-output\\
       SL & sidelink\\
       SLAC & simultaneous localization and communication\\
       SIM & stacked intelligent metasurfaces \\
       SLAM & simultaneous localization and mapping\\
       STAR & Simultaneously transmitting and reflecting \\
       SNR & signal-to-noise ratio\\
       THz & terahertz band\\
       TOA & time-of-arrival\\
       TDOA & time-difference-of-arrival\\
 %      ToF & time-of-flight\\
        UAV & unmanned ariel vehicle\\
       UE & user equipment\\
       UL & uplink\\
       UWB & ultra wide band\\
       VLP & visible light positioning\\

    \end{tabular}

    \label{abb}
\end{table}

 The unique aspects of our work are stated below:
\begin{itemize}
    \item Via the survey, answers to the 3 research questions are provided in the following sections, where Section II answers research question \ref{q1}, Section III answers research question \ref{q2} and Section IV answers  research question \ref{q3}.
    
    \item We  present the potential applications of RIS-assisted localization from the use case families in RISE-6G project, to motivate the reader towards the topic. We  also discuss the RIS-assisted radio localization taxonomy, which has not been comprehensively presented anywhere in literature before.
    
%    \item \textcolor{red}{  }
    %Within the taxonomy subsection we discuss application scenarios (indoor, outdoor), localization measurements(time of arrival, angle of arrival, received signal strength, channel state information), RIS configuration and deployment(phase configuration, placement, density, control and adaptation, and synchronization), localization algorithms and optimization(model based, learning based) and performance evaluation(accuracy and precision, robustness, energy efficiency, resolution, and computational complexity). 
    
    \item We conducted our survey systematically and outlined the review method in Section \ref{reviewmethod}, ensuring the entire review process is reproducible for validation and further research. Our method for collating, analyzing, and interpreting data is explicitly stated as part of systematic approach. Consequently, our work covers all significant advances in RIS-assisted localization, including the latest studies and their implications.
    
    %\item \textcolor{red}{Our work addresses significant advances in RIS-assisted localization, for which we discuss the latest studies and their implications in detail.} 
    
    \item Finally, we provide a detailed, updated, and diverse perspective on what are the future research directions in RIS-assisted radio localization in 6G networks, particularly in light of recent advancements.
\end{itemize}

\begin{scriptsize}
\begin{figure}[t!]
\centering
\begin{tikzpicture}[node distance=1cm]

\node (start) [startstop, draw=black, fill=ashgrey] {\textbf{\ref{sec1} Introduction}\\
\begin{customframe}
\textit{A. Related Work and Motivation } \\ 
\textit{B. Review Method}
\end{customframe}
};

\node (third) [startstop, below of=start,yshift=-1.7cm,draw=black, 
fill=aliceblue] {\textbf{\ref{sec3} \acrshort{ris} In Radio Localization}\\
\begin{customframe}
\textit{A. Overview of \acrshort{ris} Operation} \\
\textit{B. \acrshort{ris}-Assisted Localization in 6G Networks} \\   
\textit{C. \acrshort{ris} Support for Radio Localization}\\ 
%\textit{C. Potential Applications}\\
  \textit{ D. \acrshort{ris}-Assisted Localization Taxonomy} 
\end{customframe}
};

\node (fourth) [startstop, below of=third, yshift=-2.2cm, draw=black, fill=almond] {\textbf{\ref{sec4} State of the Art in \acrshort{ris}-Assisted Radio Localization}\\
\begin{customframe}
\textit{A. Outdoor and Far-field Localization} \\
\textit{B. Indoor Localization } \\     \textit{C. Near-field Localization  }
\end{customframe}
};
\node (fifth) [startstop, below of=fourth, yshift=-1.25cm,draw=black, 
fill=asparagus!50] {\textbf{\ref{sec5} Challenges and Research Outlook}
};
\node (seventh) [startstop, below of=fifth,yshift=-0.3cm,draw=black, 
fill=chestnut!40] {\textbf{\ref{sec6} Conclusion}};

\draw [arrow] (start) -- (third);
%\draw [arrow] (second) -- (third);
\draw [arrow] (third) -- (fourth);
%\draw [arrow] (fourth) -- (fourthhalf);
\draw [arrow] (fourth) -- (fifth);
\draw [arrow] (fifth) -- (seventh);
%\draw [arrow] (sixth) -- (seventh);
\end{tikzpicture}
\caption[c]{The overall outline of the article.}
\label{outline}
\end{figure}
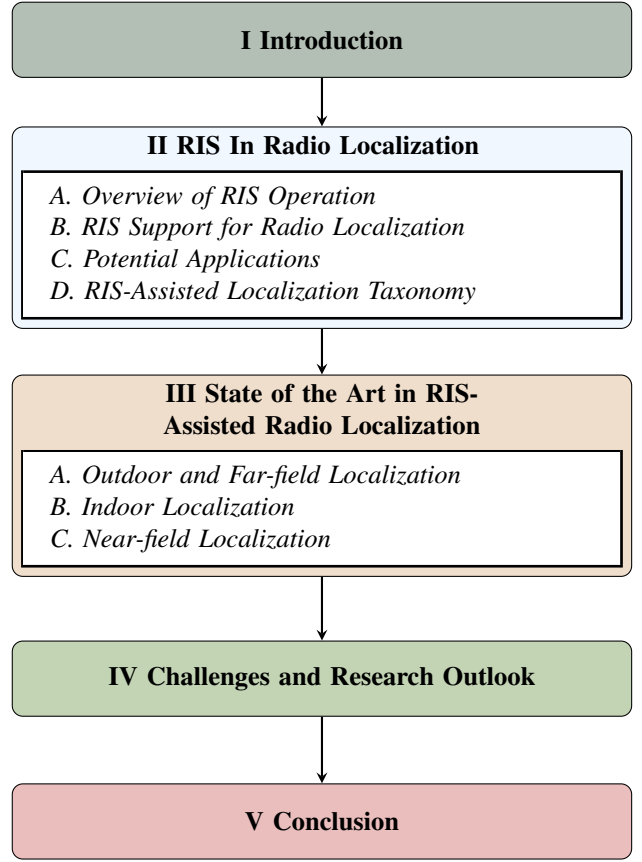
\end{scriptsize}

In the subsequent sections, we present the results of our review. In Section \ref{sec3}, we give the comprehensive background of \acrshort{ris} technology and discuss it from the perspective of its potential for radio localization. In Section \ref{sec4}, the developments in \acrshort{ris}-assisted radio localization are consolidated in terms of research in various frequency bands, deployment scenarios, and \acrshort{ris} placement for enhanced localization. In Section \ref{sec5}, we outline the limitations and unexplored research directions of \acrshort{ris} for localization in 6G networks, followed by conclusions in Section \ref{sec6}. The sections and main topics of this article are shown in Figure \ref{outline}. A list of definitions of frequently used abbreviations is given in Table \ref{abb}.

\section{\acrshort{ris} In Radio Localization}
\label{sec3}
In this section, we briefly discuss the \acrshort{ris}  and the different types of its working operations. We focus on the reflective \acrshort{ris} signal and channel modeling, as it is the most commonly researched type of \acrshort{ris} operation. We provide the brief overview of the system model and channel model for \acrshort{ris}-assisted radio localization networks to demonstrate how \acrshort{ris} is used in 6G radio localization. Subsequently, we highlight why \acrshort{ris} is an advantageous technology for 6G radio localization, followed by a discussion on the taxonomy of \acrshort{ris}-assisted localization. 

\subsection{Overview of \acrshort{ris} Operation}
\label{sec2}
\begin{figure}
    \centering
    \subfigure[Passive Reflective \acrshort{ris}]{
        \includegraphics[width=0.45\textwidth]{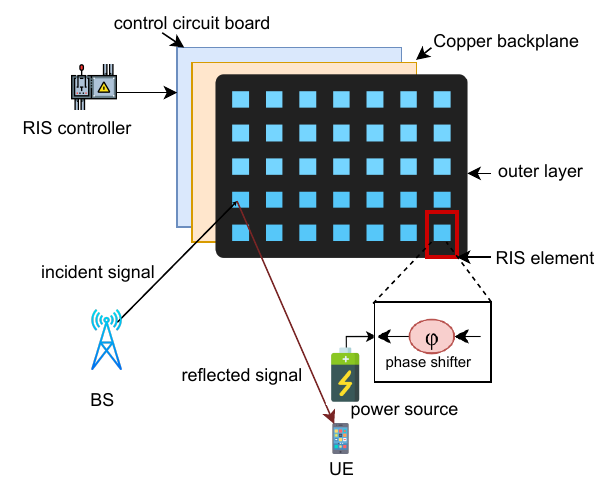}
        \label{fig:sub1}
    }
    \hspace{5mm}%\hfill % Space between figures
    \subfigure[Active Reflective RIS]{
        \includegraphics[width=0.45\textwidth]{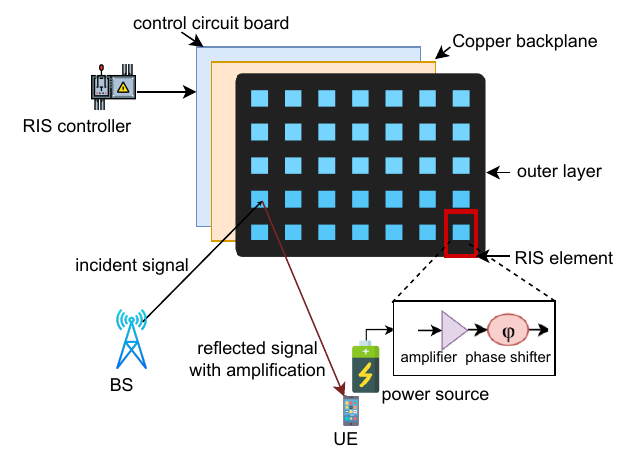}
        \label{fig:sub2}
    }
    \hspace{5mm}%\hfill %Space between figures
    \subfigure[Transmissive/STAR \acrshort{ris} which can be active or passive]{
        \includegraphics[width=0.45\textwidth]{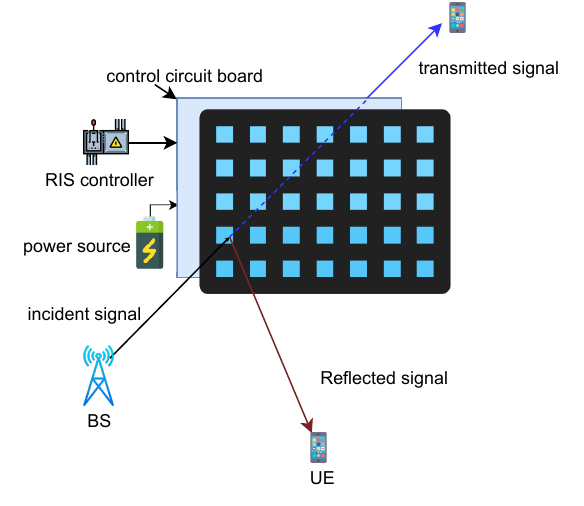}
        \label{fig:sub3}
    }
    \caption{Comparative structure of \acrshort{ris} under different modes of operation, (a) Passive Reflective \acrshort{ris} can alter the phase of the incident signal only, (b) Active Reflective \acrshort{ris} can amplify and alter the phase of the incident signal, (c) Transmissive \acrshort{ris} passes the signal through while STAR \acrshort{ris} can perform both transmission and reflection simultaneously. The reflection coefficient of each \acrshort{ris} element is reconfigurable in real time via the controller.}
    \label{ris}
\end{figure}

%\subsection{Fundamentals of RIS Operation}
\acrshort{ris} represents a novel electromagnetic surface capable of altering the behavior of wireless channel, paving the way for more efficient and adaptable wireless communication systems. A \acrshort{ris} comprises a multitude of tunable elements designed to regulate the amplitude, phase, and polarization of electromagnetic waves either passing through or reflecting off its surface \cite{Zhang2020TowardsUP}. These individual elements are generally small, cost-effective, and programmable to accommodate varying channel conditions and modulation methods.
%\acrshort{ris} were first introduced by Huanyang Zheng et al. in 2018 in \cite{8811733}. The authors proposed using an RIS to enhance the wireless network's performance by jointly optimizing the active and passive beamforming weights. 
The structure of \acrshort{ris} operating across different modes is depicted in Figure \ref{ris}. 

The \acrshort{ris} operates by modifying the propagation of electromagnetic waves by reflecting, refracting, or scattering them \cite{9215972}. Typically, the tunable elements within \acrshort{ris} are diminutive antennas or resonators that can be electrically manipulated to adjust their electromagnetic properties, including resonant frequency, impedance, and polarization \cite{8466374}. By altering the impedance of each tunable element, the \acrshort{ris} can govern the amplitude and phase of the reflected or transmitted wavefront, effectively directing the wave towards a desired direction or concentrating it on a specific location \cite{9330512}. Entrusted with the role of tweaking the electrical parameters of these tunable elements in real-time, the \acrshort{ris} controller uses feedback from the wireless channel conditions, modulation scheme, and performance targets. It employs feedback from the receiver to fine-tune the settings of the tunable elements, optimizing signal quality while curbing interference and noise \cite{chepuri2022integrated}. Various hardware and software technologies, ranging from analog, digital, or hybrid to software-defined or \acrfull{ai}-based solutions, can implement the \acrshort{ris} controllers \cite{8466374}. The selection of a controller hinges on the \acrshort{ris}'s application, complexity, and performance prerequisites \cite{8910627}.

Here, we briefly discuss some of the \acrshort{ris} working operations:
\begin{itemize}
  \item \textit{Reflective \acrshort{ris}}: This is the most common type, shown in Figure \ref{fig:sub1}, where elements of the surface can alter the phase of the incident signal. These essentially act as programmable mirrors that shape and direct the radio waves toward a specific direction \cite{9424177}. Each \acrshort{ris} element reflects the incoming signal due to the copper backplane \cite{9365009}. 

    An \acrshort{ris} can reflect electromagnetic signals independently in reflection mode using its $N$ reflection units. The magnitude, $\alpha_{n} \in [0,1]$, and phase, $\phi_n \in [0,2\pi)$, of the reflection coefficient of each reflection unit are reconfigurable via the controller. This leads to a baseband signal model of, $\mathbf{y}_n = \alpha_{n}e^{j\phi_n}\mathbf{x}_n,$  for each unit, where $\mathbf{x}_n$ is the incident signal and $\mathbf{y}_n$ is the reflected signal, respectively. For the entire \acrshort{ris} surface, the relationship between the incident and reflected signals can be represented by a diagonal matrix, as the reflection units are independent, given as,  $\mathbf{y} = \text{diag} (\alpha_1e^{j\phi_1} ,...,\alpha_1e^{j\phi_n} ,...,\alpha_1e^{j\phi_N})\mathbf{x} = \boldsymbol{\Omega} \mathbf{x}, $ where $\boldsymbol{\Omega}$ is the reflection coefficient matrix of \acrshort{ris}. The \acrshort{ris} is designed to reflect incident signals maximally, i.e., ideally $\alpha =1$. However, $\alpha$ in practice may not be equal to 1. It is usually a constant with a value dependent on the specific circuit \cite{ozturk2023ris}. The magnitude and phase,  $\alpha$ and $\phi$, can be varied within an interval with the limitation on cost and complexity. This leads to three practical reflection coefficient types: constant amplitude with continuous phase shift, optimized amplitude with continuous phase shift, and constant amplitude with discrete phase shift. Continuous phase shift is assumed in some papers, but it is limited by high hardware costs. Thus, a discrete phase shift is often used to increase cost-effectiveness. It is worth mentioning that the amplitude and phase control are not necessarily independent, i.e., when the phase is varied, this also varies the amplitude.
    \begin{figure*}[t!]
    \centering
    \includegraphics[width=0.7\textwidth]{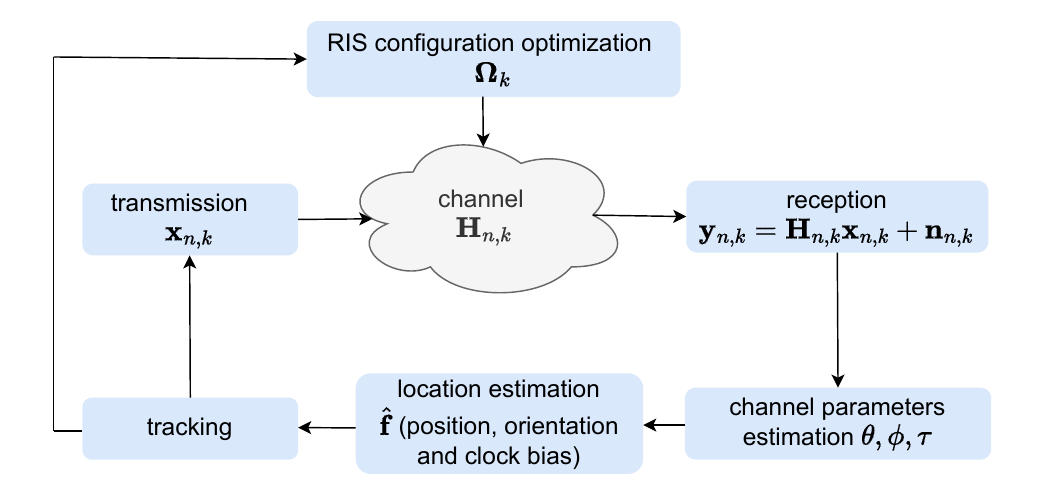}
    \caption{Flowchart demonstrating the typical processing phases in RIS-assisted localization in 6G networks. In the downlink scenario, the transmitted signal received at the UE is used for channel parameters estimation and UE localization. Tracking information is utilized to optimize the RIS configuration. }
    \label{flowchart} 
\end{figure*}
    \item \textit{Transmissive \acrshort{ris}}:   This type allows signals to pass through the surface, modifying their characteristics in the process. This provides an additional degree of flexibility in controlling the wave propagation. Incident signal penetrates the \acrshort{ris} elements due to the absence of copper backplane as  shown in Figure \ref{fig:sub3} \cite{9365009}.

    \item \textit{Hybrid \acrshort{ris}}: Common \acrshort{ris} designs feature metasurfaces made of passive meta-atoms (building blocks of the meta materials) that can reflect incoming waves in adjustable ways. However, this exclusive reflection method poses considerable coordination challenges in wireless networks. For instance, \acrshort{ris} don't possess the needed data to modify their reflection patterns independently; this data must be gathered by other network components and then relayed to the \acrshort{ris} controller. Moreover, gauging the communication channel, vital for coherent \acrshort{ris}-aided communication, is problematic when using existing \acrshort{ris} models. Hybrid Reflecting and Sensing \acrshort{ris} offer a solution by allowing metasurfaces to not only adjustably reflect the incoming signal but also sense a fraction of it \cite{alexandropoulos2023hybrid}. This sensing ability of hybrid \acrshort{ris} supports many network management tasks, like estimating channel parameters and pinpointing locations, paving the way for potentially self-regulating and self-setting metasurfaces.
    
    %This type combines both reflective and transmissive properties, capable of both reflecting and transmitting signals. Just like the transmissive variant, the hybrid type of RIS also involves the removal of the copper backplane. However, adjustments made to the RIS structures result in different induced currents within the surface. This difference leads to both the reflection and transmission of the incident signals \cite{9365009}. In this context, it is typically assumed that the energy is equally distributed for both transmission and reflection. 

    \item \textit{\acrfull{star} \acrshort{ris}}: This variant allows the \acrshort{ris} to perform both transmission and reflection simultaneously, making it highly efficient and versatile for various communication needs \cite{9437234}. Conventional \acrshort{ris}, due to their hardware design, are capable of only reflecting incident signals, serving wireless devices situated on the same side. This restricts their deployment adaptability and coverage span \cite{8811733, 8930608}. To overcome these limitations, a new type of metamaterial called \acrshort{star}-\acrshort{ris} has been introduced \cite{Liu2021STARST, 9570143}. It supports both electric-polarization and magnetization currents, enabling it to reflect and/or transmit the incident signals \cite{news}. In contrast to conventional \acrshort{ris}, it can offer full-space service coverage (i.e., 360 degrees), leading to enhanced deployment flexibility.

    \item \textit{\acrshort{ris} with Non-Diagonal Control}: In conventional \acrshort{ris} structures, it is assumed that a signal hitting a specific element can only be reflected from that same element after the phase shift adjustment. There was no deliberate association between the \acrshort{ris} elements. The phase shift matrix in such designs was diagonal, such that each \acrshort{ris} element is connected to the load disassociated from the other elements on the surface, leaving untapped potential for system performance enhancement through \acrshort{ris}. On the contrary, \acrshort{ris} with non-diagonal control has a design based on non-reciprocal connections, allowing the signal impinging on one element to be reflected from a different element after phase shift adjustment \cite{li2022reconfigurable}. Consequently, the phase shift matrix can be non-diagonal. This allows for greater adaptability in configuring the \acrshort{ris} structure to optimize system performance. They can increase reflected power, enhance aggregate data rate, and provide versatility in a variety of deployment scenarios.

    \item \textit{Active \acrshort{ris}} \label{activeris}: In the case of passive \acrshort{ris}, the path loss between the transmitter-\acrshort{ris}-receiver connection is determined by multiplying, rather than adding, the path losses of the transmitter-\acrshort{ris} and \acrshort{ris}-receiver connections. This value is typically many times greater than the direct link's path loss \cite{najafi2020physics}. Consequently, this ``multiplicative fading" phenomenon often renders it highly challenging for passive \acrshort{ris} to realize significant capacity gains in numerous wireless settings \cite{9833921}. It is, thus, a significant performance hindrance to passive \acrshort{ris} operation \cite{9998527}. Active \acrshort{ris} was introduced as a solution. Like its passive counterpart, it can reflect incident signals with adjustable phase shifts, but it can also amplify these signals, as shown in Figure \ref{fig:sub2}. Its hardware architecture is, thus, different from the passive \acrshort{ris} such that its design involves reflection-type amplifiers in addition to the phase shift circuits. Active \acrshort{ris} needs considerably more power to operate \cite{article}.
\end{itemize} 
\subsection{\acrshort{ris}-Assisted Localization in 6G Networks}
    \begin{figure*}[t!]
    \centering
    \subfigure[Conventional non-\acrshort{ris}-assisted localization network. ]{
    \includegraphics[width=0.46\textwidth]{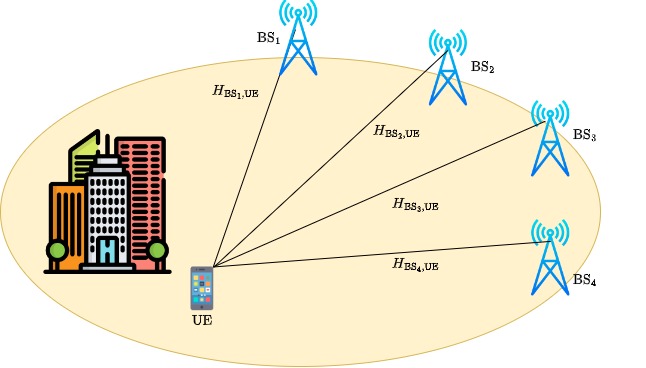}
    \label{fig31}}
    \hfill
    \subfigure[\acrshort{ris}-assisted localization network. ]{
    \includegraphics[width=0.5\textwidth]{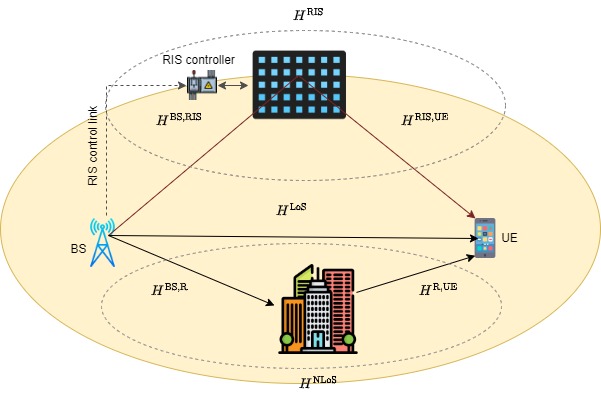}
        \label{fig:my_label}}
    \caption{Illustration of non-\acrshort{ris}-assisted and \acrshort{ris}-assisted localization networks. Localization without \acrshort{ris} requires multiple BSs. The use of \acrshort{ris} makes localization possible with lesser infrastructure with the added advantage of energy efficiency, minimal deployment, and maintenance cost.}
    \label{fig:overall_fig} 
\end{figure*}
In this subsection, we present a fundamental example for RIS-assisted radio localization, providing a foundational framework upon which more advanced and specialized methods can be developed to address a range of complex problems. The flowchart presenting the typical processing phases is shown in Figure \ref{flowchart}. To further illustrate the functioning of  \acrshort{ris} in collaboration with \acrshort{bs} for \acrshort{ue} localization, consider the scenario showing the \acrshort{ris}-assisted radio localization environment in Figure \ref{fig:my_label} where a multiantenna BS equipped with $N_{\text{BS}}$ antennas is located at $\mathbf{p}_{BS}= (x_{\text{BS}}, y_{\text{BS}}, z_{\text{BS}}$). \acrshort{ris} has $N_{\text{RIS}}$ reflecting elements with its center located at ($x_{\text{RIS}}, y_{\text{RIS}}, z_{\text{RIS}}$) and the UE with $N_{\text{UE}}$ antennas is at $\mathbf{p}_{UE}= (x_{\text{UE}}, y_{\text{UE}}, z_{\text{UE}}$). In the context of localization, the UE has an unknown state, i.e., location, orientation and clock bias that needs to be determined based on received downlink signals. The BS, with a known state transmits signals on distinct subcarriers. These signals create observations \( \mathbf{y}_{n,k} \) at the UE receiver over the channel \( \mathbf{H}_{n,k} \). By analyzing these downlink observations, the UE can estimate its location based on the known location of the BS and the \acrshort{ris}.
\subsubsection{Signal Model at UE}
We consider a generic frequency domain representation of the received signal and channel model for $N$ samples spaced $\Delta f $ apart \cite{9893187}. The received signal at the UE for frequency $n \in \{0,...., N-1\}$ and symbol $k \in \{0,...., k-1\}$ can be represented as, 
\begin{equation}
\centering
    \mathbf{y}_{n,k} = \mathbf{H}_{n,k}\mathbf{x}_{n,k} + \mathbf{n}_{n,k},
\end{equation}
where $\mathbf{y}_{n,k}$ is the received pilot signal at the UE, $\mathbf{x}_{n,k}$ is the transmitted pilot signal from the BS, and $\mathbf{n}_{n,k}$ is the additive Gaussian noise. 
\subsubsection{Channel Model}
Two types of propagation paths are available from the BS to the UE in this scenario, i.e., a direct, \acrfull{los}, path from the BS to the UE and reflected, \acrshort{nlos}, paths through the \acrshort{ris} or other objects in the environment.   
In the illustrated downlink scenario, considering a far-field channel with planar wave assumptions (as opposed to the spherical wave structure of near-field channels), the combined channel response, encompassing both the direct and NLOS reflected paths, is given as, 
\begin{equation}
\label{eq2}
\centering
    \mathbf{H}_{n,k}= \mathbf{H}_{n,k}^{\text{direct}} + \mathbf{H}_{n,k}^{\text{NLoS}}.
\end{equation}
%The $\mathbf{H}_{n,k}= \mathbf{H}_{n,k}^{\text{direct}} + \mathbf{H}_{n,k}^{\text{RIS}}$, is the total channel response of the direct and \acrshort{ris} reflected path. 
%\paragraph{\textcolor{blue}{LoS Channel Response}}
Here, $\mathbf{H}_{n,k}^{\text{direct}} \in \mathbb{C}^{N_{\text{BS}} \times N_{\text{UE}}}$ is the channel frequency response of the direct path, given as \cite{9893187}
\begin{equation}
\centering
    \mathbf{H}_{n,k}^{\text{direct}} = \sum_{l=1}^{L} \alpha_{l} \mathbf{a}_{\text{UE}}(\mathbf{\theta}_{l}) \mathbf{a}_{\text{BS}}(\mathbf{\phi}_{l}) e^{-j2\pi n \Delta f \tau_{l}},
\end{equation}
%$\mathbf{H}_{n,k}^{\text{direct}} = \sum_{l=1}^{L} \alpha_{l} \mathbf{a}_{\text{UE}}(\theta_{l}) \mathbf{a}_{\text{BS}}(\phi_{l}) e^{-j2\pi n \Delta f \tau_{l}}$, 
where 
\begin{equation}
    \centering
    \left| \alpha_{l} \right|^2 = \frac{\lambda^2}{(4 \pi)^2} \frac{G_{\text{UE}} \left( \mathbf{\theta}_l \right) G_{\text{BS}} \left( \mathbf{\phi}_l \right)}{\left\| d_{BS-UE} \right\|^2},
\end{equation}
is the complex channel gain with $L$ being the number of direct signal propagation paths, $\mathbf{a}_{\text{UE}}(\mathbf{\theta}) \in \mathbb{C}^ {N_{\text{UE}}}$ is the UE array response as the function of \acrfull{aoa} $\mathbf{\theta} \in \mathbb{R}^ {2}$ in azimuth and elevation, $\mathbf{a}_{\text{BS}}(\mathbf{\phi})\in \mathbb{C}^ {N_{\text{BS}}}$ is the BS array response as function of \acrfull{aod} $\mathbf{\phi} \in \mathbb{R}^ {2}$ in azimuth and elevation, $ \lambda $ is the carrier wavelength,  \( G_{\text{BS}}(\cdot) \) and \( G_{\text{UE}}(\cdot) \) are the antenna gains at the BS and UE, and $\tau_l$ is the \acrfull{toa} given by 
\begin{equation}
    \centering
    \tau_l= \frac{d_{BS-UE}}{c} + K
\end{equation}
where $d_{BS-UE}$ is the distance from BS antenna element to the UE antenna element, $c$ is the speed of light and $K$ is the clock synchronization offset.
%\textcolor{blue}{The paramters $\theta$, $\phi$ and $\tau$ contain the required geometric information to determine the UE location.  }
%$\nu$ is the Doppler shift and $T_s$ is the duration of the symbol \cite{9893187}. 
%\paragraph{\textcolor{blue}{NLoS Channel Response}}

From \eqref{eq2}, $\mathbf{H}_{n,k}^{\text{NLoS}}$ is the combination of reflected signal channel frequency response at the UE from the intended \acrshort{ris} reflectors as well as the random obstacles in the environment. It can, thus, be expressed as, 
\begin{equation}
\label{eq5}
    \centering
    \mathbf{H}_{n,k}^{\text{NLoS}}=  \mathbf{H}_{n,k}^{\text{RIS}} + \sum_{i=1}^{I}\mathbf{H}_{n,k}^{\text{NLoS}_{(i)}}, 
\end{equation}
where $I$ is the number of reflectors between BS and UE, other than \acrshort{ris}. The channel response from such reflectors is defined as \cite{Trevlakis2023LocalizationAA}
\begin{equation}
\centering
    \mathbf{H}_{n,k}^{\text{NLoS}} = \sum_{i=1}^{I} \alpha_{i} \mathbf{a}_{\text{UE}}(\mathbf{\theta}_{i}) \mathbf{a}_{\text{BS}}(\mathbf{\phi}_i)\mathbf{a}_{\text{R}} e^{-j2\pi n \Delta f \tau_{i}},
\end{equation}
%    \mathbf{H}_{n,k}^{\text{NLoS}} = \sum_{i=1}^{I} \alpha_{i} \mathbf{a}_{\text{UE}}(\theta_{i}) \mathbf{a}_{\text{R}}(\phi_{i}) e^{-j2\pi n \Delta f \tau_{i}},
such that 
\begin{equation}
    \centering
    \left| \alpha_{i} \right|^2 = \frac{\lambda^2}{(4 \pi)^2} \frac{\Gamma_{R_{i}-UE}G_{\text{UE}} \left( \mathbf{\theta}_i \right) G_{\text{BS}} \left( \mathbf{\phi}_i \right)}{ (d_{BS-R_i}+ d_{R_{i}-UE})^2},
\end{equation}
and 
\begin{equation}
    \centering
    \tau_i= \frac{d_{BS-R_i}+ d_{R_{i}-UE}}{c} + K,
\end{equation}
where $d_{BS-R_i}$ and $d_{R_{i}-UE}$ are the distances from BS to reflector $R_i$ and from $R_i$ to UE, $\mathbf{a}_{\text{R}}$ is the steering vector of the reflector R and $\Gamma_{R_{i}-UE}$ is the attenuation coefficient. Given the point of incidence of signal on a reflector  be \( \mathbf{p}_{\text{inc},m}\in \mathbb{R}^3\), the UE array response \( \mathbf{a}_{\text{UE}}(\mathbf{\theta}_i) \) (similarly for BS array response vector \( \mathbf{a}_{\text{BS}}(\mathbf{\phi}_i) \) consists of entries given by \cite{9893114},  
\begin{equation}
    \centering
    [\mathbf{a}_{\text{UE}}(\mathbf{\theta}_i)]_k = \exp\left(j\frac{2\pi{(\mathbf{p}_{\text{UE},k} - \mathbf{p}_{\text{UE}})^\top \mathbf{u}(\mathbf{\theta}_i) }}{{\lambda}}\right),
\end{equation}
where \( \mathbf{p}_{\text{UE},k} - \mathbf{p}_{\text{UE}} \) represents the position of the \( k \)-th antenna element relative to the reference UE location, and \( \mathbf{u}(\mathbf{\theta}_i) = {\mathbf{p}_{\text{inc},m} - \mathbf{p}_{\text{UE}}}/{\| \mathbf{p}_{\text{inc},m} - \mathbf{p}_{\text{UE}} \|} \), all defined in the UE's frame of reference. These NLoS signals become increasingly important when the LoS between the BS and UE is absent.

The $\mathbf{H}_{n,k}^{\text{RIS}}$ in  \eqref{eq5} is the RIS incident and reflected signal channel response such that $\mathbf{H}_{n,k}^{\text{BS,RIS}}\in \mathbb{C}^{N_{\text{BS}} \times N_{\text{RIS}}}$ is the channel response of the path from the BS to the \acrshort{ris} and $\mathbf{H}_{n,k}^{\text{RIS,UE}}\in\mathbb{C}^{N_{\text{RIS}} \times N_{\text{UE}}}$ is the channel response from the \acrshort{ris} to the UE, collectively given as \cite{9893114}, 
\begin{align}
            \mathbf{H}_{n,k}^{\text{RIS}} &= \mathbf{H}_{n,k}^{\text{BS,RIS}}\mathbf{H}_{n,k}^{\text{RIS,UE}}\\
     &=\alpha^{\text{RIS}}_k \mathbf{a}_{\text{UE}}(\mathbf{\theta}_{\text{RIS}}) \mathbf{a}_{\text{BS}}(\mathbf{\phi}_{\text{RIS}}) e^{-j2\pi n \Delta f \tau_{\text{RIS}}},
\end{align}
where 
\begin{equation}
    \centering
    \alpha_{k}^{\text{RIS}} = \alpha_{\text{BS-RIS}} \alpha_{\text{RIS-UE}} \mathbf{a}^{\top}_{\text{RIS}}(\mathbf{\phi}_{\text{RIS-UE}}) \boldsymbol{\Omega}_{k} \mathbf{a}_{\text{RIS}}(\mathbf{\theta}_{\text{BS-RIS}}).
\end{equation}
Here, $\alpha_{k}^{\text{RIS}}$ is controllable, $\alpha_{\text{BS-RIS}}$ is the complex gain from BS to \acrshort{ris}, $\alpha_{\text{RIS-UE}}$ is the complex gain from \acrshort{ris} to the UE, $\mathbf{a}_{\text{RIS}}(.)$ is the \acrshort{ris} response function as the function of \acrshort{aoa} from BS, $\mathbf{\theta}_{\text{BS-RIS}}$, and the \acrshort{aod} to the UE, $\mathbf{\phi}_{\text{RIS-UE}}$ and $\boldsymbol{\Omega}_{k}$ 
determines the \acrshort{ris} configuration \cite{9893114}. A \acrshort{ris} with a known location can provide valuable geometric information, such as \acrshort{toa}, an angle at the UE, and an angle at the RIS. Essentially, the RIS operates as an auxiliary, synchronized BS with a phased array, transmitting the same signal as the primary BS \cite{9893114}. The RIS configurations $\boldsymbol{\Omega}_{k}$ can be optimized based on prior location information. 
\subsubsection{UE Localization}
For the localization of the UE, its position, orientation, and clock bias information are inferred from the received signal $\mathbf{y}_{n,k}$, details of which are well summarized in \cite{9893187, 9893114} and discussed briefly below. The process encompasses three steps: first, the channel parameters (\acrshort{toa}s, \acrshort{aoa}s, \acrshort{aod}s) are estimated. Second, if the  \acrshort{los} is available then its parameters, \acrshort{nlos} and \acrshort{ris} path parameters are extracted, and finally, the \acrshort{ue} is localized.
\paragraph{Channel Parameter Estimation}
In wireless communication, channel estimation methods like FFT/Periodograms \cite{braun2014ofdm}, ESPRIT \cite{roemer2014analytical}, and orthogonal matching pursuit \cite{8240645} use principles of sparsity or harmonic retrieval \cite{9893187}. A common approach is to first obtain a least-squares channel estimate \( \hat{\mathbf{H}}_{n,k} \), which can then be vectorized for compressive sensing to estimate path counts and delays. %When path resolution is high, results align well with the \acrfull{crb}; otherwise, biases may occur. }
%
%\textcolor{blue}{The \acrshort{crb} sets a theoretical lower bound on estimator variance:
%
%\begin{equation}
 %   \centering
%    \text{Var}(\hat{\eta}) \geq \text{CRB}(\eta) = \mathbf{F}^{-1}(\eta)
%\end{equation}}
%
%\textcolor{blue}{where \( \mathbf{F}(\mathbf{\eta}) \) is the \acrfull{fim} and \( \mathbf{\eta} \) represents the vector of unknown parameters. }
%
Another way is to arrange least-square estimates of channel parameters in tensor form which allows multi-dimensional harmonic retrieval. Initial path estimates, gains, and geometry can be refined by maximizing the likelihood function \( \log p(\textbf{y} | \kappa) \), where \( \kappa \) includes parameters like path gains, angles, and delays. 
%This refinement yields estimates near the CRB, with uncertainty given by the inverse \acrshort{fim} \( \boldsymbol{\Sigma}(\hat{\mathbf{\eta}}) \).}
%
\paragraph{Location Estimation}
To estimate the UE's location, the channel parameters          \( \hat{\boldsymbol{\eta} }\) 
(e.g., angles, delays), whose estimator uncertainties need to be known, 
%\( \boldsymbol{\Sigma}(\hat{\mathbf{\eta}}) \) 
are mapped to the UE state \( \mathbf{f} \) through the relationship \( \hat{\mathbf{\eta}} = \mathbf{h}(\mathbf{f}) + \mathbf{n} \), where  \( \mathbf{h}(\mathbf{f}) \) represents the geometric model linking the UE state to channel parameters $(\tau, \theta, \phi)$ and \( \mathbf{n} \) represents measurement noise with statistics that depend on the channel parameter estimation. The parameter \( \hat{\mathbf{\eta}} \) provides estimated angles, and delays with associated noise. We estimate \( \mathbf{f} \) by minimizing the cost function \cite{9893187} given by  
\begin{equation}
    \centering
    \hat{\mathbf{f}} = \arg \min_\mathbf{f}  \left( \hat{\eta} - \mathbf{h}(\mathbf{f}) \right)^\top \boldsymbol{\Sigma}^{-1}(\hat{\boldsymbol{\eta}}) \left( \hat{\boldsymbol{\eta}} - \mathbf{h}(\mathbf{f}) \right).
\end{equation}
After obtaining an initial estimate, local optimization refines \( \hat{\mathbf{f}} \). The uncertainty, \( \boldsymbol{\Sigma}(\hat{\mathbf{f}}) \), is then calculated as
\begin{equation}
    \centering
    \Sigma^{-1}(\hat{\mathbf{f}}) = \left( \frac{\partial \mathbf{\eta}}{\partial \mathbf{f}} \right)^\top \boldsymbol{\Sigma}^{-1}(\hat{\mathbf{\eta}}) \frac{\partial \mathbf{\eta}}{\partial \mathbf{f}} \Big|_{\mathbf{f}=\hat{\mathbf{f}}}.
\end{equation}
This estimated location and uncertainty, \( (\hat{\mathbf{f}}, \boldsymbol{\Sigma}(\hat{\mathbf{f}})) \), can be further integrated into tracking algorithms for more robust localization and RIS configuration optimization.
\subsubsection{Extensions}
The aforementioned channel and signal modeling represent the most simple version of the RIS-assisted localization problem, setting the stage for many extensions. For instance, the channel model can be generalized to include wavefront curvature, the RIS model can be generalized to hybrid, STAR, or Non-Diagonal Control RIS, each with their own localization benefits and challenges.  
%, where the array response can be expressed as 
%
%\begin{equation}
%    \centering
%[\mathbf{a}_{\text{UE}}(\mathbf{p}_{\text{inc},m})]_k = \exp\left(-j \frac{2 \pi (d_i - d_{\text{ref}})}{\lambda}\right),
%\end{equation}}
%
%\textcolor{blue}{where \( d_{\text{ref}} = \|\mathbf{p}_{\text{inc,m}} - \mathbf{p}_{\text{UE}}\| \) is the distance from the source to the UE array’s phase reference, and \( d_i = \|\mathbf{p}_{\text{inc,m}} - \mathbf{p}_{\text{UE},k}\| \) is the distance from the source to the \( k \)-th element of the array \cite{10529957, 9893187}. Accounting for this wavefront curvature opens up new possibilities for enhancing localization performance, discussed in detail in Section \ref{nearfield}.

\subsection{\acrshort{ris} Support for Radio Localization}
In a standard \acrshort{siso} non-\acrshort{ris}-assisted localization network, shown in Figure \ref{fig31}, four synchronized single antenna \acrshort{bs}s transmit pilots, generating \acrfull{tdoa} or \acrshort{toa}. The UE position is estimated by intersecting hyperboloids (\acrshort{tdoa}s) or calculating \acrshort{toa}s for clock bias synchronization \cite{10044963}. Using \acrfull{rtt} with three \acrshort{bs}s allows localization by intersecting spheres centered at the \acrshort{bs}s. The \acrshort{ue}'s velocity is determined from four Doppler shifts. The limitations of such conventional localization include: (i) the need for several \acrshort{bs}s for localization, whereas only one is needed for communication, leading to network over-provisioning to support localization systems; (ii) even massive \acrshort{mimo} \acrshort{bs}s offer limited angular resolution, making localization constrained by multipath; and (iii) time-based measurements, i.e., \acrshort{toa} and \acrshort{tdoa}, which provide the most information for localization, require \acrshort{bs}s to be time-synchronized at the sub-nanosecond level. In the following subsection, we show how the inclusion of RIS can address these limitations by reducing reliance on infrastructure, offering high angle resolution, and eliminating extreme synchronization requirements. We delve into the multifaceted role of RIS in enhancing radio localization capabilities, examining how RIS transforms cellular networks into intelligent environments, how it works in tandem with \acrshort{bs} for effective \acrshort{ue} localization, and exploring applications of RIS-assisted localization in emerging technologies.

\subsubsection{RIS-Enhanced Localization in Smart Radio Environments}
%Advancements in RIS Technology for Enhanced Cellular Network Localization
Next-generation cellular networks are evolving towards ``smart radio environments" enhanced by \acrshort{ris}-covered walls and objects, enabling them to reconstruct and modify radio signal properties like transmission direction and polarization. This innovation leads to an intelligent transmission environment, offering new possibilities in communication, sensing, and localization \cite{8466374, 9247315, 8108330}. The integration of \acrshort{ris} promises improved localization accuracy and extended coverage, contingent upon the development of suitable models and algorithms \cite{9215972}. Compared to traditional \acrshort{mimo} systems, \acrshort{ris} offers superior localization due to its large surface area, which improves its signal transmitting, receiving, or reflecting capabilities. The \acrfull{crlb}, a measure of the lower limit of the variance for an unbiased estimator, for \acrshort{ue} localization decreases as the \acrshort{ris} area increases, barring certain positions \cite{8288263, 8264743}. Distributed \acrshort{ris} deployments are found to enhance \acrshort{crlb} and broaden localization coverage. Further, \acrshort{ris} allows for high-precision radio localization and sensing, significantly estimating \acrshort{ue} and device positions \cite{8815412}. The \acrshort{ris}-aided downlink localization offers better coverage and accuracy than traditional reflecting surfaces and scatter points \cite{9148744}. These advancements highlight the role of \acrshort{ris} in enhancing radio localization systems.
\FloatBarrier
\begin{figure*}
        \centering
            \subfigure[Collaboration among humans and robots in an intelligent factory]{
        \includegraphics[width=0.3\textwidth]{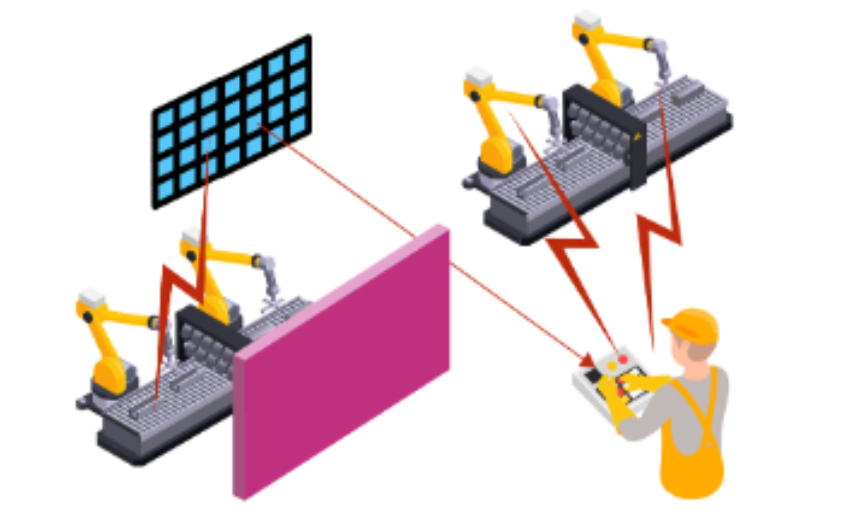}
        \label{fig:sub73}
    }
    %\hspace{2mm}
    \hfill % Space between figures
        \subfigure[Traffic monitoring and landscape sensing for an immersive smart city]{
        \includegraphics[width=0.3\textwidth]{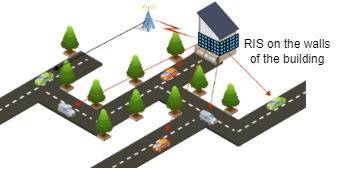}
        \label{fig:sub77}
    }
    %\hspace{2mm}%
    \hfill %Space between figures
        \subfigure[Local coverage/infrastructure less network extensions]{
        \includegraphics[width=0.3\textwidth]{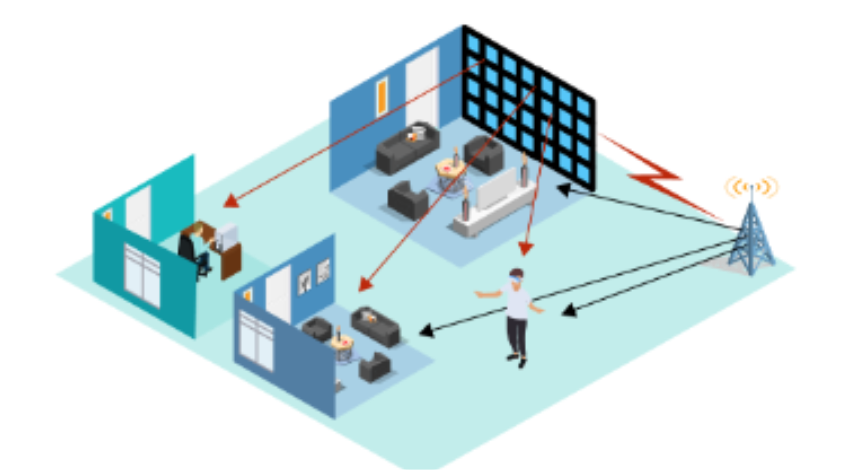}
        \label{fig:sub75}
    }
    %\hspace{2mm}%
    \hfill %Space between figures
        \subfigure[Telepresence and multisensory extended reality for enhanced interactions ]{
        \includegraphics[width=0.3\textwidth]{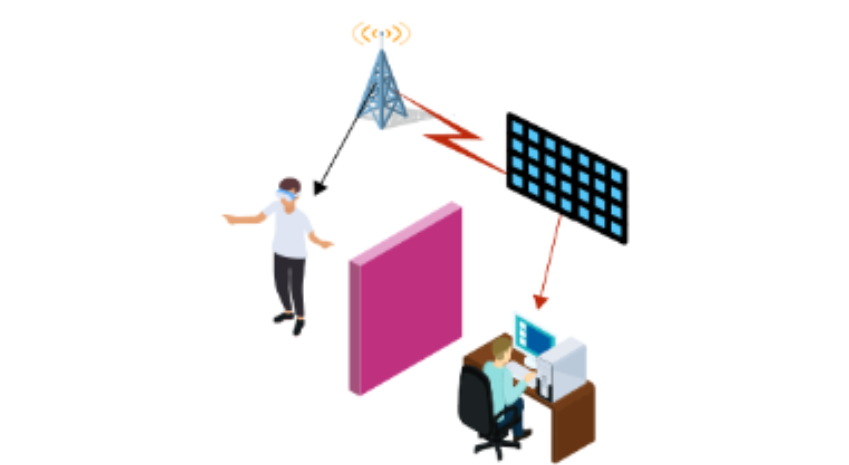}
        \label{fig:sub76}
    }
    %\hspace{2mm}%
    \hfill %Space between figures
    \subfigure[Interactive IoT networks for local trust zones for humans and machines]{
        \includegraphics[width=0.3\textwidth]{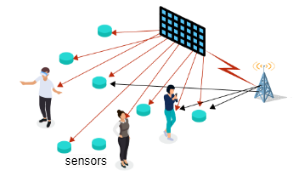}
        \label{fig:sub74}
    }
    %\hspace{2mm}%
    \hfill %Space between figures
    \subfigure[Wireless brain-computer interfaces for patient tracking and monitoring in local trust zones for human and machines]{
        \includegraphics[width=0.3\textwidth]{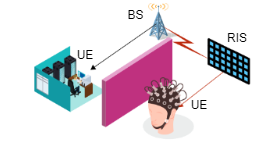}
        \label{fig:sub71}
    }
    %\hspace{2mm}%
    \hfill %Space between figures
        \subfigure[Telecontrol and gesture recognition in human-machine interface for enhanced interactions]{
        \includegraphics[width=0.3\textwidth]{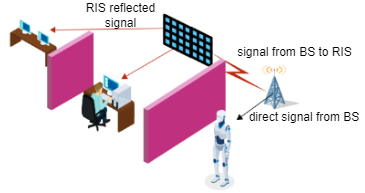}
        \label{fig:sub72}
    }
    \caption{Illustration of \acrshort{ris}-assisted localization applications in 6G networks from the use case families in \cite{hexa2}. Use cases prominently include sustainable development, massive twinning, telepresence, robots to collaborative robots, and local trust zones.}
    \label{figapps}
\end{figure*}
Accurately quantifying phase and amplitude is essential for effective localization with \acrshort{ris}, but full-resolution measurements can be expensive. Exploring the impact of phase and amplitude quantization on localization is therefore important. The comparison of full-resolution measurements with varying quantization resolutions shows that phase quantization has a more significant effect on the \acrshort{crlb} of localization than amplitude quantization, but both do not greatly differ in \acrshort{crlb} loss \cite{8264743, 9048973}. This insight is pivotal for practical \acrshort{ris}-assisted localization, suggesting enhanced phase resolution at the \acrshort{ris} for better accuracy. 
%Theoretical investigations into the CRLB of location estimates have been conducted using the LOS propagation assumption in \acrshort{ris}-assisted UE localization \cite{8288263, 8264743, 9048973}.}
%Additionally, a novel online wireless \acrshort{ris} phase design method, employing deep learning, has been proposed to optimize RSS at the UE's projected position \cite{8815412}.
Utilizing explicit geometric information in wireless channels below 6 GHz is challenging due to limited delay and angle resolution and weak path connectivity to environment geometry. In contrast, frequencies above \acrshort{mmw} link more closely to environment geometry and offer better resolution \cite{9893187, 7426565}. The high bandwidth and large antenna arrays in \acrshort{mmw} and \acrshort{thz} bands provide high spatial and temporal resolution, leading to a focus on these higher frequency bands in the majority of the studies in
contemporary literature \cite{8240645, 8752016, 8647451, 8515231, 9129075}. Moreover, often indoor scenarios are considered due to the limited transmission range of \acrshort{mmw} or \acrshort{thz} frequencies. While \acrfull{gnss} can offer acceptable outdoor localization, its effectiveness is reduced indoors where signal strength is low. In such environments, \acrshort{ris} not only aids in precise localization but also helps alleviate communication congestion caused by obstructions.

When considering algorithms for \acrshort{ris}-assisted localization, it is important to distinguish between far-field and near-field assumptions \cite{9215972}. In the far-field, energy travels away from the source, and the plane wave assumption is valid, while in the near-field, energy is periodically stored and returned to the source, with radiation patterns varying based on distance from the source, hence the spherical wave model applies. The near-field region's extent in \acrshort{ris}-assisted networks is linked to the \acrshort{ris}'s surface area. Near-field propagation, characterized by wavefront curvature, becomes significant at moderate distances to the \acrshort{ris} and must be accurately represented in the communication system. The near-field area of \acrshort{ris} inversely correlates with the wavelength of the incident signal, and its size increases with both operation frequency and surface area, making the \acrshort{ue} likely to be in this region. Therefore, particularly in indoor environments at mmW and THz frequencies, far-field models are generally inapplicable \cite{9140329}.

\subsubsection{Applications of RIS-Assisted Localization in Emerging Technologies}
\acrshort{ris}-assisted localization offers significant benefits for various applications requiring high accuracy and low latency. It revolutionizes indoor and outdoor localization in IoT networks, smart cities, and automated factories, as shown in Figure \ref{figapps} \cite{hexa2}. \acrshort{ris} enhances IoT network performance by optimizing signal propagation in different communication scenarios, contributing to advanced multisensory extended reality applications and tele-control technologies by ensuring precise device tracking and location accuracy (Figures \ref{fig:sub76} and \ref{fig:sub72}) \cite{9874802}. It also plays a crucial role in wireless brain computer interfaces, improving signal quality and reliability for patient tracking and tele-surgery applications (Figure \ref{fig:sub71}) \cite{9148610, Cao2021SumRateMF}. In smart indoor services, \acrshort{ris} overcomes NLoS challenges and improves privacy protection, crucial for secure and efficient local coverage (Figure \ref{fig:sub75}). It is pivotal in smart transportation, enhancing autonomous driving and vehicle-to-vehicle communications by providing accurate \acrfull{3d} mapping and robust localization in dynamic environments (Figure \ref{fig:sub77}). For automated factories and connected robotics and autonomous systems (CRAS), \acrshort{ris} facilitates cooperative localization, essential for IoT device interaction and efficient production processes (Figure \ref{fig:sub73}) \cite{9369969, 9148610}. However, fulfilling these applications in 6G networks presents challenges as outlined in Section \ref{sec5}. The integration of
\acrshort{ris} into communication networks can significantly enhance localization and sensing performance. By intelligently manipulating
wireless signals, \acrshort{ris} can optimize signal paths, minimize delay, and improve accuracy, thereby enabling intelligent
interactions across various applications.
\subsection{\acrshort{ris}-Assisted Localization Taxonomy}
\begin{figure*}[t!]
    \centering
    \includegraphics[scale=0.8]{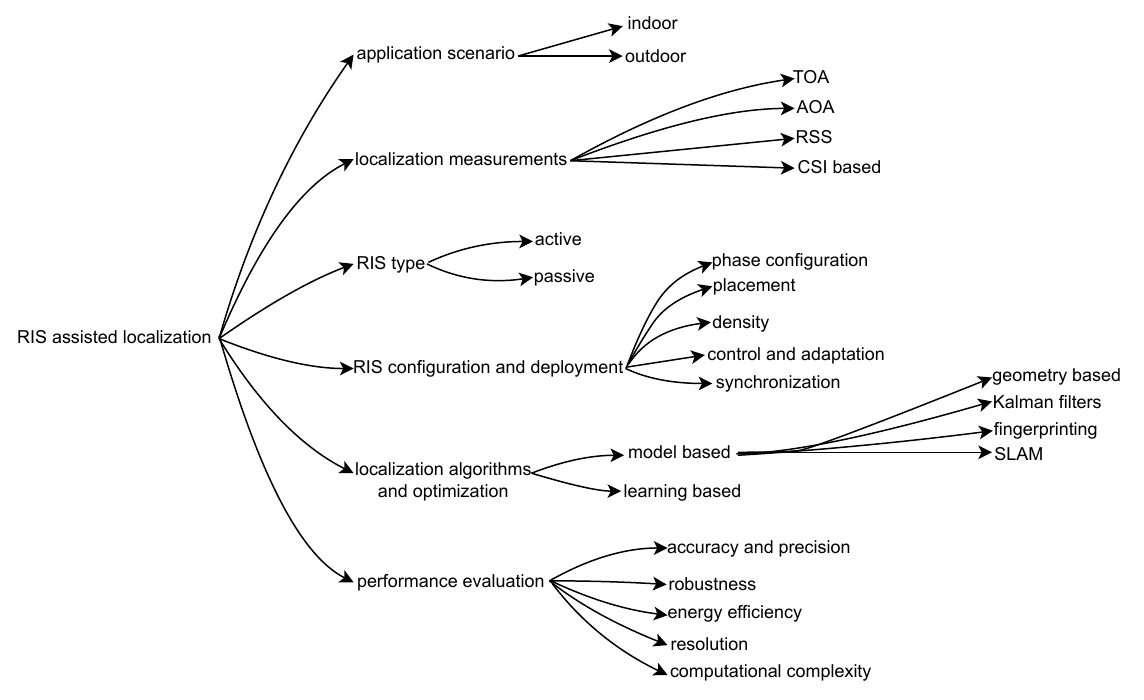}
    \caption{Taxonomy of \acrshort{ris}-assisted localization. Signal processing involving localization measurements and algorithms could take place at the UE or BS depending on the mode of communication, i.e., uplink or downlink.}
    \label{locatax}
\end{figure*}

\acrshort{ris}-assisted localization works by estimating the location and orientation of a UE with the help of anchor nodes (BS and \acrshort{ris}) provided the location of anchor nodes is already known \cite{9215972}. To locate itself, UE sends out a known uplink pilot signal to the BS or receives a downlink pilot signal from the BS. The signal's behavior is influenced by the propagation channel, which depends on the location and orientation of the BS and UE, as well as the environment surrounding them. The level of distortion in the received signal is determined by reflections from the \acrshort{ris} and other objects in the vicinity \cite{9893114}. The direct unobstructed link from the BS to the UE is called the \acrshort{los} path, the path of the signal reflected by the \acrshort{ris} is \acrshort{ris} path, and all the other \acrshort{nlos} paths from walls and objects in the environment are classified as the reflected paths and the scattered paths, respectively. Successful modeling of the pilot signal and the channel allows us to estimate the \acrfull{csi} and identify the parameters for the signal paths \cite{9782674}. These parameters, which aid in localization, include \acrshort{toa}/delay, \acrshort{aoa}, and \acrshort{aod}. UE location can be estimated based on these parameters and their geometrical relationships with the BS and \acrshort{ris} locations \cite{9893187}. \acrshort{ris}-assisted radio localization scenarios can be either in  indoor or outdoor environments, based on which the channel modeling is different. 

\begin{figure*}[t!]
    \centering
    \includegraphics[scale=0.66]{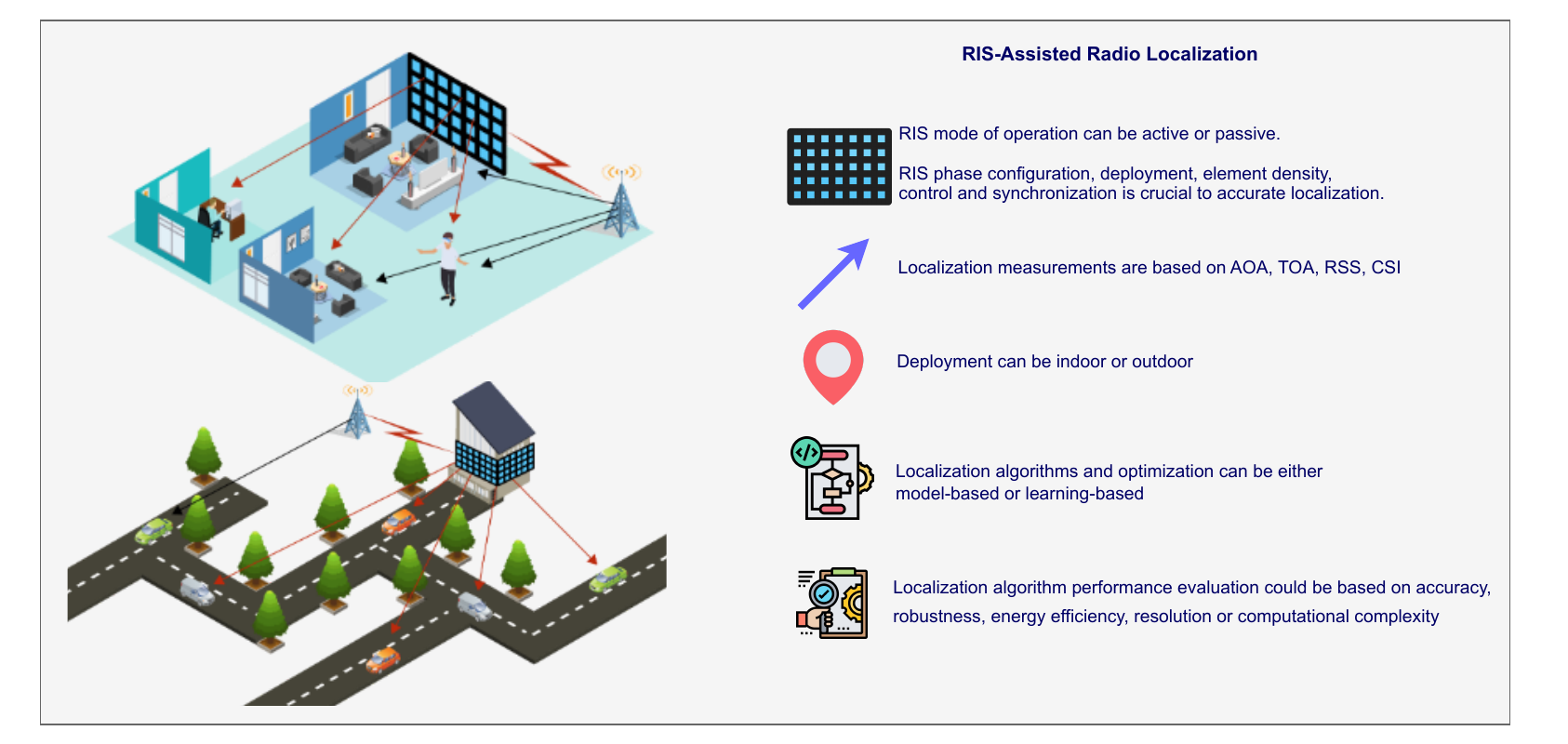}
    \caption{Illustration of \acrshort{ris}-assisted localization taxonomy. Signal processing involving localization measurements and algorithms could take place at the UE or BS depending on the mode of communication, i.e., uplink or downlink.}
    \label{locataxillus}
\end{figure*}

\acrshort{ris}-assisted localization can also be classified on the basis of the application scenario, localization technique, functionality of \acrshort{ris} employed and its configuration and deployment details, localization method as well the localization performance matrices. A brief taxonomy of \acrshort{ris}-assisted localization systems is shown in Figure \ref{locatax} and illustrated in Figure \ref{locataxillus}, respectively.

\subsubsection{Localization Measurements} 
Geometry-based techniques are widely used for radio localization and typically involve timing-based (\acrshort{toa}) and angle-based (\acrshort{aoa}/\acrshort{aod})  methods \cite{9356512}. The \acrshort{toa} is the time taken by the signal to travel from the BS to the UE. Timing-based localization technique, named trilateration, uses the measured \acrshort{toa} while considering the effects of \acrshort{ris} reflections, to estimate the location. The technique requires at least three BSs to get an unambiguous \acrfull{2d} estimate of the UE location when \acrshort{ris} are not considered. 
%For the TOA based methods, distance is calculated from the measured ToF. The estimation of signal propagation distance can be from the known channel state information.  
%from RSS. However, relying solely on RSS can lead to inaccurate estimations \cite{9782674}. 
An alternative is to estimate \acrshort{rtt} by recording signal transmission, processing and reception times, providing necessary \acrshort{toa} information. Well-synchronized systems can directly infer \acrshort{toa} from signals, with resolution dependent on signal bandwidth.
%Distance is estimated from the TOA measurement whose information is embedded in the channel information. 
Angle-based localization technique, named triangulation, estimates the angle from which signals arrive at the receiver, incorporating \acrshort{ris}-assisted reflections, to determine the location \cite{9082625}. It is typically employed when an antenna array is available at the BS. Some of the other estimation techniques rooted in \acrshort{toa} and \acrshort{aoa} include \acrshort{tdoa} estimation, \acrshort{aod}-based estimation, angle-difference-of-arrival location and orientation estimation.
%AOD-based estimation that is applicable when UE is equipped with an antenna array

The \acrfull{rss}-based localization techniques utilize the \acrshort{rss} measurements from \acrshort{ris}-assisted reflections to estimate the location of the receiver \cite{8815412}. It assists with the geometry-based trilateration and fingerprinting localization algorithms \cite{isprs}. This method capitalizes on the sensitivity of \acrshort{rss} to spatial variations, allowing for accurate localization even in complex indoor or outdoor urban scenarios \cite{9293395}. The reconfigurability of \acrshort{ris} enables real-time adaptation to changing propagation conditions, enhancing the precision and robustness of the localization system.

The \acrfull{csi}-based localization methods exploit the fine-grained \acrshort{csi} obtained through \acrshort{ris}-assisted reflections for accurate localization.
CSI contains valuable insights into the wireless propagation environment, including path loss, multipath components, and spatial signatures \cite{9538860}. Using advanced signal processing and \acrfull{ml} algorithms, the collected \acrshort{csi} data can be used to estimate the position of the devices \cite{pan2022risaided}. The integration of \acrshort{ris} with \acrshort{csi}-based localization enables radio localization in dynamic scenarios with changing propagation conditions.

% \begin{figure*}[t!]
%     \centering
%     \includegraphics[scale=0.6]{RIS_taxonomy_illustration-Page-4.drawio.pdf}
%     \caption{Illustration of RIS-assisted localization taxonomy. Signal processing involving localization measurements and algorithms could take place at the UE or BS depending on the mode of communication, i.e., uplink or downlink.}
%     \label{locataxillus}
% \end{figure*}

\subsubsection{\acrshort{ris} type}
\label{ristype}
As discussed in Section \ref{activeris}, in 6G radio localization, RIS are pivotal, functioning in either passive or active states. Passive RIS, requiring little external power, adjust only the phase of electromagnetic waves, offering simpler, energy-efficient deployment but limited signal control. In contrast, active RIS, equipped with elements like integrated circuits, can independently manipulate both signal phase and amplitude, enabling more complex functions like dynamic beamforming and interference cancellation\cite{9424177, 9833921, zheng2023jrcup}. 

\subsubsection{\acrshort{ris} configuration and deployment}
The configuration and deployment of \acrshort{ris} play a crucial role in \acrshort{ris}-assisted radio localization. The optimization of the phase configuration in \acrshort{ris} systems holds immense significance in unlocking its full potential. By precisely fine-tuning the phase shifts of the \acrshort{ris} elements, we can achieve remarkable control over signal propagation, allowing for unprecedented customization and optimization of wireless communication links \cite{9215972}. The optimization process involves carefully analyzing the channel characteristics, understanding the desired signal characteristics, and employing advanced algorithms to determine the optimal phase shifts for each \acrshort{ris} element \cite{9893114}. Through this optimization, we can exploit constructive interference, nullify destructive interference, and shape the signal to match specific requirements, such as maximizing coverage, minimizing signal attenuation, or focusing energy in desired directions \cite{9782674}. 

The placement and arrangement of \acrshort{ris} in the environment directly impacts the accuracy, coverage, and performance of the localization system \cite{article}. Determining the optimal locations for deploying \acrshort{ris} elements is fundamental to achieving the desired coverage area, signal propagation characteristics, and localization requirements. Proper deployment ensures optimal coverage of the desired area, minimizing blind spots and maximizing the availability of \acrshort{ris} reflections for localization purposes \cite{9215972}. Additionally, the number of \acrshort{ris} elements deployed per unit area or volume affects the granularity of control and the accuracy of localization. The configuration of the \acrshort{ris} elements, including the reflection coefficients, is vital in manipulating the signal propagation and optimizing the received signal at the receiver \cite{10012935}. The dynamic reconfigurability of \acrshort{ris} further enhances their role, allowing for real-time adaptation to changing propagation conditions and environmental dynamics \cite{10100675}. Moreover, ensuring proper synchronization among the \acrshort{ris} elements to avoid interference is equally important to configure \cite{9852985, 9771731}. By intelligently configuring and deploying \acrshort{ris}, the localization system can achieve improved accuracy, robustness, and scalability, enabling a wide range of location-based applications in various scenarios.
\subsubsection{Localization algorithms and optimization}
\acrshort{ris}-assisted localization algorithms can be broadly classified as model-based and learning-based \cite{9782674}. Model-based methods include deductive (physics-based) techniques such as geometry-based location estimation algorithms, Kalman filters, fingerprinting, and \acrshort{slam}. On the other hand, learning-based techniques are inductive (data-driven) and leverage \acrshort{ml} algorithms such as neural networks to learn and model the relationship between \acrshort{ris}-assisted signals and the receiver's location. The advantages of model-based approaches versus data-driven methods are numerous. They are supported by performance constraints that give solid assurances of optimality and dependability, rely on well-established signal processing techniques, and offer typically less complexity than data-driven systems. Some of the model-based approaches that are commonly employed in \acrshort{ris}-assisted localization are briefly discussed below.

\begin{itemize}
    \item \textit{Geometry-based methods}: Geometry-based methods rely on the \acrshort{toa} and \acrshort{aoa} measurements or their combination to determine the \acrshort{2d} or \acrshort{3d} location of the UE \cite{9893187, 9893114}. In traditional systems, such methods require a combination of measurements from multiple BSs to determine the UE location. However, the location of the UE can be estimated with the help of one BS and a \acrshort{ris} \cite{9148744}, more details in \cite{10044963}. Location estimation typically entails creating an objective function that incorporates geometric information and solving an optimization problem with geometric constraints. Geometry-based localization techniques are characterized by being free from training requirements, easily analyzable from a theoretical standpoint, and scalable across various environments.

    \item \textit{Kalman filter}: The Kalman filter is a recursive estimation algorithm that optimally fuses noisy measurements with a dynamic model to estimate the state of a system. In the context of localization, the Kalman filter predicts the device's position based on the previous state estimate and motion dynamics and then updates it using \acrshort{ris}-assisted measurements such as \acrshort{rss} or \acrshort{toa} \cite{9860999}. The Kalman filter-based approach can effectively mitigate the impact of noise, multipath, and other propagation effects on localization accuracy by iteratively updating the state estimate and incorporating \acrshort{ris}-assisted measurements \cite{9893114}. Despite its widespread use in localization, Kalman filtering has several limitations. Its assumptions of system linearity and Gaussian noise can be inaccurate in complex, real-world scenarios  \cite{9782674}. The initial state, which the filter requires, may not always be accurately known, and any errors in it can propagate, causing inaccuracies in state estimation. Furthermore, Kalman filters assume constant process and measurement noise covariances, an assumption often violated in real-world conditions. For nonlinear challenges, one can utilize an extended Kalman filter \cite{882463}. This approach estimates the state distribution by employing a Gaussian random variable and advances it through first-order linearization. Finally, Kalman filters are sensitive to model mismatches and outliers, which can significantly affect their performance \cite{9782674}.

    \item \textit{Fingerprinting}: In the fingerprinting  approach, a database of signal fingerprints is created by collecting and mapping the received signal characteristics at various locations in the environment \cite{8226757}. The fingerprint database contains information such as \acrshort{rss}, \acrshort{csi}, or signal amplitude patterns specific to each location. \acrshort{rss} has limited precision and \acrshort{csi} demands significant computational resources, spatial beam \acrfull{snr} is chosen as an intermediate channel measurement with moderate granularity \cite{9081910}. When a device needs to be localized, it measures the \acrshort{rss} or other signal characteristics at multiple \acrshort{ris}-assisted points in the environment. These measurements are then compared with the fingerprint database to find the closest match. The \acrshort{ris} play a crucial role in this process by manipulating the wireless channel to enhance the quality and reliability of the measurements. By adjusting the reflection coefficients of the \acrshort{ris} elements, the received signal can be optimized, leading to more accurate and consistent measurements.  However, if the configuration of the \acrshort{ris} changes over time, it would effectively alter the environmental characteristics that the fingerprint is based on, potentially decreasing the accuracy of location estimates. To guarantee the stationarity of the environment when employing \acrshort{ris} in fingerprinting, the configuration of the \acrshort{ris} should be kept constant during both the fingerprinting process and when using the fingerprint for location estimation. This means that the phase shifts or other manipulations applied to signals by the \acrshort{ris} should be fixed and not vary over time. In practice, this might require careful design and control of the \acrshort{ris}, and thorough testing to ensure its configuration remains stable under different conditions. This issue is resolved by including in the fingerprint the \acrshort{ris} configuration. This provides a richer set of fingerprints. The fingerprint matching process can utilize techniques such as pattern matching or \acrshort{ml} algorithms, deep learning methods such as \acrfull{dnn} and \acrfull{cnn} to find the best match between the measured signals and the fingerprints in the database \cite{ 9081910, 9782674, 8292280}. Once the closest match is found, the device's location is estimated based on the known location associated with the matched fingerprint \cite{isprs}. Fingerprinting can handle complex indoor or outdoor environments where multipath and \acrshort{nlos} conditions pose challenges for traditional localization techniques \cite{9159543}.

    \item \textit{Simultaneous localization and mapping}: \acrshort{slam} is a well-established method used to estimate the position of a device while constructing a map of its surroundings \cite{bresson2017simultaneous}. In this approach, \acrshort{ris} are strategically deployed in the environment to manipulate the wireless channel \cite{kim2022risenabled}. During the \acrshort{slam} process, the device measures parameters such as \acrshort{rss} or \acrshort{toa} at multiple \acrshort{ris}-assisted points. These measurements, along with the known positions of the \acrshort{ris}, are used to estimate the UE location and construct a map of the environment \cite{9685930}. The \acrshort{ris} plays a crucial role in improving the accuracy and reliability of the localization and mapping process by optimizing the quality of the received signals. It offers the advantage of accurate localization in complex environments where multipath propagation and \acrshort{nlos} conditions may exist. Additionally, \acrshort{ris} can adaptively adjust their reflection coefficients, enhancing the performance and robustness of the SLAM-based localization system.  
    %, enhancing the quality of received signal measurements
\end{itemize}

Explicit modeling of geometry information becomes difficult in complicated cases when there are several non-resolvable \acrshort{nlos} paths \cite{9356512}. Under such conditions the learning-based approaches are recommended \cite{9523553, mitchell1997machine}. \acrshort{ml}-based techniques need offline training, which drastically decreases online calculations, in contrast to the practical algorithms employed in geometry-based localization. To train the models, however, a significant amount of system data must be gathered, and the learned models must be updated on a regular basis to account for changes in the environment. The DNN are used to perform environmental sensing to achieve the best performance in \acrshort{ris}-assisted radio localization \cite{isprs, 9495362}. Since the channels are sparse at the higher frequency bands, therefore, most of the studies use geometry-based algorithms \cite{9782674}.

\subsubsection{Localization performance evaluation}
A radio localization system is designed on the basis of a number of performance objectives that include accuracy, coverage, latency, robustness, resolution, update rate, stability, scalability, mobility and system complexity, etc \cite{9782674}. Here we discuss most relevant ones.

Accuracy and precision are metrics to assess the accuracy of the estimated location compared to the ground truth, and  the level of precision in determining the location, often represented by the standard deviation or confidence interval. Accuracy is the most widely used localization metric in state-of-the-art studies as it accounts for localization resolution as well as identifiability. \acrshort{crlb} are the commonly used bounds on the achievable accuracy in studies. The deployment geometry, also known as \acrfull{gdop} or UE relative position with respect to BSs, determines accuracy in addition to link-level \acrshort{snr} \cite{10044963}. 

The separability of completely correlated radio propagation channels in at least one domain is referred to as resolution. Unresolvable signal paths will be treated as a single path, limiting accuracy (regardless of \acrshort{snr}), and producing worse performance than predicted by analytical bounds. The resolution is constrained by the physical resources available, such as antenna array aperture for angle resolution and bandwidth for delay/distance resolution \cite{9893187}. Despite having a high degree of resolution, radio localization can nevertheless suffer from ambiguity and non-identifiability \cite{10044963}. This indicates that the localization problem of UE might have many solutions or a continuous space of solutions. This might happen when there are barriers in the way of specific BS signals, or when infrastructure rollout or coverage is insufficient. Ambiguity, which emerges as numerous unique locations, is frequently addressed by prior knowledge or external signals. On the other hand, non-identifiability poses a more significant challenge as there are numerous equally valid solutions to the localization problem, and it is difficult to discard them based on external information.

The robustness of the localization algorithm evaluates its performance under various environmental conditions, such as multipath fading, interference, and mobility. It accounts for availability, latency, and update rate under such conditions \cite{9782674}. Likewise, assessing the energy consumption of \acrshort{ris}-assisted localization techniques, considering both \acrshort{ris} elements and the receiver device is an important measure in \acrshort{ris}-assisted localization in comparison to the systems without \acrshort{ris}. Energy-efficient techniques aim to minimize energy consumption while achieving accurate localization. Computational complexity quantifies the computational resources needed to perform \acrshort{ris}-assisted localization at the hardware and algorithm level. It includes the processing power, memory requirements, and time complexity of the localization algorithm. Lower computational complexity allows for faster and more efficient localization.

\begin{figure}[t!]
    \centering
    \includegraphics[width=0.5\textwidth]{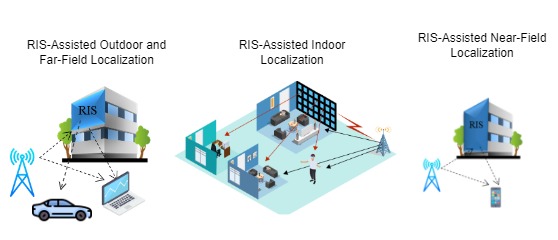}
    \caption{State of the art literature breakdown. Studies can be broadly categorized into outdoor, far-field, indoor and near-field \acrshort{ris}-assisted localization. Within these categories, we notice trend towards investigation of systems on mmW and THz frequencies. Further, the passive reflective \acrshort{ris} is the most common type of \acrshort{ris} used in literature for \acrshort{ris}-assisted localization.}
    \label{sota}
\end{figure}

\section{State of The Art in \acrshort{ris}-Assisted Radio Localization}%State of the Art Trends and Developments in RIS-assisted Radio Localization}
\label{sec4}

In this section, we discuss the recent literature on \acrshort{ris}-assisted radio localization for 6G networks. The trends in the latest studies are on developing algorithms, optimization, and investigation of \acrshort{ris}-assisted localization systems from the perspective of accuracy and availability at various frequency bands, near-field and far-field modeling, and indoor and outdoor scenarios, as summarized in Figure \ref{sota}. 
 
In the realm of \acrshort{ris}-assisted localization, we classify the frequency bands into two categories: \textit{\acrfull{crf}}- and \textit{high frequency bands}. \acrshort{crf} bands typically refer to those below 24 GHz, encompassing well-known sub-bands such as FR1. \acrshort{crf} bands are recognized for their longer wavelengths, better penetration capabilities, and widespread application in various wireless services. For outdoor localization, \acrshort{gnss} is predominantly employed, providing meter-level precision with support from long-term evolution (LTE) communication signals. Yet, this method proves ineffective for indoor environments, where intricate surroundings and LoS channel obstructions are common challenges. As alternatives, there are documented localization systems utilizing \acrfull{uwb} \cite{s16050707}, WiFi \cite{8688470}, wireless local-area network (WLAN) \cite{7949029}, and LoRA \cite{8827665} \cite{8692423}. By leveraging CRF bands systems in association with the \acrshort{ris}, advantages are gained in location-centric services, including navigation and identification of nearby amenities.

On the other hand, we refer to the localization services in the range of spectrum above 24 GHz as high frequency bands localization. This category consists of FR2 frequencies, \acrshort{mmw} and \acrshort{thz} band. The range of 100- 300 GHz in \acrshort{mmw} band is also referred to as the sub-terahertz (sub-THz) according to the deliverable D2.1 of the European HEXA-X project \cite{hexa}. These higher frequency bands offer significant advantages in terms of data rate, capacity and localization performance but come with challenges related to limited propagation and penetration. Utilizing antenna arrays at the UE enables to estimation of its orientation \cite{8240645}. Furthermore, through the use of \acrshort{nlos} paths \cite{8356190} and \acrshort{ris} \cite{9129075, 9215972} localization tasks can be accomplished with only one BS. The \acrshort{thz} systems are anticipated to complement \acrshort{mmw} systems in diverse environments, and the comparison between the two reveals distinct advantages and challenges in terms of localization \cite{9782674}. As technology progresses from CRF to \acrshort{5g} and onto \acrshort{6g}, expectations include higher frequencies, increased bandwidths, more compact footprints, and larger array sizes \cite{9997557}. These changes will affect path loss, delay estimation resolutions, and antenna array design, among other features. Challenges may arise with hardware imperfections and synchronization at \acrshort{thz} frequencies. The design of localization algorithms must also consider the specific properties of \acrshort{thz} signals, such as the beam split effect and high path loss \cite{9782674}. Ultimately, the adaptations and innovations within the \acrshort{thz} systems are expected to lead to improved localization performance in \acrshort{6g} networks.

We group the studies as the development of methods for localization of UE in outdoor and far-field, indoor and near-field, respectively. To enhance reader clarity, it is important to emphasize that the first group predominantly encompasses studies centered around far-field and outdoor scenario investigations. Distinct groupings have been established to specifically address indoor scenarios and near-field localization, both of which are considered specialized cases. Studies are presented in comprehensive detail and tabular format in the subsequent subsections.

\begin{scriptsize}
\begin{table*}[t!]
\centering
    \caption{ SUMMARY OF \acrshort{ris}-ASSISTED OUTDOOR AND FAR-FIELD LOCALIZATION ARTICLES AT CONVENTIONAL RADIO FREQUENCY BAND. HERE ``R" REFERS TO ``REFLECTIVE  \acrshort{ris}"}
   
\begin{tabular}{p{0.02\linewidth}p{0.04\linewidth}p{0.05\linewidth}p{0.02\linewidth}p{0.03\linewidth}p{0.24\linewidth}p{0.25\linewidth}p{0.14\linewidth}}
    \label{table1} \\
         \thickhline
         \textbf{\acrshort{ris}} & \textbf{Ref} & \textbf{$f_c$} & \textbf{Link} & \textbf{System}& \textbf{Purpose} & \textbf{Technique}& \textbf{Performance metric} \\
         \thickhline

        R & \cite{9724202} & 1.5 GHz &DL&MISO&  \acrshort{ris} for sensing/localizing targets in wireless networks & Self-sensing \acrshort{ris} architecture, customized \acrshort{music} algorithm, CRLB& Accuracy
         \\
         \hline

        R& \cite{ghiasvand2022miso} & 2 GHz & DL & MISO & Localization with obstructed \acrshort{los} and three \acrshort{ris} & Elimination of destructive effect of the \acrshort{aod} & Accuracy\\
         \hline

        R  & \cite{10042425}& 2.5 GHz& DL&MIMO&Joint active and passive beamforming design for \acrshort{ris}-enabled \acrshort{isac} system in consideration of the target size &  Non-convex optimization & Detection probability, SNR\\
         \hline

        R & \cite{9750199}& 2 GHz & DL& SISO& UE localization assisted by multiple \acrshort{ris} & Localization algorithm design based on nodes distances & Accuracy\\
         \hline
         
        R & \cite{9434917} & 3 GHz & DL& SISO&  \acrshort{rss} fingerprinting based multi-user outdoor localization using \acrshort{ris} and single BS & localization error minimization algorithm & Accuracy\\
         \hline

        R & \cite{9729782} & 3 GHz & &SISO & Fundamental limits of \acrshort{ris}-aided localization and communication system with \acrshort{ris} as continuous and discrete intelligent surface & \acrshort{ris} phase design, \acrfull{fim}& Accuracy, spectral efficiency, SNR
         \\
         \hline

         R & \cite{9860384} & 3 GHz & DL & SISO & Wideband localization with \acrshort{ris} & \acrshort{fim}& Accuracy\\
         \hline

         R & \cite{9860587} & 3 GHz & DL & SISO & \acrshort{jcal} system design & PEB, joint \acrshort{ris} discrete phase shifts design and subcarrier assignment using Lagrange duality and penalty-based optimization& Accuracy and data rate\\
         \hline

         R & \cite{9900336} & 3 GHz & DL & SISO & Multi-user localization system using modulated \acrshort{ris} & \acrshort{tdoa} &Accuracy\\
         \hline

        R  & \cite{s23020984} & 3.4/3.5/28 GHz & &MIMO & Localization technique that does not require \acrshort{ris} codewords for online location inference & Domain adversarial neural network, fingerprinting & Accuracy\\
         \hline

%        R& \cite{liu2021optimization} & 4.9 GHz & DL & MIMO &  Analysis of multiple-\acrshort{ris}-aided localization system & CRLB, \acrshort{ris} phase shift design using particle swarm optimization algorithm& accuracy
 %        \\       \hline

        R  & \cite{hong2023risposition} & 4.9 GHz& UL & MIMO & UE position estimation& Compressed sensing orthogonal simultaneous matching pursuit algorithm, maximum likelihood estimation, discrete Fourier transform, SAGE algorithm, CRLB& Accuracy\\
        \hline

         R & \cite{9977919} & 5 GHz& DL & MIMO &  Lower bounds on the location estimation error for multiple \acrshort{ris}-aided mmW system & \acrshort{crlb}, \acrshort{peb}, \acrshort{reb}& Accuracy\\
         \hline

        R & \cite{10113892} & 10 GHz & DL & MIMO & Bayesian analysis of the information in \acrshort{los} and \acrshort{ris} reflected signal & Bayesain analysis, \acrshort{fim} & Accuracy\\ 
         \hline

        R & \cite{9779402} & 20 GHz & UL & MISO & To localize a large number of energy-limited devices simultaneously and accurately & Triangulation-based localization framework, optimization& Accuracy, energy efficiency
         \\
         \hline
         R & \cite{10285302} & 2 GHz & UL, DL & MISO & Environment-adaptive user positioning & Federated learning & Accuracy\\
         \hline

         R & \cite{10216343} & 3 GHz & UL & MIMO & \acrshort{isac} for wireless extended reality using \acrshort{ris} & \acrshort{music}, \acrshort{crlb} & Accuracy, capacity\\
         \hline

         R & \cite{10325432} & 2.45 GHz & DL & SIMO & Theoretical and practical design of \acrshort{ris}-assisted sensing system with localization & Atomic norm minimization direction of arrival estimation network, generative adversarial network& Accuracy\\
         \hline

         R & \cite{10478036} & 1 GHz & DL& SISO &  Method for \acrshort{ris}-assisted fingerprinting localization & Fingerprinting, graph-based radio map interpolation   & Accuracy\\
         \hline

         R & \cite{10373864} & 2.5 GHz & DL & MISO & Joint beamforming and aerial \acrshort{ris} positioning design with multiple access points& Generalized benders decomposition,  mixed-integer semidefinite programming & Spectral efficiency\\
         \hline

         R & \cite{10257295} & 400 MHz & &SISO & Positioning, navigation, and timing solution & Game theory & Accuracy\\
         \hline
         \thickhline
    \end{tabular}
\end{table*}
\end{scriptsize}
  
\subsection{\acrshort{ris}-Assisted Outdoor and Far-Field Localization}
In this subsection, we explore the latest advancements and contemporary studies focused on leveraging \acrshort{ris} technology to augment outdoor and far-field localization at \acrshort{crf} and high frequency bands.
\subsubsection{\acrshort{ris}-Assisted Conventional Radio Frequency Bands Localization (Below 24 GHz)}

 Studies are briefly summarized in Table \ref{table1}. We attempt to categorize the works into thematic groups based on their core focus as follows.

\paragraph{Foundational Studies}
Authors in \cite{9434917} introduce \acrshort{ris} to enhance \acrshort{rss} fingerprinting-based outdoor localization using just one BS. By adjusting the \acrshort{ris} phase shifts, the approach creates distinct \acrshort{rss} values at the same location, optimizing this through a localization error minimization algorithm. Simulations confirm the scheme's efficacy.  Article \cite{9729782} introduces the concept of continuous intelligent surfaces and investigates the fundamental limits of \acrshort{ris}-aided \acrshort{isac} systems. The paper proposes a general signal model, derives theoretical limits on localization and communication performance, and performs Fisher information analyses. The numerical results demonstrate that optimized \acrshort{ris} can improve the \acrshort{snr} and spectral efficiency of communication, as well as enhance localization accuracy. Authors in article \cite{9779402} propose a \acrshort{ris}-assisted positioning method for simultaneously localizing multiple energy-limited \acrshort{iot} devices in location-based \acrshort{iot} services. The proposed method utilizes triangulation-based localization, estimating the propagation delay difference between the direct and reflected paths using cross-correlation. By optimizing the multi-antenna BS and \acrshort{ris} to minimize total transmission power, significant power gain and decimeter-level positioning accuracy are achieved, demonstrating the effectiveness of the proposed optimization approach compared to unoptimized \acrshort{ris}-assisted localization.

\paragraph{Novel Techniques and Systems} 
In \cite{9724202}, the authors propose a new \acrshort{ris} self-sensing system where the \acrshort{ris} controller transmits probing signals and dedicated sensors at the \acrshort{ris} are used for location and angle estimation based on the reflected signals by the target. The \acrfull{music} algorithm is applied to accurately estimate the \acrshort{aoa} of the target in the \acrshort{ris}'s vicinity, and the \acrshort{ris} passive reflection matrix is optimized to maximize the received signals' power at the \acrshort{ris} sensors, leading to minimized \acrshort{aoa} estimation \acrfull{mse}. The results demonstrate the benefits of using the \acrshort{ris} controller for probing signals and provide the \acrshort{crlb} for target \acrshort{aoa} estimation. 
%Future work includes investigating the impact of RIS impairments, NLoS components, and UE mobility on the proposed system model and estimation approach. 
In \cite{ghiasvand2022miso}, a positioning algorithm is introduced for \acrshort{ris}-assisted networks, focusing on multi-antenna BS and single-antenna UE. Leveraging three \acrshort{ris} and specific phase shifter adjustments, the method effectively overcomes \acrshort{los} obstructions and minimizes the adverse effects of the \acrshort{aod}, resulting in improved localization accuracy compared to non-\acrshort{aod} estimating algorithms.
The study \cite{10042425} introduces a joint active and passive beamforming design for \acrshort{ris}-enabled \acrshort{isac} systems, accounting for target size. Through an alternative optimization method, the paper addresses non-convex problems involving beamforming solutions and \acrshort{ris} phase shifts, with the developed algorithm showcasing superior target detection performance in simulations, especially for practical target sizes, against existing benchmarks.
The article \cite{10113892} presents a Bayesian analysis of the information contained in a signal received by a UE from a BS that includes reflections from \acrshort{ris}. The analysis considers both near and far-field scenarios and incorporates prior information about the UE and the \acrshort{ris} for localization. The results indicate that the orientation offset of the \acrshort{ris} affects the pathloss of the \acrshort{ris} paths when the \acrshort{ris} elements are spaced half a wavelength apart. In the far-field regime, an unknown phase offset in the received signal prevents the correction of the \acrshort{ris} orientation offset. However, in the near-field regime, the estimation of the \acrshort{ris} orientation offset is possible when the UE has multiple receive antennas. The article also demonstrates that accurate localization with \acrshort{ris} is only possible when there is prior knowledge of their locations. Finally, numerical analysis shows the loss of information when applying a far-field model to signals received in near-field propagation. 

Unlike the methods that rely on channel matrices or \acrshort{ris} codewords, the authors in \cite{s23020984} proposed an approach that uses a domain adversarial neural network to extract codeword-independent representations of fingerprints for online location inference in \acrshort{ris}-assisted localization network. The solution is evaluated using the DeepMIMO data set, and the results show that the proposed method performs significantly closer to the theoretical upper bound (oracle case) than the lower bound (baseline case), indicating its effectiveness and robustness. Authors in \cite{hong2023risposition} article investigate the estimation of position and angle of rotation for a UE in a MIMO system with the assistance of a \acrshort{ris}. The \acrshort{ris} creates a virtual \acrshort{los} link, along with \acrshort{nlos} links from scatterers in the environment, to aid in the estimation process. A two-step positioning scheme is utilized, where channel parameters are acquired first and then position-related parameters are estimated. Coarse estimation is performed using various algorithms, followed by joint refinement using the space-alternating generalized expectation maximization (SAGE) algorithm. The performance of the proposed algorithms is demonstrated to be superior through simulation results, and theoretical quantification is done using the \acrshort{crlb}. The authors in \cite{10285302} introduced HoloFed, a high-precision, environment-adaptive user positioning system that integrates multi-band reconfigurable holographic surface with federated learning. To enhance positioning accuracy, the system uses a calculated lower bound on error variance to guide multi-band reconfigurable holographic surface beamforming design. An federated learning framework is implemented for collaborative training of a position estimator, utilizing transfer learning to compensate for the lack of user position labels, while a scheduling algorithm optimizes user selection for training based on federated learning convergence and efficiency. Simulation results indicate that HoloFed reduces positioning error variance by 57\% compared to traditional beam-scanning methods, demonstrating significant adaptability and precision across diverse environments. A novel RIS-enabled wireless fingerprinting localization method that addresses missing \acrshort{rss} data using a graph-based radio map interpolation technique is proposed in \cite{10478036}. This approach leverages RIS configuration flexibility and encodes similarities through a multi-layer graph. Numerical results show improved radio map recovery and localization accuracy compared to other methods. 

\paragraph{Integrated Sensing and Communication}
The work presented in \cite{10216343} explores the use of \acrshort{ris} in future wireless networks for enhanced positioning and communication in extended reality applications. It introduces a positioning algorithm based on \acrshort{music} and optimized \acrshort{ris} configurations. The joint optimization of \acrshort{ue} beamformers and \acrshort{ris} phase shifters maximizes channel capacity under \acrshort{crlb} constraints, solved via alternating optimization. Authors in \cite{10325432} present a low-cost \acrshort{ris}-assisted sensing system, featuring a novel atomic norm minimization direction of arrival estimation network method for improved localization and communication reliability. The system uses a cost-effective RIS structure and a practical signal model that accounts for RIS phase shifts. The atomic norm minimization direction of arrival estimation network method combines atomic norm minimization for spatial spectra and direction of arrival estimation network for angle estimation, using generative adversarial network to generate a large dataset. Experimental results show that proposed technique effectively localizes sources in the RIS-aided sensing system.

\subsubsection{\acrshort{ris}-Assisted High Frequency Bands Localization (24 GHz and above)}
%Higher frequency bands can be broadly categorized as mmW Band (30-300 GHz) and THz (0.1-10 THz) bands of radio communication \cite{9782674}. 
Studies on \acrshort{ris}-assisted localization in this band of operation are briefly summarized in Table \ref{table2}, \ref{table5} and \ref{table6} and discussed as follows.

\paragraph{Foundational Studies}
Authors in \cite{9148744} discussed the use of \acrshort{ris} in 6G radio positioning. The authors propose a two-step optimization scheme that selects the best combination of \acrshort{ris} and controls their constituent elements' phases to improve positioning performance. Preliminary simulation results demonstrate gains in coverage and accuracy compared to natural scattering, but limitations are identified in terms of low \acrshort{snr} and inter-path interference. Assuming the LOS route between the BS and the MS is present, authors in \cite{9129075} introduced \acrshort{ris} as a reflector into the \acrshort{mmw} MIMO positioning system. The CRLB of the positioning as well as the orientation estimation error are obtained by calculating FIM, which reveals that the \acrshort{ris}-aided mmW MIMO positioning system offers better localization accuracy and coverage as compared to the conventional localization system comprising BS nodes only. It has also been demonstrated that one BS with the help of reflection from \acrshort{ris} can also achieve promising positional precision. Nevertheless, nothing is discussed about how to localize the UE in a LOS-obscured environment. To determine the absolute location of the MS under the NLoS scenario, authors in \cite{9500437} developed the \acrshort{crlb} based on \acrshort{fim}. The study suggests that, in the given setup, the localization can reach the decimeter level of accuracy by refining the reflect beamforming architecture to reduce \acrshort{crlb}. Authors in \cite{Keykhosravi2020SISORJ} present localization and synchronization in a wireless system with a single-antenna UE, a single-antenna BS, and a \acrshort{ris}. They calculate the \acrshort{crlb} and develop a low-complexity estimator to determine the \acrshort{aod}  from the \acrshort{ris}, as well as the delays of direct and reflected signals. The results indicate that efficient \acrshort{3d} localization and synchronization are achievable in the considered system, showcasing the potential of \acrshort{ris} for enabling radio localization in simple mmW wireless networks.
\begin{scriptsize}
\begin{table*}[t!]
\centering
    \caption{ SUMMARY OF \acrshort{ris}-ASSISTED OUTDOOR AND FAR-FIELD LOCALIZATION ARTICLES AT HIGH FREQUENCY BANDS. HERE "R" REFERS TO "REFLECTIVE \acrshort{ris}", , "S" REFERS TO "STAR \acrshort{ris}", "A" REFERS TO "ACTIVE \acrshort{ris}"}
\begin{tabular}{p{0.02\linewidth}p{0.04\linewidth}p{0.05\linewidth}p{0.02\linewidth}p{0.03\linewidth}p{0.24\linewidth}p{0.25\linewidth}p{0.14\linewidth}}
    \label{table2} \\
         \thickhline
         \textbf{\acrshort{ris}} & \textbf{Ref} & \textbf{$f_c$} & \textbf{Link} & \textbf{System}& \textbf{Purpose} & \textbf{Technique}& \textbf{Performance metric} \\
         \thickhline

        R & \cite{9860999} & 24 GHz & DL & MISO & \acrshort{ris} to replace the function of a remote cell in the DL-TDOA measurement & Extended Kalman filter positioning and tracking algorithm& Accuracy
         \\
         \hline

           R&\cite{9148744}& 28 GHz & DL& SISO & Analysis of a \acrshort{ris}-aided localization problem & FIM, two step optimization & Accuracy, coverage
         \\
         \hline

         R & \cite{9410435} & 28 GHz& UL& MIMO & Beam training designs to estimate optimal beams for BS and UE, \acrshort{ris} reflection pattern and link blockage & \acrfull{mle}, positioning algorithm design  & Accuracy      
         \\
         \hline

         R & \cite{9538860} & 28 GHz & DL & SISO &  Use of 3D localization technology to achieve the low-complexity channel estimation &  Reflecting unit set concept, coplanar maximum likelihood-based based 3D positioning method, CRLB& Accuracy, SNR
         \\
         \hline

        R & \cite{9777939} & 28 GHz & UL &MIMO&  Localization and channel reconstruction in extra large \acrshort{ris}-assisted \acrshort{mimo} systems & Low-overhead joint localization, channel reconstruction scheme& Accuracy
         \\
         \hline
         
         R & \cite{9782100} & 28 GHz & DL& MISO & Exploiting \acrshort{ris} with suitably designed beamforming strategies for optimized localization and synchronization performance & PEB, MLE& Accuracy
         \\
         \hline
         
         S  & \cite{He2022SimultaneousIA} & 28 GHz &UL & MISO& \acrshort{star} \acrshort{ris} potential for enhanced concurrent indoor and outdoor localization & CRLB, FIM, optimization& Accuracy
         \\
         \hline

         R & \cite{xanthos2022joint} & 28 GHz & DL &MIMO &Joint beamforming and localization for \acrshort{ris}-aided mmW localization system & Joint localization and beamforming optimization algorithm& Accuracy
         \\
         \hline
         
         R & \cite{Keykhosravi2022RISEnabledSL} & 28 GHz &&& Enabling the user to estimate its own position by transmitting \acrfull{ofdm} pilots and processing the signal reflected from the \acrshort{ris} & CRLB, low-complexity position estimation algorithm, temporal coding on \acrshort{ris} phase& Accuracy
         \\
         \hline

         R & \cite{9540372} & 28 GHz & UL & MIMO & Channel estimation and user localization & \acrshort{ris} training coefficients designs, array signal processing, atomic norm denoising techniques & Accuracy \\
         \hline

        R & \cite{10012776}&28 GHz& DL & SIMO& Joint \acrshort{ris} calibration and user positioning scheme & FIM& Accuracy \\
         \hline

        R & \cite{9827856} & 28 GHz & DL & MIMO & User localization and tracking & Bayesian user localization and tracking algorithm & Accuracy\\
         \hline

        R & \cite{9977483} & 28 GHz &  & SISO & Cooperative localization to improve accuracy in \acrshort{ris}-assisted system & Beam sweeping, optimization, neural network & Accuracy\\
        \hline

        R & \cite{10012935} & 28 GHz & & & Cooperative localization with no access point & FIM, CRLB, \acrshort{ris} configuration optimization & Accuracy\\
        \hline

        Rx& \cite{10124713} & 28 GHz & UL & & Localization of UE with partially connected receiving \acrshort{ris} only & Atomic norm minimization, \acrshort{music}, \acrshort{crlb} & Accuracy\\
         \hline
         
         A & \cite{zheng2023jrcup} & 28 GHz & UL & SIMO & Joint \acrshort{ris} calibration and user positioning problem with an active \acrshort{ris} &  Tensor-ESPRIT estimator, least-squares, 2D search-based algorithm, CRLB & Accuracy\\
         \hline

         R & \cite{10096904} & 28 GHz & DL & SISO & \acrfull{mcrb} with \acrshort{ris} geometry mismatch & Method for  pseudo-true parameter determination for \acrshort{mcrb} analysis &  Accuracy\\
         \hline

        R & \cite{he2023compressed} & 28 GHz&UL & MISO & Localization of UE using distributed passive \acrshort{ris} & \acrfull{cs} approach based on atomic norm minimization, \acrshort{mle}, CRLB & Accuracy\\
         \hline

         R & \cite{wang2023target} & 28 GHz & &MIMO&Device-free target sensing via joint location and orientation estimation &Target based method for angle estimation, gradient descent method, manifold optimization &Accuracy\\
        \hline

        A  & \cite{asif2023isac} & 28 GHz& UL& MISO& \acrshort{isac} using sparse active \acrshort{ris} & \acrshort{music} algorithm, optimization& Accuracy\\
                        \hline

        R & \cite{10100675} & 28 GHz & DL& MIMO & \acrshort{jcal} framework & Novel \acrshort{ris} optimization and channel estimation methods & Accuracy, data rate\\
                \hline

        R& \cite{chen2023multirisenabled} & 28 GHz& SL& SISO & UE localization without BS involvement & Two-stage \acrshort{3d} sidelink positioning algorithm, CRLB & Accuracy\\
        \hline
        
         R & \cite{Keykhosravi2020SISORJ} & 30 GHz &DL&SISO&  Joint \acrshort{3d} localization and synchronization for a SISO multi-carrier system & CRLB, design of low complexity estimation algorithm&Accuracy
         \\
         \hline

        R & \cite{9528041} & 30 GHz& DL& SISO & \acrshort{ris} in a multi-user passive localization scenario & Low complexity \acrshort{toa} based positioning algorithm, CRLB& Accuracy\\
         \hline

        R & \cite{9774917} & 30 GHz & DL& SISO &  Positioning UE by taking into account the its mobility spatial- wideband effects & CRLB, low-complexity estimator design& Accuracy
         \\
         \hline

          R & \cite{kim2022risenabled} & 30 GHz & && \acrshort{ris}-enabled radio \acrshort{slam} without the intervention of BS & \acrshort{ris} phase profile design, marginal Poisson multi-Bernoulli \acrshort{slam} filter modification, CRLB& Accuracy\\
         \hline

         \thickhline
    \end{tabular}
\end{table*}
\end{scriptsize}

\begin{scriptsize}
\begin{table*}[t!]
\centering
    \caption{ SUMMARY OF \acrshort{ris}-ASSISTED OUTDOOR AND FAR-FIELD LOCALIZATION ARTICLES AT HIGH FREQUENCY BANDS. HERE ``R" REFERS TO ``REFLECTIVE \acrshort{ris}", ``H" REFERS TO ``HYBRID \acrshort{ris}", ``A" REFERS TO "ACTIVE \acrshort{ris}", ``S" REFERS TO ``STAR RIS"}
\begin{tabular}{p{0.01\linewidth}p{0.04\linewidth}p{0.05\linewidth}p{0.02\linewidth}p{0.03\linewidth}p{0.25\linewidth}p{0.26\linewidth}p{0.14\linewidth}}
    \label{table5} \\
         \thickhline
         \textbf{\acrshort{ris}} & \textbf{Ref} & \textbf{$f_c$} & \textbf{Link} & \textbf{System}& \textbf{Purpose} & \textbf{Technique}& \textbf{Performance metric} \\
         \thickhline

        R & \cite{10001508} & &DL&MIMO&\acrshort{isac} with \acrshort{ris}& \acrshort{cs}, expectation-maximization algorithm, Bayesian Cramér-Rao bound& Accuracy\\
         \hline

         H& \cite{ghazalian2022joint}& 30 GHz&DL& MISO & Joint localization of a hybrid \acrshort{ris} and a user & CRLB& Accuracy\\
         \hline

         R& \cite{9771873}& 30 GHz & DL & SISO & Cooperative localization in a \acrshort{ris}-aided mmW system & FIM, CRLB,  block coordinate descent-based reflect beamforming design algorithm& Accuracy\\
         \hline

        R & \cite{10070578} & 30 GHz& &MISO&Location information assisted beamforming design without the requirement of the channel training process&Relaxed alternating optimization process& Data rate\\
         \hline

         R & \cite{9500437} & 50 GHz & DL & MIMO & \acrshort{ris} beamforming under \acrshort{nlos} & CRLB, optimization & Accuracy
         \\
         \hline
         
         R & \cite{9129075} & 60 GHz & DL & MIMO& Theoratical bounds for large intelligent surface & CRLB based PEB and OEB & Accuracy
         \\
         \hline
         
         R & \cite{9124848} & 60 GHz& DL& MIMO & Improving the positioning accuracy and date rate & Adaptive phase shifter design based on hierarchical codebook and feedback from the UE & Accuracy, data rate
         \\
         \hline
         
         R & \cite{9238887}& 60 GHz&UL&MIMO &  Localization of UE & Two stage positioning method with dual \acrshort{ris}  & Accuracy
         \\
         \hline
         
         R& \cite{9513781} & 60 GHz& DL & MIMO &Utilizing \acrshort{ris} in mmW MIMO radar system for multi-target localization & Adaptive localization algorithm utilizing the concept of hierarchical codebook design& Accuracy
         \\
         \hline
         
         R & \cite{Fascista2020RISAidedJL} & 60 GHz &DL& MISO&  Joint localization and synchronization & MLE& Accuracy
         \\
         \hline
         
         R & \cite{9782674} & 60 GHz & UL& MIMO & \acrshort{ris}-assisted localization at \acrshort{thz} band in comparison with the \acrshort{mmw} band & Geometrical modeling and simulations& Accuracy
         \\
         \hline         
         
         R & \cite{9771819} & 60 GHz &UL & MIMO & Potential of \acrshort{ris} for cooperative localization performance& CRLB, manifold optimization& Accuracy
         \\
         \hline         

        R& \cite{9794297} & 60 GHz & UL & MIMO & To optimize the worst-case localization performance by
        jointly optimizing beamforming vectors at \acrshort{ris} and UE& Joint array gain and path loss search algorithm, difference of convex-based algorithm &  Accuracy\\
        \hline 

        A & \cite{9833921} & 60 GHz & DL & MIMO & UE localization with active \acrshort{ris} & Multiple signal transmissions, particle filtering, CRLB & Accuracy \\
        \hline

        R& \cite{10044963} & 60 GHz & DL &SISO& Overview of \acrshort{ris} enabled localization scenarios  & Experimental demonstration& Accuracy
         \\
         \hline

         R & \cite{10054103} & 60 GHz & DL & MIMO & Joint optimal point of the user position/orientation estimation error bound and effective achievable data rate & Worst-case robust beamforming and time allocation optimization approach, majorize-minimization based algorithm& Accuracy, data rate\\
         \hline

        R & \cite{10052208} & 60 GHz & DL & MISO & Investigate the potential of employing \acrshort{ris} in dual-functional radar-communication vehicular networks& Codebook design for optimal phase shift of \acrshort{ris}, position-based \acrshort{csi} design & Accuracy, data rate\\
        \hline

         R & \cite{10130575} & 100 GHz& UL& MISO&  Sensing of channel and location under the unique hybrid far-near field effect and the beam squint effect & Location-assisted generalized multiple measurement vector orthogonal matching pursuit algorithm,  dictionary-based localization scheme, polar-domain gradient descent algorithm& Accuracy\\    
         \hline

         R & \cite{10193850} & 30 GHz & & MISO & Enhancing RIS-equipped \acrshort{uav}s' performance & geometry-based simultaneous localization and phase shift method &  Accuracy\\
         \hline

         R & \cite{10570712} & 30 GHz &&&Enhancing 6G user-centric sensing with \acrshort{ris} for improved localization and sensing & geomtry based modeling & Accuracy and sensing performance efficiency\\
         \hline

         R & \cite{10571090}&&DL&MISO&Target localization system utilizing \acrshort{ris} and passive radars to enhance \acrshort{isac} capabilities & Localization alogirthm based on \acrshort{aoa} and \acrshort{toa}& Accuracy, detection probability, successful detection probabilit\\
         \hline

         R & \cite{10108969} & 73 GHz & DL & MISO& Environment aware joint active/ passive beamforming & Channel knowlegde map& Energy efficinecy, spectrum efficiency\\
         \hline

         R & \cite{10234214} & 30 GHz & DL & MIMO &  Fundamental localization bounds in \acrshort{ris}-assisted \acrshort{ofdm} systems & \acrshort{fim}& \acrshort{peb}, \acrshort{oeb}\\
         \hline

         R& \cite{10233198} & 30 GHz &&&\acrshort{bs} free \acrshort{ris}-assisted \acrshort{slam}& RIS phase profile design, marginal Poisson multi-Bernoulli \acrshort{slam} filter, \acrshort{crlb}& \acrshort{peb}, gains, accuracy\\
         \hline

         R & \cite{10464638} & 26 GHz & DL & SISO& \acrshort{ris}-assisted positioning approach & \acrshort{fim}, \acrshort{mse} & Accuracy\\
         \hline

         S & \cite{10464850} & 28 GHz & UL & MISO & \acrshort{star}-\acrshort{ris}-assisted bilateral user localization & \acrshort{csi}, \acrshort{ml} & Accuracy\\
         \hline

         R & \cite{10542461} & 28 GHz & DL & MISO & Joint 3D localization and synchronization using multiple RISs & \acrshort{dl} & Accuracy, complexity\\
         \hline

         R & \cite{10471980} & & UL &MISO & Dual-RIS assisted 3D localization and beamforming design in ISAC system & \acrshort{cs}, stepwise matching pursuit  algorithm & \acrshort{mse}\\
         \hline

         H & \cite{10225368} & 30 GHz & DL & MISO & Joint user localization and hybrid \acrshort{ris}location Calibration & Geometry based channel and state estimation, \acrshort{crlb}& \acrshort{mse}  \\
         \hline
 
         \thickhline
    \end{tabular}
\end{table*}
\end{scriptsize}

\begin{scriptsize}
\begin{table*}[t!]
\centering
    \caption{ SUMMARY OF \acrshort{ris}-ASSISTED OUTDOOR AND FAR-FIELD LOCALIZATION ARTICLES AT HIGH FREQUENCY BANDS. HERE ``R" REFERS TO ``REFLECTIVE \acrshort{ris}"}
\begin{tabular}{p{0.01\linewidth}p{0.04\linewidth}p{0.05\linewidth}p{0.02\linewidth}p{0.03\linewidth}p{0.25\linewidth}p{0.26\linewidth}p{0.14\linewidth}}
    \label{table6} \\
         \thickhline
         \textbf{\acrshort{ris}} & \textbf{Ref} & \textbf{$f_c$} & \textbf{Link} & \textbf{System}& \textbf{Purpose} & \textbf{Technique}& \textbf{Performance metric} \\
         \thickhline
         R & \cite{10571832} & 100 GHz & DL & MIMO & User sensing and localization & Tensor, parallel factor method, alternating least squares algorithm & Accuracy\\
         \hline    

         R & \cite{10464904} & & &SISO&  \acrshort{ris}-assisted cooperative localization &  Cooperative localization, \acrshort{crlb}, Polyblock-based algorithm, heuristic algorithm& Accuracy\\
         \hline

         R & \cite{10188365} & 30 GHz & DL & SISO & Autonomous receiver localization and tracking &  Complex extended Kalman filter & Accuracy\\
         \hline

         R & \cite{10543050} & & UL& MISO&  \acrshort{isac} systems design & Structure-aware sparse Bayesian learning framework, simultaneous communication and localization algorithm for multiple users& Accuracy, spectral efficiency\\
         \hline

         R & \cite{10283493}&&DL & MIMO& Limitation of \acrshort{ris}-assisted localization& \acrshort{fim}, Bayesian \acrshort{fim}& Accuracy\\
         \hline

         R& \cite{10582646} & 28 GHz & & MIMO & Joint channel estimation and localization & \acrshort{cs}, simultaneous orthogonal matching pursuit & Accuracy\\
         \hline

         R & \cite{10416244} & 300 GHz && MIMO & \acrshort{isac} & Multi-group Khatri-Rao space-time coding scheme, low-complexity least squares Khatri-Rao factorization algorithm, optimized five linear alternating least squares algorithm & Accuracy, bit error rate\\
         \hline

         R & \cite{10143971} & 70 GHz & DL & MISO & Target-to-user association in \acrshort{isac} Systems & Mount RIS on the roof of the vehicular \acrshort{ue} & Accuracy\\
         \hline

         R & \cite{10436573} & & DL & MISO & \acrshort{isac}& Modeling based heterogeneous networked sensing architecture& Accuracy\\
         \hline

         R & \cite{10565756}& 28GHz, 5.9GHz& UL& MIMO & Multi-band \acrshort{isac} & $\text{X}^{2}$ Track framework design, \acrshort{dl} & Accuracy\\
         \hline

         R & \cite{10552748} & 28 GHz& UL & MIMO & ISAC & Tensor, \acrshort{dl}& Accuracy\\
         \hline

          \thickhline
    \end{tabular}
\end{table*}
\end{scriptsize}

Authors in \cite{9860999} investigate the potential of \acrshort{ris} in replacing the function of a remote cell for \acrshort{dl}-\acrshort{tdoa} measurements in 6G positioning. The study demonstrates that the TDOA between the LoS path and the reflected path through the \acrshort{ris} can effectively replace \acrshort{dl}-\acrshort{tdoa} measurements, enabling accurate localization within a single cell. Simulation results indicate that \acrshort{ris}-enabled localization achieves positioning accuracy comparable to the traditional two-cell structure, offering a cost-effective solution. Authors present an efficient \acrshort{csi} acquisition method for a \acrshort{ris}-aided communication system in \cite{9538860}. They propose a coplanar maximum likelihood-based \acrshort{3d} localization approach and utilize the concept of reflecting unit set to acquire channel information with minimal training resources. Study indicates substantial performance improvements in terms of the \acrshort{snr} of the received signal. Article \cite{9777939} presents a low-overhead joint localization and channel reconstruction scheme for extra-large \acrshort{ris}-assisted \acrshort{mimo} systems. The proposed scheme accurately identifies the visibility region of each user, achieves centimeter-level user localization accuracy, and obtains more accurate channel reconstruction results compared to existing works. The results demonstrate the potential of \acrshort{ris} for improving communication and sensing integration. 

Authors in \cite{9540372} focus on the challenges of channel estimation and user localization in a \acrshort{ris}-assisted \acrshort{mimo}-\acrshort{ofdm} system. The article proposes a unique twin-\acrshort{ris} structure that incorporates spatial rotation to extract the \acrshort{3d} propagation channel. They employ tensor factorization, sparse array processing, and atomic norm denoising techniques to design training patterns and recover the associated parameters. By decoupling the channel's angular and temporal parameters, they achieve precise channel parameter extraction and centimeter-level positioning resolution.  A two-stage method is proposed in \cite{he2023compressed}, utilizing the tunable reflection capability of passive \acrshort{ris} and the multi-reflection wireless environment. The first stage employs an off-grid \acrshort{cs} approach to estimate the angles of arrival associated with each \acrshort{ris}, followed by a maximum likelihood location estimation in the second stage. The study demonstrates the high accuracy of the proposed \acrshort{3d} localization method, consistent with the theoretical \acrshort{crlb} analysis. 

Authors in \cite{10234214} establish fundamental localization bounds for \acrshort{ofdm} systems aided by \acrshort{ris}. It derives Bayesian localization bounds using the Bayesian \acrshort{fim} and the equivalent \acrshort{fim}. The study reveals that constant \acrshort{ris} reflection coefficients across \acrshort{ofdm} symbols make angle estimation impossible, necessitating varying coefficients for effective localization. It also decomposes the \acrshort{fim} for \acrshort{ris}-related parameters into contributions from the receiver, transmitter, and RIS components, and concludes that single-antenna \acrshort{ue} localization via a single RIS reflection is infeasible in the far-field with constant RIS coefficients. A RIS-assisted positioning method for high-precision user localization, considering the incident planar wavefront on the RIS is proposed in \cite{10464638}. It derives the \acrshort{mse} to evaluate system accuracy and offers solutions for optimal RIS placement and configuration. Numerical analyses are conducted to assess performance across different scenarios, focusing on RIS location, element count, and communication bandwidth. Results indicate that RIS significantly enhances positioning accuracy in 6G networks.

\paragraph{Advanced Techniques and Systems}
In \cite{He2022SimultaneousIA}, the authors investigate the potential of \acrshort{star}-\acrshort{ris} for enhanced indoor and outdoor localization. They study the fundamental limits of \acrshort{3d} localization performance using Fisher information analysis and optimize the power splitting between refraction and reflection at the STAR-\acrshort{ris}, as well as the power allocation between the UE. The results indicate that high-accuracy \acrshort{3d} localization can be achieved for both indoor and outdoor UEs when the system parameters are well optimized, demonstrating the potential of \acrshort{star}-\acrshort{ris} in concurrent localization. Existing \acrshort{ris}-aided localization approaches assume perfect knowledge of the \acrshort{ris} geometry, which is not realistic due to calibration errors. The authors in \cite{10096904} derive the \acrshort{mcrb} for localization with \acrshort{ris} geometry mismatch and propose a closed-form solution for determining pseudo-true parameters. Numerical results validate the derived parameters and \acrshort{mcrb}, demonstrating that \acrshort{ris} geometry mismatch leads to performance saturation in high \acrshort{snr} regions. The article \cite{Keykhosravi2022RISEnabledSL} introduces a concept of 3D UE self-localization using a single \acrshort{ris}. The approach involves the UE transmitting multiple OFDM signals and processing the reflected signal from the \acrshort{ris} to estimate its position. The estimation process includes separating the \acrshort{ris}-reflected signal from the undesired multipath, obtaining a coarse position estimate, and refining the estimation through maximum likelihood techniques. The performance of the estimator is evaluated in terms of positioning error and compared to an analytical lower bound. The results demonstrate the potential of \acrshort{ris} as an enabling technology for radio localization, offering improved positioning accuracy. 

Authors in \cite{10124713} introduce the concept of partially-connected receiving \acrshort{ris} that can sense and localize users emitting electromagnetic waveforms. The receiving \acrshort{ris} hardware architecture consists of meta-atom subarrays with waveguides that direct the waveforms to reception RF chains for signal and channel parameter estimation. The focus is on far-field scenarios, and a \acrshort{3d} localization method is presented based on narrowband signaling and \acrshort{aoa} estimates using phase configurations of meta-atoms. The results include theoretical CRLBs and extensive simulations, demonstrating the effectiveness of the proposed receiving \acrshort{ris}-empowered 3D localization system, providing cm-level positioning accuracy. The impact of various system parameters on localization performance is also evaluated, such as training overhead, distance between R-\acrshort{ris} and the user, and spacing among R-\acrshort{ris} subarrays and their partitioning patterns. The joint calibration and positioning problem in an uplink system with an active \acrshort{ris} is addressed in \cite{zheng2023jrcup}. Existing approaches often assume known positions and orientations for \acrshort{ris}, which is not realistic for mobile or uncalibrated \acrshort{ris}. The proposed two-stage method includes a tensor-ESPRIT estimator followed by parameter refinement and a 2D search-based algorithm to estimate user and \acrshort{ris} positions, \acrshort{ris} orientation, and clock bias. The derived CRLBs verify the effectiveness of the algorithms, and simulations show that the active \acrshort{ris} significantly improves localization performance compared to the passive case. Blind areas that limit localization performance can be mitigated by providing additional prior information or deploying more BSs. 

A novel RIS-enabled \acrshort{slam} framework is persented in \cite{10233198} for 6G wireless systems that operates without access points. \acrshort{ris} phase profiles are designed to illuminate the likely location of \acrshort{ue}. The modified Poisson multi-Bernoulli \acrshort{slam}  filter estimates \acrshort{ue} state and maps the radio environment efficiently. Theoretical \acrshort{crlb}  are derived for channel parameters and UE state estimations. Performance evaluations show the method's effectiveness in scenarios with limited transmissions and channel coherence time considerations. This study performed in \cite{10464850} introduces a high-precision 6G positioning method using \acrshort{isac} and STAR-RIS. It employs the CsiNet-Former network to extract spatial features from CSI for accurate 3D user localization and trajectory prediction. Simulation results show that in energy splitting mode, the method achieves an average localization error of 0.75 m and a single-side error of 0.034 m, outperforming existing methods and covering a larger area. The authors in \cite{10225368} present a method for joint localization of a hybrid RIS and a user, using a single radio frequency chain for reflections and sensing. A multi-stage approach estimates hybrid RIS position, orientation, and user location where simulations confirm the method's efficiency and identify an optimal HRIS power splitting ratio.

\paragraph{Advanced Beamforming and Phase Shifter Designs}
The adaptive beamforming of \acrshort{ris}-assisted mmW MIMO placement with obstructed \acrshort{los} between the \acrshort{bs} and the UE is studied in \cite{9124848}. The authors suggest a hierarchical codebook and receiver feedback-based adaptive phase shifter architecture to optimize the phase of each of the \acrshort{ris} units and in turn optimize performance in terms of localization accuracy and data rate. Authors in \cite{9238887} have proposed a two-stage localization technique using dual \acrshort{ris}. The reflecting element's phase shift is first designed for each \acrshort{ris}, and then in the second stage, the location data is calculated and it demonstrates the localization accuracy in the range of $10^{-5}$–$10^{-4}$ meters. Article \cite{9771819} explores the potential of \acrshort{ris} for improving cooperative localization performance in mmW MIMO systems. The paper presents a study on the fundamental limits of cooperative localization using the CRLB and proposes an optimal phase design at the \acrshort{ris} to enhance position accuracy. The study demonstrates that the proposed optimal passive beamforming algorithm significantly improves localization accuracy, and  achieves near-optimal performance with minimal computational complexity. Authors in \cite{9410435} suggest a simultaneous beam training and placement technique to address the \acrshort{los} obstruction in mmW MIMO network. The UE estimates its location using the \acrshort{aod}, which is determined via beam training. The location of UE, in turn, helps to improve the beam training. Results demonstrate that the proposed approach can obtain centimeter-level multi-user localization accuracy. 

A low-complexity method for joint localization and synchronization in \acrshort{mmw} systems using \acrshort{ris} is proposed in \cite{9782100}. Their approach involves optimizing the beamforming strategies of the BS active precoding and \acrshort{ris} passive phase profiles, considering a single-antenna receiver. The results indicate that the proposed joint BS-\acrshort{ris} beamforming scheme achieves enhanced localization and synchronization performance compared to existing solutions, with the proposed estimator achieving the theoretical bounds even under challenging conditions such as low \acrshort{snr} and uncontrollable multipath propagation. The authors in \cite{xanthos2022joint} focused on the successive localization and beamforming design of a \acrshort{ris}-aided \acrshort{mmw} communication system. They formulated the problem as a multivariable coupled non-convex problem and proposed an alternating optimization algorithm to solve it. The results showed that their proposed scheme, called joint localization and beamforming optimization, significantly improved the performance compared to existing joint localization and beamforming methods, as demonstrated through simulation results. 

The authors in \cite{10193850} proposed simultaneous localization and phase Shift method for enhancing RIS-equipped \acrshort{uav}s in 6G networks. It focuses on integrating accurate localization with phase shift adjustments to improve coverage and address blockage issues. The method utilizes geometry-based localization to optimize passive beam-steering, benefiting aerial line-of-sight communication, especially in vehicle-to-vehicle scenarios. Simulation results validate the effectiveness of the method, showing significant improvements in communication performance and safe navigation in complex urban environments. This study \cite{10108969} explores environment-aware active and passive beamforming for \acrshort{ris}-aided communication using Channel Knowledge Map, which eliminates the need for online training and reduces overhead in mmWave systems while maintaining high energy efficiency. Simulation results show that CKM significantly enhances beamforming performance compared to traditional training-based methods and remains robust against errors in \acrshort{ue} location.

\paragraph{Multiuser and Joint Communication and Localization Approaches}
Authors in \cite{9513781} examined a multiuser localization method based on an hierarchical codebook design in light of the \acrshort{los} obstruction scenario. The study results under various \acrshort{snr} situations demonstrate that, with the right hierarchical codebook design, the suggested approach has the ability to provide multiuser localization in mmW MIMO radar systems.  In \cite{10054103}, authors present an \acrshort{ris}-aided mmW-MIMO system for \acrshort{jcal}. They derive closed-form expressions of CRLB for position/orientation estimation errors and effective achievable data rate based on \acrshort{ris} phase shifts. They propose a joint optimization algorithm to balance the trade-off between the two metrics, and simulation results demonstrate the effectiveness of the algorithm in terms of estimation accuracy and effective achievable data rate, even in the presence of estimation errors and user mobility. Authors in article \cite{Fascista2020RISAidedJL} address the problem of joint localization and synchronization in a mmW MISO system using a \acrshort{ris}. They formulate the joint maximum likelihood estimation problem in the position domain and propose a reduced-complexity decoupled estimator for position and clock offset. Simulation results demonstrate that their approach achieves high accuracy in localization and synchronization, even in low SNR scenarios, without the need for optimizing transmit beamforming, \acrshort{ris} control matrix, or prior knowledge of the clock offset.
%\FloatBarrier

%%%%%%%%%%%%%%%%%%%%%%%%%%%%%%%%%%%%%%%%%%%%%%%%%%%%5

%\fontsize{7}{8}\selectfont {
 
\paragraph{Localization in Special Scenarios}
Authors in \cite{9774917} address the challenge of positioning a single-antenna user in \acrshort{3d} space by considering the received signal from a single-antenna \acrshort{bs} and the reflected signal from an \acrshort{ris}. They take into account both user mobility and spatial-wideband effects. Initially, a spatial-wideband channel model is derived under the assumption of far-field conditions, focusing on \acrshort{ofdm} signal transmission with a user of constant velocity. \acrshort{crlb} are derived as a benchmark, and a low-complexity estimator is developed to achieve these bounds under high \acrshort{snr} ratios. The proposed estimator compensates for user mobility by estimating radial velocities and iteratively accounting for their effects. The results indicate that spatial-wideband effects can have a detrimental impact on localization accuracy, especially for larger \acrshort{ris} sizes and signal bandwidths deviating from the normal of the \acrshort{ris}. However, the proposed estimator demonstrates resilience against spatial-wideband effects up to a bandwidth of 140 MHz for a 64x64 \acrshort{ris}. Notably, user velocity does not significantly affect the bounds or accuracy of the estimator, indicating that high-speed users can be localized with similar precision as static users. 

The potential of 6G THz systems for localization and comparison with mmW localization systems is performed  in \cite{9782674}. They compare various aspects including system properties, channel modeling, localization problem formulation, and system design. Preliminary simulations demonstrate the potential of THz localization in terms of PEB and OEB compared to mmW systems. The article provides recommendations for efficient localization algorithm design for \acrshort{ris}-assisted adaptive optics-based spatial modulation MIMO systems and highlights the anticipated applications in future communication systems, such as intelligent networks, autonomous transportation, and tactile internet. A framework for \acrshort{ris}-enabled \acrshort{slam} without the need for access points is proposed in \cite{kim2022risenabled}. They design \acrshort{ris} phase profiles based on prior information about the UE, allowing for uniform signal illumination in the UE's probable location. They also modify the Poisson multi-Bernoulli SLAM filter to estimate the UE state and landmarks, facilitating efficient mapping of the radio propagation environment. Theoretical CRLB are derived for the estimators of channel parameters and UE state. The proposed method is evaluated under scenarios with a limited number of transmissions and considering the channel coherence time. Study demonstrates that \acrshort{ris} can solve the radio \acrshort{slam} problem without the need for access points, and incorporating the Doppler shift improves UE speed estimates.

\paragraph{High Frequency Sensing Operations}
The authors in \cite{10130575} perform the sensing of the user's uplink channel and location in \acrshort{thz} extra-large \acrshort{ris} systems. The authors propose a joint channel and location sensing scheme that includes a location-assisted generalized multiple measurement vector orthogonal matching pursuit algorithm for channel estimation and a complete dictionary-based localization scheme. They address challenges such as the hybrid far-near field effect and beam squint effect caused by the extra large array aperture and extra large bandwidth. The proposed schemes demonstrate superior performance compared to existing approaches, as indicated by simulation results. They also introduce a partial dictionary-based localization scheme to reduce sensing overhead, where the \acrshort{ris} serves as an anchor for user localization using \acrshort{tdoa}. An \acrshort{isac} scenario using \acrshort{ris} is investigated in \cite{10001508} where multiple devices communicate with a BS in full-duplex mode while simultaneously sensing their positions. \acrshort{ris} are mounted on each device to enhance reflected echoes, and device information is passively transferred to the BS through reflection modulation. The problem of joint localization and information retrieval is addressed by constructing a grid-based parametric model and formulating it as a CS problem. An expectation-maximization algorithm is applied to tune the grid parameters and mitigate model mismatch. The efficacy of various CS algorithms is analyzed using the Bayesian Cramér-Rao bound. Numerical results demonstrate the feasibility of the proposed scenario and the superior performance of the EM-tuning method. 

The challenge of enhancing user-centric sensing capabilities in future 6G networks is studied in \cite{10570712}. The study explores the use of \acrshort{ris} to support monostatic sensing, improving localization accuracy and sensing performance despite the limited hardware capabilities of user equipment. The proposed method leverages RIS to create a more effective sensing environment, showcasing significant enhancements in sensing precision and efficiency through simulation results. Authors in \cite{10571090} have introduced a novel target localization system that combines \acrshort{ris} and passive radars for enhanced \acrshort{isac}. The system leverages the preamble of communication signals to perform detection and localization tasks. RIS helps passive radar detect targets obscured by obstacles within the propagation channel. The proposed algorithm is designed to uniquely position targets by estimating their \acrshort{aoa} and \acrshort{toa}, and it can detect the number of targets present. Simulation results show that the localization method performs effectively across various detection metrics and with increasing RIS sizes.

Authors in \cite{10471980} address high-frequency fading in 6G ISAC systems using dual-RIS for 3D localization and beamforming. They proposes a stepwise matching pursuit algorithm for accurate, low-complexity localization and uses this data for RIS beamforming to maximize system rate with a triangle inequality-based alternating optimization algorithm. Simulations show the proposed method achieves near-optimal rates, confirming its effectiveness. The authors in \cite{10416244} present a nested tensor-based algorithm for \acrshort{isac} in \acrshort{ris}-assisted THz MIMO systems. It uses Khatri-Rao space-time coding and RIS phase shifts to model the received signal as nested tensors. The algorithm includes low-complexity factorization for joint channel estimation and symbol detection, and optimized estimation of sensing parameters. Simulations show it outperforms current methods with more relaxed parameter conditions.

\paragraph{Practical Localization Scenarios and Applications}
The authors in \cite{10100675} focus on leveraging \acrshort{ris} to enhance communication performance when the \acrshort{los} path between the UE and BS is blocked. The authors propose a novel framework that integrates localization and communications by fixing \acrshort{ris} configurations during location coherence intervals and optimizing BS precoders every channel coherence interval. This approach reduces pilot overhead and the need for frequent \acrshort{ris} reconfiguration. The framework utilizes accurate location information from multiple \acrshort{ris}, along with novel \acrshort{ris} optimization and channel estimation methods. The results indicate improved localization accuracy, reduced channel estimation error, and increased achievable rate, demonstrating the effectiveness of the proposed approach.
Authors in \cite{chen2023RISs} focus on the requirements of localization and sensing in the context of smart cities and highlight the limitations of traditional communication infrastructure for meeting localization and sensing demands. The authors argue that \acrshort{ris} and sidelink communications are promising technologies that can address the localization and sensing needs of smart cities. They propose and evaluate anchor point-coordinated and self-coordinated \acrshort{ris}-enabled localization and sensing architectures, considering different application scenarios such as low-complexity beacons, cooperative localization, and full-duplex transceivers. The article also discusses practical issues and research challenges associated with implementing these localization and sensing systems. 

Authors in \cite{chen2023multirisenabled} address the importance of localization in intelligent transportation systems and explore the use of reflective \acrshort{ris} to enhance high-precision localization. The authors propose a two-stage 3D sidelink positioning algorithm that utilizes at least two \acrshort{ris} and sidelink communication between UEs to achieve localization without the involvement of BSs. They evaluate the effects of multipath and \acrshort{ris} profile designs on positioning performance, analyze localizability in various scenarios, and propose solutions to eliminate blind areas. The study demonstrates the promising accuracy of the proposed BS-free sidelink communication system in challenging intelligent transportation systems scenarios. A low-complexity, two-stage method for 3D localization and synchronization using multiple RISs in multipath scenarios is proposed in \cite{10542461}. First, signals are preprocessed and a \acrshort{dl} model estimates \acrshort{aod}. An initial user position is then estimated using an asynchronous \acrshort{aod} \acrshort{toa} approach. Finally, an optimization refines the position using estimated delays and clock offset. This method improves accuracy and robustness against multipath effects.
\begin{scriptsize}
\begin{table*}[!]
    \centering
    \caption{ SUMMARY OF \acrshort{ris}-ASSISTED INDOOR LOCALIZATION ARTICLES. HERE ``R" REFERS TO "REFLECTIVE \acrshort{ris}, ``T" REFERS TO ``TRANSMISSIVE \acrshort{ris}", ``H" REFERS TO ``HYBRID \acrshort{ris}"} 
\begin{tabular}{p{0.02\linewidth}p{0.04\linewidth}p{0.05\linewidth}p{0.02\linewidth}p{0.03\linewidth}p{0.24\linewidth}p{0.25\linewidth}p{0.14\linewidth}}
    \label{table4} \\
         \thickhline
         \textbf{\acrshort{ris}} & \textbf{Ref} & \textbf{$f_c$} & \textbf{Link} & \textbf{System}& \textbf{Purpose} & \textbf{Technique}& \textbf{Performance metric} \\
         \thickhline

 %       R& \cite{nguyen2020reconfigurable} & 2.4 GHz & &SISO& Exploiting \acrshort{ris} to generate and select easily differentiable radio maps for use in wireless fingerprinting & Machine Learning and fingerprinting& accuracy       \\ \hline

         R & \cite{9193909} & 2.4 GHz & UL & SISO &  Enhancing the accuracy of \acrshort{rss} based localization & Configuration optimization iterative algorithm & Accuracy
         \\
         \hline
         R & \cite{9456027} & 2.4 GHz & DL& SISO &  Enhancing the accuracy of RSS based positioning & particle swarm optimization algorithm & Accuracy
         \\
         \hline
         
         R& \cite{9201330} & 2.4 GHz & DL  & SIMO &Employment of \acrshort{ris} in indoor localization & \acrshort{uwb} technique, CRLB& Accuracy
         \\
         \hline

         R & \cite{9548046} & 2.4 GHz& DL& SISO& Fingerprinting
        localization estimation using \acrshort{ris} & Supervised learning feature selection method, localization heuristics states selection framework& Accuracy\\
        \hline

         R & \cite{isprs} & 2.4 GHz &UL  & MISO&Investigating multiple RSS fingerprint based localization & CRLB, projected gradient descent optimization, DNN& Accuracy
         \\
         \hline

         R& \cite{10118610} & 2.4 GHz &DL &SIMO & Passive person localization & Phase control optimization algorithm, side-lobe Cancellation Algorithm & Accuracy\\
         \hline

         R & \cite{zhang2023multi} & 2.4 GHz & DL & SIMO & Passive multi person localization & Phase control optimization algorithm, side-lobe Cancellation Algorithm & Accuracy\\
         \hline

        R & \cite{8815412} & 2.6 GHz& UL &MISO & Method for efficient online wireless configuration of \acrshort{ris} & \acrshort{dnn} & Throughput
         \\
         \hline
         
         R & \cite{vaca2023radio} & 3.5 GHz&&& Radio sensing & ML and computer vision (clustering, template matching and component labeling) & Accuracy\\
         \hline

        R & \cite{9827014} & 5 GHz & DL & SISO & Fingerprint-based indoor localization system using \acrshort{ris} & \acrshort{ris} configuration design & Accuracy \\
         \hline

        R & \cite{10124023}& 5 GHz& DL & MISO & \acrshort{ris}-enabled fingerprinting-based localization& Deep \acrshort{rl}& Accuracy\\
         \hline

         R & \cite{9836317} & 5 GHz & UL &  MISO & Distributed \acrshort{ris} assisted localization & Two-step positioning approach, CRLB, theoretical analysis& Accuracy \\
         \hline
         
        R & \cite{9685930} & 10 GHz &DL &SIMO& Indoor wireless SLAM system assisted by a \acrshort{ris}& \acrshort{ris}-assisted indoor SLAM optimization problem and design of error minimization algorithm& Accuracy
         \\
         \hline

         R& \cite{9726785}& 30 GHz&&&User localization with multiple \acrshort{ris}& \acrshort{mle}, least squares line intersection technique & Accuracy\\
         \hline

         R& \cite{9838395}& 60 GHz& DL& MIMO& \acrshort{ris}-assisted downlink mmW indoor localization framework&  Coarse-to-fine localization algorithm with low-complexity
         grid design& Accuracy\\
         \hline

        R & \cite{10123105}& 90 GHz& UL& MIMO & \acrshort{ris} aided UE localization & Space-time channel response vector, residual convolution network regression learning algorithm & Accuracy\\
         \hline
         
         R & \cite{9903366} & 150 GHz & DL & SISO & Optimal \acrshort{ris} placement with respect to position and orientation & Analytical modeling & Received power\\
         \hline

         T & \cite{10333393} & 28 GHz & &&Improve the accuracy of indoor positioning& Fingerprinting, near-field multi-focal focusing method& Accuracy, throughput\\
         \hline

         H & \cite{umer2024} & & DL & SISO& RIS configuration optimization for localization in dynamic rich scattering environment & Physics based simulator, bi-directional long short term memory & Accuracy\\
         \hline

         R &\cite{10188251} &&DL&&Enhancing indoor localization through \acrshort{ris}-Based \acrshort{rss} optimization &  \acrshort{tdoa} based static reconfiguration algorithm & Accuracy\\
         \hline

         R& \cite{10440181} &30Ghz& & SISO & Passive localization and tracking of a RIS-equipped object using monostatic sensing& Temporal coding & Accuracy\\
         \hline

         %R& \textbf{\cite{10124023}} & 5GHz & DL & MISO &\acrshort{ris}-assisted fingerprinting&  deep \acrshort{rl}& accuracy\\

         \thickhline
    \end{tabular}
\end{table*}
\end{scriptsize}

\subsection{\acrshort{ris}-Assisted Indoor Localization}
Studies on \acrshort{ris}-assisted indoor localization are briefly summarized in Table \ref{table4} and discussed as follows. 

\subsubsection{Foundational Techniques}
The multipath signal traveling through each \acrshort{ris} may be labeled, which offers a workable concept for processing them, provided each of the \acrshort{ris} elements has a unique phase, i.e., $\phi_{1} \neq \phi_{2}......\neq \phi_{n} \in [0,2\pi)$. Taking advantage of the high multipath resolution of \acrshort{uwb} signals and the capability of \acrshort{ris} to identify multipath channels, authors in \cite{9201330} created a unique indoor \acrshort{ris}-assisted localization technique.
The suggested localization scheme's CRLB is calculated, demonstrating how \acrshort{ris} has the ability to provide precise location with just one access point. Also, the suggested system offers a more precise and economical option for indoor placement because it only calls for a single access point and a few inexpensive \acrshort{ris} devices. \acrshort{ris} may be used to complement the RSS-based localization technique in many ways, such as,  strengthening the signal received, diminishing co-channel interference, and providing additional propagation paths. As a result, \acrshort{ris} can significantly improve the RSS-based localization algorithms that rely on it. Nevertheless, because it is challenging to tell apart nearby RSS data, the accuracy of such algorithms is constrained. A deep learning method for efficient online wireless configuration of \acrshort{ris} in indoor communication environments is proposed in \cite{8815412}. They use a database of coordinate fingerprints to train a DNN that maps user location information to the optimal phase configurations of the \acrshort{ris}, maximizing the RSS at the intended location. Simulations in a 3D indoor environment show that the proposed DNN-based configuration method effectively increases the achievable throughput at the target user location in all considered cases. 

A \acrshort{ris}-assisted localization scheme utilizing multiple RSS fingerprints and a DNN is presented in \cite{isprs}. The scheme utilizes RSS values obtained under different \acrshort{ris} configurations as fingerprints and employs an optimization method based on the CRLB to find the optimal \acrshort{ris} configurations. A DNN regression network is trained for localization. The simulation results demonstrate that the proposed scheme achieves robust and accurate location estimation, with an accuracy of approximately 0.5 meters in the \acrshort{nlos} scenario. %The authors in \cite{nguyen2020reconfigurable} propose and evaluate a novel \acrshort{ml} method for wireless fingerprinting localization in \acrshort{ris}-assisted environments. The approach combines off-the-shelf components such as k-nearest neighbors localization and genetic algorithms, leveraging the capabilities of \acrshort{ris} to create a smart reconfigurable radio environment. The results demonstrate that this approach achieves excellent localization accuracy, eliminating the need for multiple access nodes and extensive fingerprint grid sample points. %The study suggests that \acrshort{ris} and smart radio environments have the potential to enable sub-meter localization accuracy, and future research should explore more challenging scenarios involving mixed LoS and NLoS environments, higher frequencies, multiple \acrshort{ris} elements, and multiple \acrshort{ris} deployments.

\subsubsection{Accuracy Improvement with \acrshort{ris}}
By theoretical analysis and practical testing, it has been shown in \cite{9293395} that \acrshort{ris} can in fact customize the wireless environment. Authors have clearly shown with the help of measurements that \acrshort{ris} configuration changes the RSS at a particular location. Thus, the issue of similar RSS values from nearby sites can be resolved in smart radio environments enabled by the \acrshort{ris}. Authors in \cite{Zhang2020TowardsUP} have designed an \acrshort{ris}-assisted localization algorithm that is focused on enhancing localization accuracy. To do so, an iterative configuration optimization algorithm is proposed whose purpose is to select the \acrshort{ris} configuration that improves the localization accuracy. The localization accuracy of the suggested technique is substantially higher than that of the localization method without \acrshort{ris}. The authors also designed a Phase shift optimization technique to address the same issue in \cite{9456027}. This approach offers a unique solution to the multiuser localization problem and can minimize localization error by at least three times when compared to the conventional RSS-based solution. 

Work presented in \cite{10188251} focuses on improving the accuracy of \acrshort{rss}-based indoor positioning system by addressing the limitations caused by temporal and spatial uncertainties in the indoor wireless environment. An algorithm is introduced that utilizes \acrshort{ris} technology to optimize \acrshort{rss} values at specific reference points. This is achieved by adjusting the \acrshort{ris} reflection coefficients to enhance the differentiation of \acrshort{rss} values between reference points. Simulation results indicate that this method, even with a relatively small number of RIS elements, significantly boosts the efficiency and accuracy of indoor positioning. Authors in \cite{10124023} present an \acrshort{ris}-enabled fingerprinting-based localization method enhanced by deep \acrshort{rl}. The methodology involves creating a database of \acrshort{rss} lists through periodic \acrshort{ris} configurations, which are then used to estimate receiver positions by comparing them with online-measured \acrshort{rss} data via the nearest neighbor algorithm. Additionally, a deep \acrshort{rl}-based \acrshort{ris} configuration selector is developed to optimize \acrshort{ris} configurations and minimize localization error. A communication protocol for the system's operation is also proposed. Extensive simulations demonstrate that the localization accuracy improves with an increasing number of \acrshort{ris} configurations and is further enhanced by over 40\% with the incorporation of deep \acrshort{rl} compared to previous methods.

\subsubsection{Wireless Indoor Simultaneous Localization and Mapping} Researchers developed a \acrshort{ris}-assisted wireless indoor SLAM system in \cite{9685930}. Channel models incorporating \acrshort{ris} are proposed, and a \acrshort{ris}-aided SLAM protocol is introduced to coordinate the \acrshort{ris} and the agent. An optimization problem for SLAM is formulated and solved using a particle filter-based localization and mapping algorithm. The study demonstrates that the \acrshort{ris} significantly enhances channel amplitudes compared to scattered environments. Furthermore, the \acrshort{ris}-assisted SLAM system reduces agent estimation errors by 0.1 meters compared to non-\acrshort{ris} wireless SLAM systems. The article \cite{10118610} presents a framework for passive human localization using WiFi signals enhanced by \acrshort{ris}. The \acrshort{ris}, consisting of controllable reflective elements, overcomes the limited spatial resolution of WiFi devices to achieve accurate localization. The proposed framework includes a phase control optimization algorithm to maximize the discrepancy between human reflection and multipath interference. Simulation results demonstrate sub-centimeter accuracy in locating moving individuals passively, even in the presence of noise and multipath interference. As an extension, the article \cite{zhang2023multi}  addresses the challenge of achieving accurate passive multi-human localization using commodity WiFi devices. A side-lobe cancellation algorithm is introduced to achieve accurate localization iteratively. Results indicate that the proposed framework enables sub-centimeter accurate localization of multiple moving individuals without modifications to existing WiFi infrastructure, even in the presence of multipath interference and random noise. 

A new application for \acrshort{ris}-enabled passive localization and tracking is introduced in \cite{10440181}. It utilizes a single-antenna full duplex transceiver to send multiple \acrshort{ofdm} signals and analyze the reflections from a \acrshort{ris}-equipped object in an efficient use of the spectrum. The RIS phase profile is designed to separate desired signal reflections from unwanted multipath signals, allowing for the estimation of the object's location and velocity with low-complexity estimators. Simulations show that this method can achieve centimeter-level location accuracy with just 6 MHz of bandwidth.

\subsubsection{Wireless Indoor Integrated Sensing and Communication}
Authors in \cite{10333393} introduced a \acrshort{ris}-aided method to enhance \acrshort{rss} fingerprinting-based indoor positioning accuracy in 6G networks. By utilizing \acrshort{ris} diversity and requiring \acrshort{rss} from a single access point, it achieves centimeter-level precision. The proposed near-field multi-focal focusing method maintains communication performance while supporting high-precision positioning. Simulations confirm the scheme's effectiveness, offering improved positioning accuracy with only a 10\% reduction in average throughput.

%\subsubsection{\acrshort{ris}'s Impact on Indoor Localization Algorithms and Systems}
%Article \cite{9201330} presents a general model for ultra-wideband (UWB)-aided \acrshort{ris}-assisted indoor positioning and derives the CRLB for position estimation. The \acrshort{ris} acts as a channel marker, while UWB signals are used for multipath recognition. Theoretical derivations and simulation results demonstrate that the combination of \acrshort{ris} and UWB signals enables accurate indoor positioning with a single access point. Compared to conventional methods using virtual anchors, the proposed \acrshort{ris} positioning scheme offers wider applicability, significant cost reduction, and a more accurate and cost-effective solution for indoor positioning. 

\subsection{\acrshort{ris}-Assisted Near-Field Localization}
\label{nearfield}
Studies on \acrshort{ris}-assisted near-field localization are briefly summarized in Table \ref{table9} and discussed as follows. 

%By leveraging the wavefront curvature, authors in \cite{9500663} explore the potential of localization with minimal infrastructure and hardware complexity. Specifically, the focus is on utilizing \acrshort{ris} as lens receivers for localization. The authors conduct a thorough analysis, including Fisher information analysis and evaluation of different lens configurations, and propose a two-stage localization algorithm. The simulation results demonstrate that positional beamforming yields superior performance with prior location information, while random beamforming is more suitable in the absence of prior information. The achieved accuracy at 28 GHz, using a moderate-sized lens, is on the decimeter level within a proximity of 3 meters to the lens.

%Article \cite{9508872} focuses on the use of \acrshort{ris} in smart radio environments for communication and localization purposes. The authors present the localization performance limits for scenarios where a single base station infers the position and orientation of a UE in a \acrshort{ris}-assisted environment. They derive the CRLB and propose a closed-form \acrshort{ris} phase profile, showing that the proposed scheme achieves remarkable performance, even in asynchronous signaling, and approaches the optimal phase design for minimizing the CRLB.
\begin{scriptsize}
\begin{table*}[t!]
\centering
    \caption{ SUMMARY OF \acrshort{ris}-ASSISTED NEAR-FIELD LOCALIZATION ARTICLES. HERE ``R" REFERS TO ``REFLECTIVE \acrshort{ris}", ``H" REFERS TO ``HYBRID \acrshort{ris}", ``A" REFERS TO ``ACTIVE RIS"} \begin{tabular}{p{0.02\linewidth}p{0.04\linewidth}p{0.05\linewidth}p{0.02\linewidth}p{0.03\linewidth}p{0.24\linewidth}p{0.25\linewidth}p{0.14\linewidth}}
    \label{table9} \\
         \thickhline
         \textbf{\acrshort{ris}} & \textbf{Ref} & \textbf{$f_c$} & \textbf{Link} & \textbf{System}& \textbf{Purpose} & \textbf{Technique}& \textbf{Performance metric} \\
         \thickhline
         R& \cite{9674914}& 3 GHz&DL&SISO& \acrshort{rss} based localization algorithms & Weighted least square and alternate iteration methods&Accuracy\\
         \hline
         
        R & \cite{9625826}& 3.5/28 GHz & DL & SISO & UE localization under \acrshort{los} and \acrshort{nlos} conditions & Practical signaling and positioning algorithms design based on an OFDM, \acrshort{ris} time-varying reflection coefficients design& Coverage, accuracy\\
        \hline
         
         R & \cite{9500663} & 28 GHz &DL &SISO & Localization of a transmitter using a \acrshort{ris}-based lens & FIM, two stage localization algorithm& Accuracy
         \\
         \hline
         R & \cite{9508872} & 28 GHz  &UL &MIMO&  Localization performance limits in single BS and \acrshort{ris}-assisted UE localization & CRLB, signaling model design applicable for near and far-field localization & Accuracy
         \\
         \hline
         R & \cite{9593200} & 28 GHz & DL&SISO& Potential to exploit wavefront curvature in geometric near-field conditions & FIM&Accuracy
         \\
         \hline
         R & \cite{9593241} & 28 GHz & DL &SISO &Localization of UE under \acrshort{nlos} & Propose a low complexity algorithm& Accuracy, latency, robustness, coverage
         \\
         \hline

         R & \cite{Gong2022AsynchronousRL} & 28 GHz & UL& MISO & Performance limits of the \acrshort{ris}-based near-field localization in the asynchronous scenario, impact of cascaded channel on the localization performance & FIM, PEB & Accuracy
         \\
         \hline
         
         R & \cite{9650561} & 28 GHz &DL & SISO &   Near-field regional target localization with the \acrshort{ris}-assisted system &  Iterative entropy regularization based algorithm for \acrshort{ris} phase design, near-field target localization algorithm& Accuracy
         \\
         \hline 

         R & \cite{9838889}& 28 GHz & DL& SISO & \acrshort{ris}-assisted near-field localization system under hardware impairment & MCRB, PEB, mismatched maximum likelihood estimator & Accuracy\\
         \hline

         R&\cite{9860413}& 28 GHz& DL& SISO &  Suitable phase profiles design at a reflective \acrshort{ris} to enable \acrshort{nlos} localization & PEB, localization-optimal phase profile design & Accuracy\\
         \hline

        R & \cite{10000689} &28 GHz & UL & MIMO & UE localization in near-field& Atomic norm minimization& Accuracy\\
        \hline

        R & \cite{9940381} & 28 GHz& DL& SISO & Multiuser localization using \acrshort{ris} and cooperative links & CRLB, iterative searching algorithm & Accuracy, power allocation\\
        \hline

        R & \cite{9902991} & 28 GHz  & &MISO& Integration of holographic \acrshort{ris} into mmW localization system & FIM, CRLB, iterative entropy regularization based \acrshort{ris} phase optimization& Accuracy
         \\
         \hline

         R & \cite{10017173} & 28 GHz & DL& SISO & Near-field localization of a UE under phase-dependent amplitude variations at each \acrshort{ris} element & Low-complexity approximated mismatched maximum likelihood estimator, iterative refinement algorithm to update individual parameters of the \acrshort{ris} amplitude model, MCRB& Accuracy
         \\
         \hline

         R &\cite{rahal2023performance} & 5.15/28 GHz & DL & SISO &  Optimizing the precoders that control \acrshort{ris} under hardware constraints & Low-complexity algorithm design for \acrshort{ris} configuration, FIM, PEB& Accuracy\\
         \hline

        R & \cite{ozturk2023risaided}& 28 GHz & DL & SISO & \acrshort{ris}-aided Localization under Pixel Failures & MCRB, joint localization and failure diagnosis method& Accuracy \\
        \hline

        H & \cite{9921216} & 28 GHz& UL & MISO & Hybrid \acrshort{ris}-assisted UE localization& CRLB, automatic differentiation-based gradient descent approach& Accuracy\\
        \hline

        R & \cite{9941256}& 30 GHz & DL & SIMO/ SISO & Design \acrshort{ris} coefficient to convert planar waves into spherical waves and cylindrical wave & MLE, focus scanning method, PEB & Accuracy, energy leakage\\
        \hline

        R& \cite{10001209}&30 GHz&&SISO& \acrshort{ris} localization&FIM, multistage low-complexity \acrshort{ris} localization algorithm,  quasi-Newton method & Accuracy\\
        \hline

        R & \cite{10001107} & 45 GHz & UL& MISO & Near-field localization & Second-order Fresnel approximation, \acrshort{ris} training phase shifts and pilots design & Accuracy\\
        \hline

        R & \cite{pan2022risaided} & 90 GHz &UL & MISO& Spherical wavefront propagation in the near-field of the sub\acrshort{thz} system with the assistance of a \acrshort{ris} & Near-field channel estimation and localization algorithm based on second-order Fresnel approximation of the near-field channel&Accuracy
         \\
         \hline

        R & \cite{9940428} & 100 GHz & UL & MIMO & UE localization under beam squint effect & Polar-domain gradient descent algorithm, MUSIC algorithm & Accuracy \\
        \hline

        R & \cite{10149471} & 320/325/ 330GHz & UL& MISO & Spherical wavefront propagation in the near-field& Proposed  near-field channel estimation and localization  algorithm&Accuracy
         \\
         \hline

         A & \cite{10436860} & 28 GHz & SL& SISO & \acrshort{sl} localization through a single active \acrshort{ris} &  Time-orthogonal random codebook, \acrshort{crlb}& Accuracy\\
         \hline
         R & \cite{10140033} & & DL & MIMO & Fundamental limits of RIS-assisted \acrfull{ofts} systems for localization & \acrshort{fim} & \acrshort{peb}, \acrshort{oeb}\\
         \hline
        R & \cite{10557490} & 28 GHz & DL & SISO & Near-field joint position and velocity estimation under user mobility & Optimization & Accuracy\\
        
         \hline
         
         \thickhline      
   \end{tabular}
\end{table*}
\end{scriptsize}

\subsubsection{Foundational Studies}
A two-stage positioning technique for determining the transmitter's location with a \acrshort{ris}, employed as a lens, running at mmW frequency demonstrates the capability of decimeter-level localization accuracy in the near-field region \cite{9500663}. A generic model of near-field as well as the far-field placement was constructed in \cite{9508872}, and it suggests an SNR-based \acrshort{ris} phase design algorithm for CLRB reduction. The suggested technique can reduce PEB and directional error bounds significantly compared to the conventional system without \acrshort{ris}. Both of these \acrshort{ris}-assisted near-field localization studies disregard the scenario of  \acrshort{los} obstruction, however, it is necessary to take care of to cater for the successful localization in real-world scenarios. Authors in \cite{9650561} propose a general framework for \acrshort{ris}-assisted regional localization, including \acrshort{ris} phase design and position determination. The results demonstrate the effectiveness of the proposed framework, showing that the designed \acrshort{ris} phase schemes lead to near-optimal localization performance.
Authors in article \cite{pan2022risaided} investigate the localization and \acrshort{csi} estimation scheme for a near-field sub-THz system with a \acrshort{ris}. The authors propose a near-field joint channel estimation and localization algorithm, which demonstrates superior performance in terms of localization and CSI estimation root \acrshort{mse} compared to conventional far-field algorithms. The complexity of near-field CSI estimation is influenced by the array steering vector formulation, which takes into account the reflection elements and their coupling effects, leading to higher resolution accuracy. The study highlights the importance of considering the near-field effects and angle separations between UEs for achieving high-precision localization with a single \acrshort{ris} panel. Additionally, it is emphasized that the inclusion of a large \acrshort{ris} panel with more elements must consider the spherical wavefront feature to avoid performance degradation. 

A RIS-assisted \acrshort{ofts} system in both near-field and far-field conditions is studied in \cite{10140033}. It derives signal expressions in the Doppler-delay domain and uses the \acrshort{fim} to find localization error limits. Simulations show that localization error decreases with more antennas and RIS elements. The paper also investigates estimating RIS orientation from UE signals, finding it feasible only in near-field conditions. A major finding is that RIS orientation errors significantly impact localization accuracy. Authors in \cite{10529957} describe the near-field features of new 6G technologies, such as the spherical wave model, spatial non-stationarity, and beam squint effect, and explores their implications for localization, sensing, and communication. 
\subsubsection{LoS Blockages}
Researchers in \cite{9593200} have investigated the SISO system's near-field localization capabilities in the presence of a significant \acrshort{los} blockage. A two-step localization algorithm based on \acrshort{toa} was presented in \cite{9593241}, and the results supported the feasibility of retaining high localization accuracy even under the situation of significant blockage in the near-field region of \acrshort{ris}. Using the Jacobi-Anger expansion and taking into account the \acrshort{ris} amplitude, a low complexity near-field localization approach, termed approximation mismatched maximum likelihood, has been devised in \cite{10017173}. It also suggests an iterative refinement approach for joint localization and \acrshort{ris} amplitude model parameter updating, using the result as the initial location estimate. The suggested low-complexity localization technique performs well in simulations, and the iterative algorithm's localization accuracy is asymptotically approaching CRLB.  

\subsubsection{Performance and Considerations}
    \begin{figure*}
        \centering
    \subfigure[Absence of LoS path]{
        \includegraphics[width=0.45\textwidth]{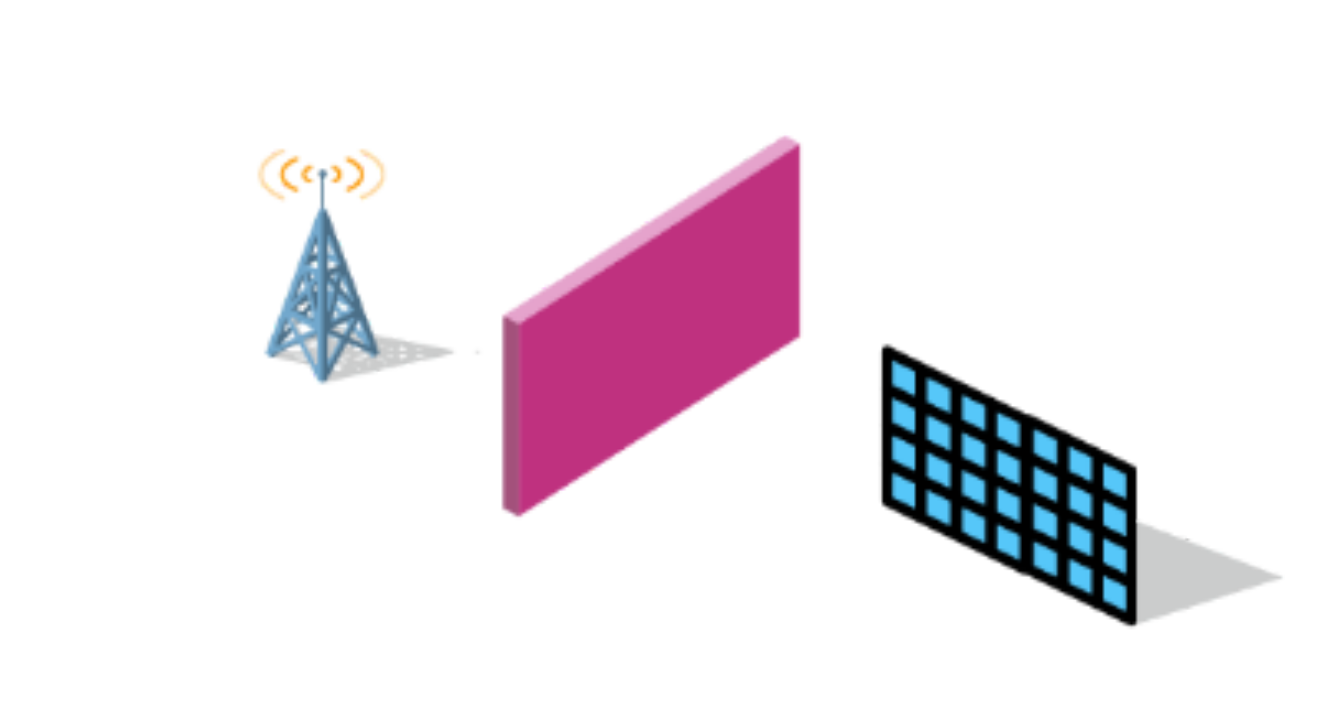}
        \label{fig:sub11}
    }
    \hspace{5mm}%\hfill % Space between figures
    \subfigure[Installation and maintenance cost]{
        \includegraphics[width=0.45\textwidth]{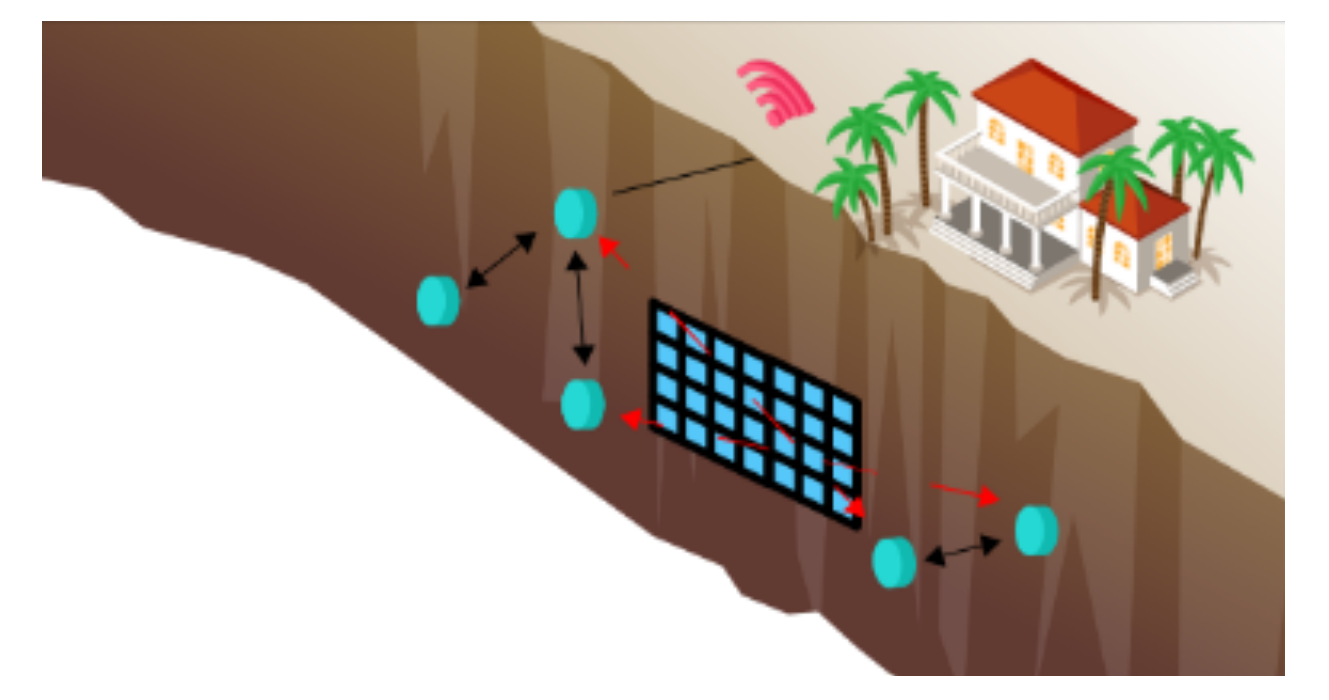}
        \label{fig:sub22}
    }
    \hspace{5mm}%\hfill %Space between figures
    \subfigure[Power supply, calibration and reliable connectivity]{
        \includegraphics[width=0.45\textwidth]{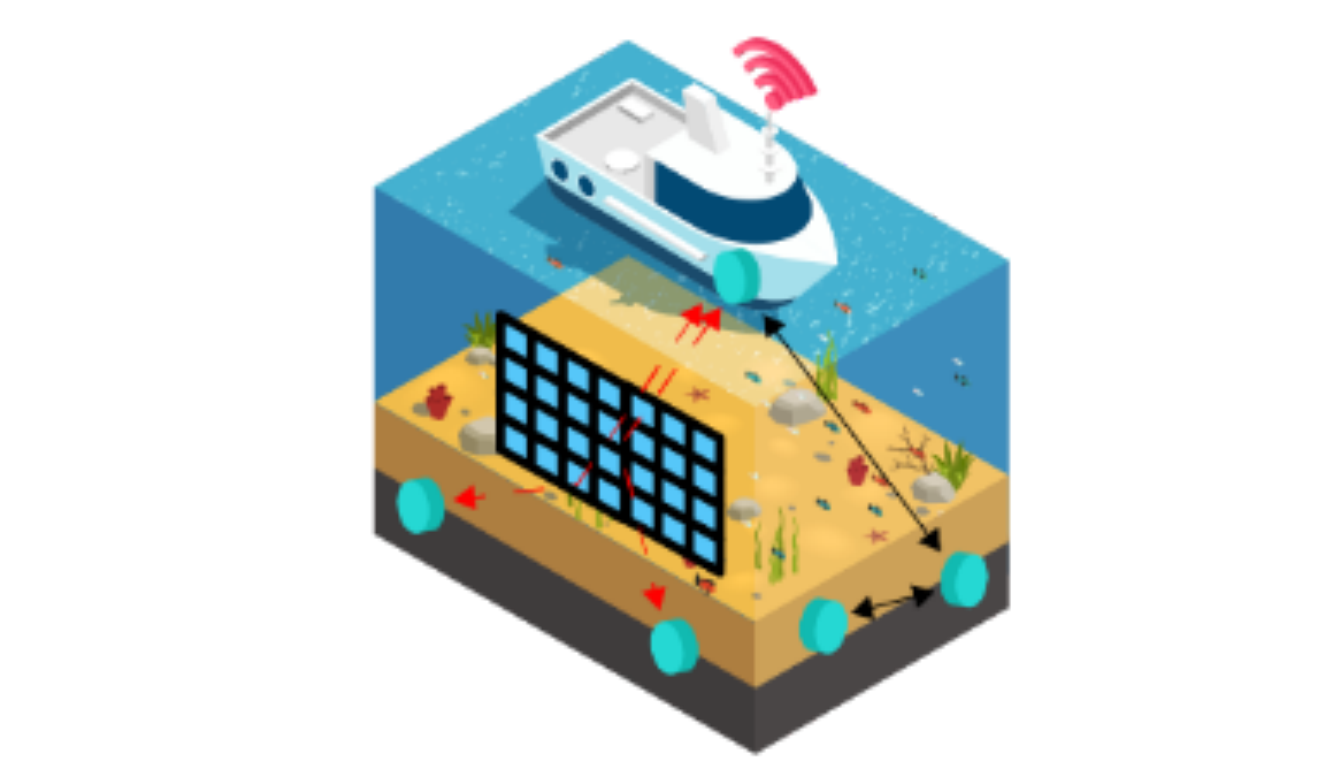}
        \label{fig:sub33}
    }
    \hspace{5mm}%\hfill %Space between figures
    \subfigure[Scalability, interference and coexistence challenge]{
        \includegraphics[width=0.45\textwidth]{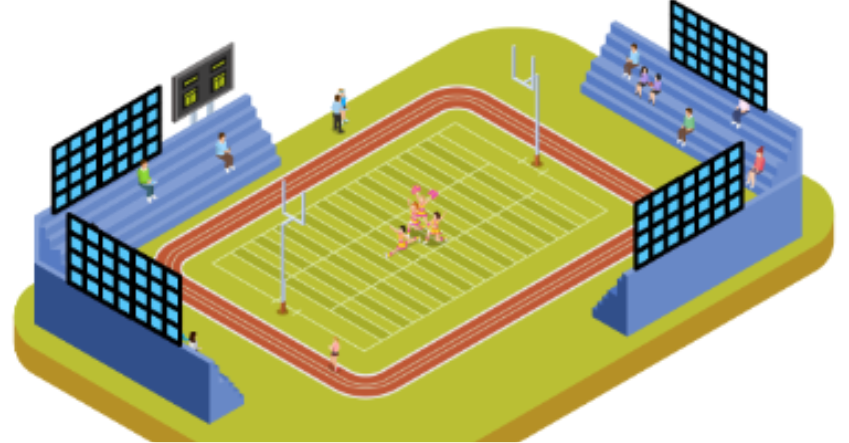}
        \label{fig:sub44}
    }
    \caption{Illustration of limitations to \acrshort{ris}-assisted radio localization. (a) NLoS challenges in complex environments
with reflective or obstructive surfaces can be challenging, (b)The deployment, installation and maintenance of \acrshort{ris} infrastructure can be challenging and require careful planning (c) Ensuring an adequate power supply, \acrshort{ris} calibration and reliable connectivity to each \acrshort{ris} device can be a logistical challenge, (d) when multiple \acrshort{ris} devices are deployed in proximity, potential interference and coexistence challenges may arise.}
    \label{figlim}
\end{figure*}
\acrshort{ris}-based asynchronous localization is studied in \cite{Gong2022AsynchronousRL} by examining the PEB and equivalent Fisher information for the intermediate parameters involved. The study considers multi-paths between the BS and the \acrshort{ris}, taking into account amplitude differences. The results indicate that near-field spherical wavefront modeling enables UE localization in the asynchronous scenario, but the equivalent Fisher information decreases as the distance between the UE and \acrshort{ris} increases. The study also highlights the performance difference between spatial gain and power gain in the BS-\acrshort{ris} channel, and cautions against using the SNR-maximizing focusing control scheme for \acrshort{ris} in localization applications. In article \cite{9902991}, the performance of a holographic \acrshort{ris} assisted \acrshort{mmw} near-field localization system is investigated. The FIM and CRLB are derived, considering the radiation pattern of antennas. The theoretical analysis demonstrates that the position accuracy improves quadratically with the size of the \acrshort{ris}. An iterative entropy regularization-based method is proposed to minimize the worst-case CRLB by optimizing the holographic \acrshort{ris} phases. 
%Simulation results confirm the impact of system parameters on the PEB and validate the effectiveness of the proposed algorithm. 
Authors in \cite{rahal2023performance} propose a low-complexity approach to optimize the precoders that control the \acrshort{ris}, considering hardware constraints. The method approximates desired beam patterns using pre-characterized reflection coefficients. The evaluation includes beam fidelity for different \acrshort{ris} hardware prototypes and theoretical analysis of the impact on near-field downlink positioning in NLoS conditions. Results demonstrate the effectiveness of the proposed optimization scheme in producing desired \acrshort{ris} beams within hardware limitations, while also highlighting the sensitivity to hardware characteristics and the specific requirements of \acrshort{ris}-aided localization applications. 

Authors address the problem of near-field localization using \acrshort{ris} in the presence of phase-dependent amplitude variations at each \acrshort{ris} element in \cite{ozturk2023ris}. The authors analyze the performance limitations using a MCRB and demonstrate that performance penalties can occur, particularly at high SNRs, when the UE is unaware of the amplitude variations. They propose a low-complexity approximated mismatched maximum likelihood estimator, leveraging Jacobi-Anger expansion, to mitigate performance loss. The method shows fast convergence and performance close to the \acrshort{crlb}, indicating the effectiveness of the proposed method in recovering performance and calibrating the \acrshort{ris} amplitude model. The issue of \acrshort{ris} pixel failures is studied in \cite{ozturk2023risaided}, which can severely impact localization accuracy. The paper investigates the impact of pixel failures on accuracy and develops two strategies for joint localization and failure diagnosis to detect failing pixels while accurately locating the UE. The proposed algorithms demonstrate significant performance improvements over conventional failure-agnostic approaches, enabling successful localization in the presence of pixel failures. The sidelink positioning in near-field wireless communication using a single active RIS is studies in \cite{10436860}. It jointly estimates transmitter and receiver positions and clock bias. A time-orthogonal random codebook is used for channel separation, followed by coarse positioning and least-squares refinement. The \acrshort{crlb} validates the algorithm's efficiency. Simulations show that positioning accuracy depends on RIS size and transmission power, with an optimal power allocation ratio that minimizes \acrshort{crlb}.

\section{Challenges and Research Outlook}
\label{sec5}
Based on the review performed in the previous section, in this section, we present a detailed overview of the limitations, open areas of research and challenges in \acrshort{ris}-assisted radio localization that need to be investigated to make it a practical and feasible solution to radio localization in 6G networks. 
%\subsection{Research Directions and Challenges}
%\label{sec5a}
While \acrshort{ris} offers significant benefits for localization, there are some associated limitations, as shown in Figure \ref{figlim} \cite{9424177}. Implementing \acrshort{ris}-assisted localization may involve significant costs, including the installation and maintenance of \acrshort{ris} devices throughout the target environment. The deployment of \acrshort{ris} infrastructure can be challenging and require careful planning. Inferring from the reviewed literature, \acrshort{ris} devices typically rely on LoS communication with the devices they assist in localizing. This means that obstacles, such as walls or objects, can obstruct the signal path and potentially degrade the accuracy or reliability of localization. While \acrshort{ris} can optimize signal propagation, there may still be scenarios where NLoS signal paths exist, leading to potential inaccuracies in localization. Overcoming NLoS challenges in complex environments with reflective or obstructive surfaces can be a limitation for \acrshort{ris}-assisted localization. \acrshort{ris} devices require power and connectivity to function effectively. Ensuring an adequate power supply and reliable connectivity to each \acrshort{ris} device can be a logistical challenge, particularly in large-scale deployments or areas with limited infrastructure. \acrshort{ris}-assisted localization may face scalability limitations when applied to larger or more complex environments. As the number of devices and users increases, coordinating and optimizing the \acrshort{ris} network can become more challenging. Additionally, adapting the \acrshort{ris} configuration to accommodate changes in the environment or user requirements may require significant adjustments and maintenance. In situations where multiple \acrshort{ris} devices are deployed in proximity, potential interference and coexistence challenges may arise \cite{9852985}. Careful planning and coordination are necessary to ensure that \acrshort{ris} devices do not interfere with each other or with other wireless communication systems operating in the same frequency band \cite{9771731}. These limitations should be carefully considered and addressed during the planning, deployment, and operation of \acrshort{ris} systems to maximize their effectiveness.

Here we discuss the research directions related to the limitations and open areas of research in the path of practical applications of \acrshort{ris}-assisted radio localization in 6G networks that includes the technical as well as the deployment challenges. 

 %Some key research directions for \acrshort{ris} based radio localization, that have not been explored yet, are listed below.    
 \begin{itemize}
   \item\textit{{Availability, Scalability, Privacy and Security}}: It is observed in the previous section that the theoretical approaches being developed in the literature are primarily focused on the accuracy of localization. While devising new methods for \acrshort{ris}-assisted localization in 6G networks, it is an important factor to consider also the coverage area and availability of the service \cite{7762095}. Techniques must be developed in a fashion that it is scalable without any major hardware as well as the software limitations, as shown in Figure \ref{fig:sub11} and \ref{fig:sub44}. Lastly, user privacy in such networks is an interesting area of investigation since the transmission and processing of data within \acrshort{ris} systems can be susceptible to eavesdropping, leading to the leakage of sensitive information \cite{10044963}.

    \item\textit{{Mobile User Localization}}: Mobility of the UE is an important factor in a real-world scenario that needs to be considered in addition to the 3D position and 3D orientation to localize the user with maximum accuracy \cite{9893114}. If the Doppler delay effects due to user mobility are ignored, it will negatively impact the location estimates. It is, thus, necessary to account for the velocity of the user and its relative impact on UE position and orientation in a \acrshort{ris}-assisted localization scenario. Continuous monitoring of mobile UEs would gain advantages by incorporating \acrshort{nlos} channel identification to ensure optimal activation of \acrshort{ris}, and the ability to control \acrshort{ris} with low-latency location-based capabilities. Consequently, this necessitates the availability of accurate UE location and uncertainty information at all times. Adapting localization algorithms to changing propagation conditions would be required. 
   
    \item\textit{Multi-User Localization}: Methods scalable to multi-user localization need to be developed for both LoS and obstructed LoS scenarios for more realistic and holistic designs for 6G networks \cite{9330512}. It would require the development of algorithms for managing interference and optimizing resource allocation.
    
    \item\textit{{Modeling and Analysis of \acrshort{ris}-assisted Localization at Multiple Frequencies}}: In practical scenarios, access points operating at different frequency bands, i.e., conventional, mmW and THz, will coexist in future 6G networks. The operation of \acrshort{ris} when interacting with BS operating at different frequency bands needs to be modeled and analyzed. What kind of element configuration is required at \acrshort{ris} to how would the phase and amplitude change of the \acrshort{ris} be modeled to successfully allow multi-band radio localization. 
    
    \item \textit{{Integrated Localization, Sensing and Communication}}: Convergence of hardware as well as the technical design of radio localization, sensing and communication is one of the major agenda of 6G network design \cite{9625032, 9569364}. In light of this design requirement, it would be an interesting study direction to devise and analyze the methods for \acrshort{ris}-assisted joint localization, sensing and communication such that a unique trade-off is worked out between their performance matrices, thus, they complement one another depending on the scenario at hand. The introduction of these services within a wireless environment enabled by \acrshort{ris} presents new challenges related to optimizing \acrshort{ris} for multiple purposes. These challenges involve striking a suitable balance between configurations that prioritize localization, communication, and sensing. It entails selecting the appropriate protocols, managing resource sharing among multiple users and operators in complex ecosystems, achieving synchronization between BS and \acrshort{ris}, and seamlessly integrating \acrshort{ris} into open radio access network (RAN) architectures. Such \acrshort{ris}-based solutions also need to be cost-effective for supporting localization and sensing functionalities together with communication. These alternatives include vehicle-mounted reflective \acrshort{ris}, approaches resembling BS-free or multi-static radar systems, and hybrid \acrshort{ris} that can operate in the receiving mode to sense both connected UEs and passive objects.
     
    \item\textit{Deployment and Optimization of \acrshort{ris}-Assisted Localization Radio Network}:  Most of the contemporary literature quoted in the previous section is based on the development of theoretical approaches where there have been little to no practical campaigns to study the practical design and deployment perspectives of \acrshort{ris}-assisted radio localization. It is, therefore, an important area to explore the practicality of the methods proposed in the literature. The optimization of both the number and positioning of \acrshort{ris} is essential to achieve optimal performance in terms of communication metrics, localization/sensing accuracy, and coverage. Additionally, it is crucial to ensure that the optimized \acrshort{ris} deployment indeed offers advantages, when compared to traditional BS deployments, in terms of overall power consumption and coordination efforts. This optimization process also includes addressing the challenge of accurately calibrating the location and orientation of the \acrshort{ris} and its synchronization with BSs. 
    
    \item \textit{{AI Controlled \acrshort{ris}}}: In the age of AI, model-based signal processing is being replaced with data-driven approaches as it leads to more robust algorithms \cite{9893114}. Based on this, developing AI-driven methods for \acrshort{ris}-assisted radio localization can prove to improve the radio localization performance manifolds. Control of \acrshort{ris} using AI can empower their design manifolds, it is thus an important direction to study. 

    \item \textit{{Low Latency Control}}:
    Efficient radio localization with \acrshort{ris} requires low-latency control capabilities. This necessitates real-time knowledge of UE location and uncertainty. Developing location-based \acrshort{ris} control mechanisms that offer low-latency control while maintaining accuracy and reliability is a challenge that must be overcome.

    \item \textit{\acrshort{ris} Standardization}: In order to analyze the theoretical methods by practical experimentation as well as to develop more suitable methods while considering the practical deployment scenarios, standardization of \acrshort{ris} hardware is necessary. Global standardization of \acrshort{ris} is in very early stages and the process is well summarized in \cite{9679804}. \acrshort{ris} hardware is not yet available but the efforts for the development of \acrshort{ris} hardware prototypes are underway \cite{Tang2019MIMOTT, 9020088, Pei2021RISAidedWC, 9473842, alexandropoulos2023risenabled, 10032155, 10525242}. Standardized \acrshort{ris} hardware platforms will contribute to the acceleration of development progress. Researchers can build upon existing work, leveraging the availability of standardized platforms to iterate and refine their ideas. The collective efforts of researchers using standardized platforms can lead to the development of best practices, optimization techniques, and benchmark datasets that drive innovation and efficiency in \acrshort{ris}-related research.

    \item \textit{{Multi-Operator and Multi-\acrshort{ris} Localization}}: The coordination and communication between multiple operators can be complex, especially in heterogeneous network environments, potentially leading to increased latency and decreased system efficiency. The synchronization of signals between different operators is another hurdle, as it requires precise timing control to avoid interference and ensure accurate localization. Moreover, the deployment and management of multiple \acrshort{ris} introduces further complexities, from determining optimal placement and density of the \acrshort{ris} to managing their phase configurations, as shown in Figure \ref{fig:sub44} \cite{strinati2021reconfigurable}. Additionally, privacy and security concerns may arise with multiple operators, necessitating robust protocols to protect data integrity. Lastly, as the number of operators and \acrshort{ris} increases, so does the computational complexity of localization algorithms, potentially impacting system performance and energy-efficiency.

    \item \textit{{\acrshort{ris} Control and User Mobility}}: The process of controlling a \acrshort{ris} typically involves adjusting the electromagnetic properties of the \acrshort{ris} elements to optimize signal reflection and transmission. However, this process can be relatively slow, which poses significant challenges for high-mobility applications. As devices move rapidly across the coverage area, the channel conditions change quickly. By the time the \acrshort{ris} has collected enough information and adjusted its properties for optimal performance, the device may have already moved to a different location with entirely different channel conditions. Thus, the slow control of \acrshort{ris} may lead to outdated or ineffective configurations that fail to improve, or even degrade, the system performance. This lag in \acrshort{ris} control poses a major challenge in realizing the full potential of \acrshort{ris} technology in high-mobility applications such as autonomous vehicles, drones, and high-speed trains.

    \item \textit{{\acrshort{ris} Hardware Limitations and Pixel Failures}}: The hardware components of \acrshort{ris} bring about several challenges that can impact their performance and efficacy \cite{9424177}. \acrshort{ris} are composed of numerous smaller elements or ``pixels" that each need to be individually controlled to manipulate the phase and amplitude of incoming electromagnetic waves. However, these pixel-level controls can be limited by hardware constraints such as processing speed, energy consumption, and design complexity. Additionally, the risk of pixel failures is a significant concern. Given the high density of pixels in an \acrshort{ris}, even a small percentage of pixel failures can lead to significant degradation in the overall performance of the \acrshort{ris}. Furthermore, identifying and repairing these failed pixels can be a complex and time-consuming task, especially when the \acrshort{ris} is deployed in hard-to-reach locations, as shown in Figure \ref{fig:sub22}. These hardware limitations and pixel failures pose substantial challenges to the reliable and effective deployment of \acrshort{ris} technology.

    \item \textit{{\acrshort{ris} Calibration}}: The calibration of \acrshort{ris} is a challenging phase due to the inherent complexities of these devices \cite{9360709}. \acrshort{ris} calibration involves adjusting each individual element, or ``pixel", on the surface to manipulate the phase and amplitude of incident signals. Given that a \acrshort{ris} can consist of hundreds or thousands of these elements, this process can be highly complex and time-consuming. In addition, each element may respond differently to adjustments due to manufacturing variances, further complicating the calibration process. Real-world environmental factors, such as temperature and humidity, can also cause drift in the performance of the elements over time, necessitating frequent recalibration. Given that \acrshort{ris} are often deployed in inaccessible or hard-to-reach locations, performing this recalibration can be logistically challenging and costly, as shown in Figure \ref{fig:sub33}. Consequently, achieving precise and efficient \acrshort{ris} calibration remains a major hurdle in the wider adoption of \acrshort{ris} technology.

    \item \textit{{New \acrshort{ris} Antenna Technologies}}:

    Developing and integrating \acrshort{ris} antenna technologies into everyday items, such as clothing \cite{9424177, 9360709}, is an area ripe with potential but also rife with challenges. For instance, creating antennas thin and flexible enough to be woven into fabric without sacrificing performance is a significant technical obstacle. The material used in clothing also presents difficulties as it must be able to withstand regular wear and tear, washing, and various weather conditions while maintaining the antenna's functionality. Furthermore, it is critical to ensure that these \acrshort{ris} antennas do not negatively affect the wearer's health, particularly given concerns around prolonged exposure to electromagnetic fields. This necessitates strict control of the emitted power levels. From a design perspective, seamlessly incorporating the antennas in a way that is aesthetically pleasing and unobtrusive is also a major challenge. Given that each piece of clothing may be shaped and sized differently, custom calibration of these antennas could be needed for each garment, presenting further complexities. 
 \item \textit{{From \acrshort{ris} to SIM}}: 
    The introduction of \acrfull{sim} opens new possibilities in the context of radio localization \cite{10158690}. Unlike traditional single-layer RIS, \acrshort{sim} consist of multiple reconfigurable metasurface layers, providing enhanced control over electromagnetic wave propagation in \acrshort{3d}. This added dimensionality enables \acrshort{sim} to manipulate complex multi-path environments more effectively, improving the robustness of localization or providing means for replacing digital with analog processing. Additionally, \acrshort{sim} can support multi-functional operations, such as \acrshort{slam} and \acrshort{isac} \cite{10557708}, by leveraging their ability to perform spatial filtering, polarization control, and adaptive beamforming across multiple layers. The inter-layer coupling in \acrshort{sim} also allows for greater flexibility in addressing near-field localization challenges, particularly at higher frequency bands like \acrshort{mmw} and \acrshort{thz}. These capabilities suggest that \acrshort{sim} could serve as a transformative enabler for high-accuracy and resilient localization in 6G networks, warranting further exploration and experimental validation.
    \end{itemize}

\section{Conclusion}
\label{sec6}
We presented a comprehensive overview of the utilization of \acrshort{ris} technology for radio localization in 6G networks. We discussed the \acrshort{ris}-assisted localization taxonomy, and recent advancements in theoretical approaches for \acrshort{ris}-assisted localization, identified opportunities, explored challenges, and examined various applications alongside the limitations of \acrshort{ris}-assisted localization. Recent advancements are based primarily on modeling and \acrshort{ml}-based techniques where the focus is on improving the accuracy of the user location estimate. \acrshort{ris} can optimize wireless signals for improved localization in smart indoor services, smart transportation, and automated factories. However, there are some limitations to its use that need to be overcome such as the line-of-sight dependency, scalability, and interference. There are technical challenges and open areas of research that need to be addressed such as multi-user and mobile user localization, integrated localization, sensing, and communication algorithms, \acrshort{ris} standardization for practical experimentation as well as the investigation of availability, scalability, and privacy in \acrshort{ris}-assisted localization. Therefore, further research is required to fully realize the potential of \acrshort{ris} technology for localization in 6G.

% \section*{Acknowledgment}
% This work has received funding partly from the European Union’s Horizon 2020 Research and Innovation Program under Grant 951867 ‘5G-ROUTES’ and Grant 101058505 ‘5G-TIMBER’.
% Henk Wymeersch was supported by the European Commission through the H2020 project RISE-6G (grant agreement no.  101017011).
\bibliographystyle{IEEEtran} 
\bibliography{references}

\begin{thebibliography}{100}
\providecommand{\url}[1]{#1}
\csname url@samestyle\endcsname
\providecommand{\newblock}{\relax}
\providecommand{\bibinfo}[2]{#2}
\providecommand{\BIBentrySTDinterwordspacing}{\spaceskip=0pt\relax}
\providecommand{\BIBentryALTinterwordstretchfactor}{4}
\providecommand{\BIBentryALTinterwordspacing}{\spaceskip=\fontdimen2\font plus
\BIBentryALTinterwordstretchfactor\fontdimen3\font minus
  \fontdimen4\font\relax}
\providecommand{\BIBforeignlanguage}[2]{{%
\expandafter\ifx\csname l@#1\endcsname\relax
\typeout{** WARNING: IEEEtran.bst: No hyphenation pattern has been}%
\typeout{** loaded for the language `#1'. Using the pattern for}%
\typeout{** the default language instead.}%
\else
\language=\csname l@#1\endcsname
\fi
#2}}
\providecommand{\BIBdecl}{\relax}
\BIBdecl

\bibitem{9215972}
H.~Wymeersch, J.~He, B.~Denis, A.~Clemente, and M.~Juntti, ``{Radio
  Localization and Mapping With Reconfigurable Intelligent Surfaces:
  Challenges, Opportunities, and Research Directions},'' \emph{IEEE Vehicular
  Technology Magazine}, vol.~15, no.~4, pp. 52--61, 2020.

\bibitem{9330512}
C.~De~Lima, D.~Belot, R.~Berkvens, A.~Bourdoux, D.~Dardari, M.~Guillaud,
  M.~Isomursu, E.-S. Lohan, Y.~Miao, A.~N. Barreto, M.~R.~K. Aziz,
  J.~Saloranta, T.~Sanguanpuak, H.~Sarieddeen, G.~Seco-Granados, J.~Suutala,
  T.~Svensson, M.~Valkama, B.~Van~Liempd, and H.~Wymeersch, ``{Convergent
  Communication, Sensing and Localization in {6G} Systems: An Overview of
  Technologies, Opportunities and Challenges},'' \emph{IEEE Access}, vol.~9,
  pp. 26\,902--26\,925, 2021.

\bibitem{chepuri2022integrated}
S.~P. Chepuri, N.~Shlezinger, F.~Liu, G.~C. Alexandropoulos, S.~Buzzi, and
  Y.~C. Eldar, ``{Integrated Sensing and Communications With Reconfigurable
  Intelligent Surfaces: From signal modeling to processing},'' \emph{IEEE
  Signal Processing Magazine}, vol.~40, no.~6, pp. 41--62, 2023.

\bibitem{9765815}
M.~Jian, G.~C. Alexandropoulos, E.~Basar, C.~Huang, R.~Liu, Y.~Liu, and
  C.~Yuen, ``{Reconfigurable intelligent surfaces for wireless communications:
  Overview of hardware designs, channel models, and estimation techniques},''
  \emph{Intelligent and Converged Networks}, vol.~3, no.~1, pp. 1--32, 2022.

\bibitem{9665421}
A.~Albanese, V.~Sciancalepore, and X.~Costa-Pérez, ``{First Responders Got
  Wings: UAVs to the Rescue of Localization Operations in Beyond 5G Systems},''
  \emph{IEEE Communications Magazine}, vol.~59, no.~11, pp. 28--34, 2021.

\bibitem{bresson2017simultaneous}
G.~Bresson, Z.~Alsayed, L.~Yu, and S.~Glaser, ``{Simultaneous localization and
  mapping: A survey of current trends in autonomous driving},'' \emph{IEEE
  Transactions on Intelligent Vehicles}, vol.~2, no.~3, pp. 194--220, 2017.

\bibitem{7927385}
A.~F.~G. Gonçalves~Ferreira, D.~M.~A. Fernandes, A.~P. Catarino, and J.~L.
  Monteiro, ``{Localization and Positioning Systems for Emergency Responders: A
  Survey},'' \emph{IEEE Communications Surveys \& Tutorials}, vol.~19, no.~4,
  pp. 2836--2870, 2017.

\bibitem{8306879}
S.~Kuutti, S.~Fallah, K.~Katsaros, M.~Dianati, F.~Mccullough, and
  A.~Mouzakitis, ``{A Survey of the State-of-the-Art Localization Techniques
  and Their Potentials for Autonomous Vehicle Applications},'' \emph{IEEE
  Internet of Things Journal}, vol.~5, no.~2, pp. 829--846, 2018.

\bibitem{8409950}
C.~Laoudias, A.~Moreira, S.~Kim, S.~Lee, L.~Wirola, and C.~Fischione, ``{A
  Survey of Enabling Technologies for Network Localization, Tracking, and
  Navigation},'' \emph{IEEE Communications Surveys \& Tutorials}, vol.~20,
  no.~4, pp. 3607--3644, 2018.

\bibitem{8226757}
J.~A. del Peral-Rosado, R.~Raulefs, J.~A. López-Salcedo, and G.~Seco-Granados,
  ``{Survey of Cellular Mobile Radio Localization Methods: From 1G to 5G},''
  \emph{IEEE Communications Surveys \& Tutorials}, vol.~20, no.~2, pp.
  1124--1148, 2018.

\bibitem{8692423}
F.~Zafari, A.~Gkelias, and K.~K. Leung, ``{A Survey of Indoor Localization
  Systems and Technologies},'' \emph{IEEE Communications Surveys \& Tutorials},
  vol.~21, no.~3, pp. 2568--2599, 2019.

\bibitem{8734111}
N.~Saeed, H.~Nam, T.~Y. Al-Naffouri, and M.-S. Alouini, ``{A State-of-the-Art
  Survey on Multidimensional Scaling-Based Localization Techniques},''
  \emph{IEEE Communications Surveys \& Tutorials}, vol.~21, no.~4, pp.
  3565--3583, 2019.

\bibitem{WEN201921}
F.~Wen, H.~Wymeersch, B.~Peng, W.~P. Tay, H.~C. So, and D.~Yang, ``{A survey on
  5G massive MIMO localization},'' \emph{Digital Signal Processing}, vol.~94,
  pp. 21--28, 2019.

\bibitem{9159543}
X.~Zhu, W.~Qu, T.~Qiu, L.~Zhao, M.~Atiquzzaman, and D.~O. Wu, ``{Indoor
  Intelligent Fingerprint-Based Localization: Principles, Approaches and
  Challenges},'' \emph{IEEE Communications Surveys \& Tutorials}, vol.~22,
  no.~4, pp. 2634--2657, 2020.

\bibitem{9523553}
V.~F. Miramá, L.~E. Díez, A.~Bahillo, and V.~Quintero, ``{A Survey of Machine
  Learning in Pedestrian Localization Systems: Applications, Open Issues and
  Challenges},'' \emph{IEEE Access}, vol.~9, pp. 120\,138--120\,157, 2021.

\bibitem{9328120}
A.~Motroni, A.~Buffi, and P.~Nepa, ``{A Survey on Indoor Vehicle Localization
  Through {RFID} Technology},'' \emph{IEEE Access}, vol.~9, pp.
  17\,921--17\,942, 2021.

\bibitem{9356512}
O.~Kanhere and T.~S. Rappaport, ``{Position Location for Futuristic Cellular
  Communications: 5G and Beyond},'' \emph{IEEE Communications Magazine},
  vol.~59, no.~1, pp. 70--75, 2021.

\bibitem{xiao2022overview}
Z.~Xiao and Y.~Zeng, ``{An overview on integrated localization and
  communication towards 6G},'' \emph{Science China Information Sciences},
  vol.~65, pp. 1--46, 2022.

\bibitem{s22010247}
J.~Laconte, A.~Kasmi, R.~Aufrère, M.~Vaidis, and R.~Chapuis, ``{A Survey of
  Localization Methods for Autonomous Vehicles in Highway Scenarios},''
  \emph{Sensors}, vol.~22, no.~1, 2022.

\bibitem{Liu2021ASO}
A.~Liu, Z.~Huang, M.~Li, Y.~Wan, W.~Li, T.~X. Han, C.~Liu, R.~Du, D.~K.~P. Tan,
  J.~Lu, Y.~Shen, F.~Colone, and K.~Chetty, ``{A Survey on Fundamental Limits
  of Integrated Sensing and Communication},'' \emph{IEEE Communications Surveys
  \& Tutorials}, vol.~24, pp. 994--1034, 2021.

\bibitem{Trevlakis2023LocalizationAA}
S.~E. Trevlakis, A.-A.~A. Boulogeorgos, D.~Pliatsios, J.~Querol, K.~Ntontin,
  P.~Sarigiannidis, S.~Chatzinotas, and M.~Di~Renzo, ``Localization as a key
  enabler of 6g wireless systems: A comprehensive survey and an outlook,''
  \emph{IEEE Open Journal of the Communications Society}, vol.~4, pp.
  2733--2801, 2023.

\bibitem{9495362}
S.~Zhang, M.~Li, M.~Jian, Y.~Zhao, and F.~Gao, ``{AIRIS: Artificial
  intelligence enhanced signal processing in reconfigurable intelligent surface
  communications},'' \emph{China Communications}, vol.~18, no.~7, pp. 158--171,
  2021.

\bibitem{9782674}
H.~Chen, H.~Sarieddeen, T.~Ballal, H.~Wymeersch, M.-S. Alouini, and T.~Y.
  Al-Naffouri, ``{A Tutorial on Terahertz-Band Localization for 6G
  Communication Systems},'' \emph{IEEE Communications Surveys \& Tutorials},
  vol.~24, no.~3, pp. 1780--1815, 2022.

\bibitem{9874802}
R.~Chen, M.~Liu, Y.~Hui, N.~Cheng, and J.~Li, ``{Reconfigurable Intelligent
  Surfaces for 6G IoT Wireless Positioning: A Contemporary Survey},''
  \emph{IEEE Internet of Things Journal}, vol.~9, no.~23, pp. 23\,570--23\,582,
  2022.

\bibitem{9847080}
C.~Pan, G.~Zhou, K.~Zhi, S.~Hong, T.~Wu, Y.~Pan, H.~Ren, M.~D. Renzo,
  A.~Lee~Swindlehurst, R.~Zhang, and A.~Y. Zhang, ``{An Overview of Signal
  Processing Techniques for RIS/IRS-Aided Wireless Systems},'' \emph{IEEE
  Journal of Selected Topics in Signal Processing}, vol.~16, no.~5, pp.
  883--917, 2022.

\bibitem{9124848}
J.~He, H.~Wymeersch, T.~Sanguanpuak, O.~Silven, and M.~Juntti, ``{Adaptive
  Beamforming Design for mmWave RIS-Aided Joint Localization and
  Communication},'' \emph{2020 IEEE Wireless Communications and Networking
  Conference Workshops (WCNCW)}, pp. 1--6, 2020.

\bibitem{8288263}
S.~Hu, F.~Rusek, and O.~Edfors, ``{Cramér-Rao Lower Bounds for Positioning
  with Large Intelligent Surfaces},'' \emph{2017 IEEE 86th Vehicular Technology
  Conference (VTC-Fall)}, pp. 1--6, 2017.

\bibitem{9201413}
S.~Zeng, H.~Zhang, B.~Di, Z.~Han, and L.~Song, ``{Reconfigurable Intelligent
  Surface (RIS) Assisted Wireless Coverage Extension: RIS Orientation and
  Location Optimization},'' \emph{IEEE Communications Letters}, vol.~25, no.~1,
  pp. 269--273, 2021.

\bibitem{8741198}
C.~Huang, A.~Zappone, G.~C. Alexandropoulos, M.~Debbah, and C.~Yuen,
  ``{Reconfigurable Intelligent Surfaces for Energy Efficiency in Wireless
  Communication},'' \emph{IEEE Transactions on Wireless Communications},
  vol.~18, no.~8, pp. 4157--4170, 2019.

\bibitem{8710259}
R.~C. Shit, S.~Sharma, D.~Puthal, P.~James, B.~Pradhan, A.~v. Moorsel, A.~Y.
  Zomaya, and R.~Ranjan, ``{Ubiquitous Localization (UbiLoc): A Survey and
  Taxonomy on Device Free Localization for Smart World},'' \emph{IEEE
  Communications Surveys \& Tutorials}, vol.~21, no.~4, pp. 3532--3564, 2019.

\bibitem{10565792}
H.~Sallouha, S.~Saleh, S.~D. Bast, Z.~Cui, S.~Pollin, and H.~Wymeersch, ``On
  the ground and in the sky: A tutorial on radio localization in
  ground-air-space networks,'' \emph{IEEE Communications Surveys \& Tutorials},
  pp. 1--1, 2024.

\bibitem{8358018}
M.~F. Keskin, A.~D. Sezer, and S.~Gezici, ``{Localization via Visible Light
  Systems},'' \emph{Proceedings of the IEEE}, vol. 106, no.~6, pp. 1063--1088,
  2018.

\bibitem{9721205}
E.~Björnson, H.~Wymeersch, B.~Matthiesen, P.~Popovski, L.~Sanguinetti, and
  E.~de~Carvalho, ``{Reconfigurable Intelligent Surfaces: A signal processing
  perspective with wireless applications},'' \emph{IEEE Signal Processing
  Magazine}, vol.~39, no.~2, pp. 135--158, 2022.

\bibitem{10044963}
K.~Keykhosravi, B.~Denis, G.~C. Alexandropoulos, Z.~S. He, A.~Albanese,
  V.~Sciancalepore, and H.~Wymeersch, ``{Leveraging RIS-Enabled Smart Signal
  Propagation for Solving Infeasible Localization Problems: Scenarios, Key
  Research Directions, and Open Challenges},'' \emph{IEEE Vehicular Technology
  Magazine}, vol.~18, no.~2, pp. 20--28, 2023.

\bibitem{9815617}
G.~C. Alexandropoulos, M.~Crozzoli, D.-T. Phan-Huy, K.~D. Katsanos,
  H.~Wymeersch, P.~Popovski, P.~Ratajczak, Y.~Bénédic, M.-H. Hamon, S.~H.
  Gonzalez, R.~D’Errico, and E.~C. Strinati, ``{Smart Wireless Environments
  Enabled by RISs: Deployment Scenarios and Two Key Challenges},'' \emph{2022
  Joint European Conference on Networks and Communications \& 6G Summit
  (EuCNC/6G Summit)}, pp. 1--6, 2022.

\bibitem{8264743}
S.~Hu, F.~Rusek, and O.~Edfors, ``{Beyond Massive MIMO: The Potential of
  Positioning With Large Intelligent Surfaces},'' \emph{IEEE Transactions on
  Signal Processing}, vol.~66, no.~7, pp. 1761--1774, 2018.

\bibitem{9048973}
J.~V. Alegría and F.~Rusek, ``{Cramér-Rao Lower Bounds for Positioning with
  Large Intelligent Surfaces using Quantized Amplitude and Phase},'' \emph{2019
  53rd Asilomar Conference on Signals, Systems, and Computers}, pp. 10--14,
  2019.

\bibitem{8815412}
C.~Huang, G.~C. Alexandropoulos, C.~Yuen, and M.~Debbah, ``{Indoor Signal
  Focusing with Deep Learning Designed Reconfigurable Intelligent Surfaces},''
  \emph{2019 IEEE 20th International Workshop on Signal Processing Advances in
  Wireless Communications (SPAWC)}, pp. 1--5, 2019.

\bibitem{9148744}
H.~Wymeersch and B.~Denis, ``{Beyond 5G Wireless Localization with
  Reconfigurable Intelligent Surfaces},'' \emph{ICC 2020 - 2020 IEEE
  International Conference on Communications (ICC)}, pp. 1--6, 2020.

\bibitem{9129075}
J.~He, H.~Wymeersch, L.~Kong, O.~Silvén, and M.~Juntti, ``{Large Intelligent
  Surface for Positioning in Millimeter Wave MIMO Systems},'' \emph{2020 IEEE
  91st Vehicular Technology Conference (VTC2020-Spring)}, pp. 1--5, 2020.

\bibitem{9500437}
Y.~Liu, E.~Liu, R.~Wang, and Y.~Geng, ``{Reconfigurable Intelligent Surface
  Aided Wireless Localization},'' \emph{ICC 2021 - IEEE International
  Conference on Communications}, pp. 1--6, 2021.

\bibitem{9513781}
E.~Čišija, A.~M. Ahmed, A.~Sezgin, and H.~Wymeersch, ``{Ris-Aided mmWave MIMO
  Radar System for Adaptive Multi-Target Localization},'' \emph{{2021 IEEE
  Statistical Signal Processing Workshop (SSP)}}, pp. 196--200, 2021.

\bibitem{9410435}
W.~Wang and W.~Zhang, ``{Joint Beam Training and Positioning for Intelligent
  Reflecting Surfaces Assisted Millimeter Wave Communications},'' \emph{IEEE
  Transactions on Wireless Communications}, vol.~20, no.~10, pp. 6282--6297,
  2021.

\bibitem{9238887}
J.~Zhang, Z.~Zheng, Z.~Fei, and X.~Bao, ``{Positioning with Dual Reconfigurable
  Intelligent Surfaces in Millimeter-Wave MIMO Systems},'' \emph{2020 IEEE/CIC
  International Conference on Communications in China (ICCC)}, pp. 800--805,
  2020.

\bibitem{9193909}
H.~Zhang, H.~Zhang, B.~Di, K.~Bian, Z.~Han, and L.~Song, ``{Towards Ubiquitous
  Positioning by Leveraging Reconfigurable Intelligent Surface},'' \emph{IEEE
  Communications Letters}, vol.~25, no.~1, pp. 284--288, 2021.

\bibitem{9456027}
H.~Zhang, H.~Zhang, B.~Di, K.~Bian, and Z.~Han, ``{MetaLocalization:
  Reconfigurable Intelligent Surface Aided Multi-User Wireless Indoor
  Localization},'' \emph{IEEE Transactions on Wireless Communications},
  vol.~20, no.~12, pp. 7743--7757, 2021.

\bibitem{9201330}
T.~Ma, Y.~Xiao, X.~Lei, W.~Xiong, and Y.~Ding, ``{Indoor Localization With
  Reconfigurable Intelligent Surface},'' \emph{IEEE Communications Letters},
  vol.~25, no.~1, pp. 161--165, 2021.

\bibitem{9500663}
Z.~Abu-Shaban, K.~Keykhosravi, M.~F. Keskin, G.~C. Alexandropoulos,
  G.~Seco-Granados, and H.~Wymeersch, ``{Near-field Localization with a
  Reconfigurable Intelligent Surface Acting as Lens},'' \emph{ICC 2021 - IEEE
  International Conference on Communications}, pp. 1--6, 2021.

\bibitem{9508872}
A.~Elzanaty, A.~Guerra, F.~Guidi, and M.-S. Alouini, ``{Reconfigurable
  Intelligent Surfaces for Localization: Position and Orientation Error
  Bounds},'' \emph{IEEE Transactions on Signal Processing}, vol.~69, pp.
  5386--5402, 2021.

\bibitem{9593200}
M.~Rahal, B.~Denis, K.~Keykhosravi, B.~Uguen, and H.~Wymeersch, ``{RIS-Enabled
  Localization Continuity Under Near-Field Conditions},'' \emph{2021 IEEE 22nd
  International Workshop on Signal Processing Advances in Wireless
  Communications (SPAWC)}, pp. 436--440, 2021.

\bibitem{9593241}
D.~Dardari, N.~Decarli, A.~Guerra, and F.~Guidi, ``{Localization in NLOS
  Conditions using Large Reconfigurable Intelligent Surfaces},'' \emph{2021
  IEEE 22nd International Workshop on Signal Processing Advances in Wireless
  Communications (SPAWC)}, pp. 551--555, 2021.

\bibitem{10017173}
C.~Ozturk, M.~F. Keskin, H.~Wymeersch, and S.~Gezici, ``Ris-aided near-field
  localization under phase-dependent amplitude variations,'' \emph{IEEE
  Transactions on Wireless Communications}, vol.~22, no.~8, pp. 5550--5566,
  2023.

\bibitem{10054103}
R.~Wang, Z.~Xing, E.~Liu, and J.~Wu, ``{Joint Localization and Communication
  Study for Intelligent Reflecting Surface Aided Wireless Communication
  System},'' \emph{IEEE Transactions on Communications}, pp. 1--1, 2023.

\bibitem{9473890}
K.~Dovelos, S.~D. Assimonis, H.~Quoc~Ngo, B.~Bellalta, and M.~Matthaiou,
  ``{Intelligent Reflecting Surfaces at Terahertz Bands: Channel Modeling and
  Analysis},'' \emph{2021 IEEE International Conference on Communications
  Workshops (ICC Workshops)}, pp. 1--6, 2021.

\bibitem{guo2022dynamic}
H.~Guo, B.~Makki, M.~Åström, M.-S. Alouini, and T.~Svensson, ``{Dynamic
  Blockage Pre-Avoidance using Reconfigurable Intelligent Surfaces},''
  \emph{arXiv}, 2022.

\bibitem{Keykhosravi2020SISORJ}
K.~Keykhosravi, M.~F. Keskin, G.~Seco-Granados, and H.~Wymeersch, ``{SISO
  RIS-Enabled Joint 3D Downlink Localization and Synchronization},'' \emph{ICC
  2021 - IEEE International Conference on Communications}, pp. 1--6, 2020.

\bibitem{Fascista2020RISAidedJL}
A.~Fascista, A.~Coluccia, H.~Wymeersch, and G.~Seco-Granados, ``{RIS-Aided
  Joint Localization and Synchronization with a Single-Antenna Mmwave
  Receiver},'' \emph{ICASSP 2021 - 2021 IEEE International Conference on
  Acoustics, Speech and Signal Processing (ICASSP)}, pp. 4455--4459, 2020.

\bibitem{Gong2022AsynchronousRL}
Z.~Gong, L.~Wu, Z.~Zhang, J.~Dang, Y.~Wu, and J.~Wang, ``Asynchronous
  ris-assisted localization: A comprehensive analysis of fundamental limits,''
  \emph{IEEE Transactions on Wireless Communications}, pp. 1--1, 2024.

\bibitem{9538860}
Y.~Cui and H.~Yin, ``{Channel Estimation for RIS-aided mmWave Communications
  via 3D Positioning},'' \emph{2021 IEEE/CIC International Conference on
  Communications in China (ICCC Workshops)}, pp. 399--404, 2021.

\bibitem{9779402}
J.~Zhang, Z.~Zheng, Z.~Fei, and Z.~Han, ``{Energy-Efficient Multiuser
  Localization in the RIS-Assisted IoT Networks},'' \emph{IEEE Internet of
  Things Journal}, vol.~9, no.~20, pp. 20\,651--20\,665, 2022.

\bibitem{pan2022risaided}
Y.~Pan, C.~Pan, S.~Jin, and J.~Wang, ``{RIS-Aided Near-Field Localization and
  Channel Estimation for the Terahertz System},'' \emph{IEEE Journal of
  Selected Topics in Signal Processing}, pp. 1--14, 2023.

\bibitem{9777939}
Y.~Han, S.~Jin, C.-K. Wen, and T.~Q.~S. Quek, ``{Localization and Channel
  Reconstruction for Extra Large RIS-Assisted Massive MIMO Systems},''
  \emph{IEEE Journal of Selected Topics in Signal Processing}, vol.~16, no.~5,
  pp. 1011--1025, 2022.

\bibitem{9729782}
Z.~Wang, Z.~Liu, Y.~Shen, A.~Conti, and M.~Z. Win, ``{Location Awareness in
  Beyond 5G Networks via Reconfigurable Intelligent Surfaces},'' \emph{IEEE
  Journal on Selected Areas in Communications}, vol.~40, no.~7, pp. 2011--2025,
  2022.

\bibitem{isprs}
Z.~Zhang, L.~Wu, Z.~Zhang, J.~Dang, B.~Zhu, and L.~Wang, ``{Multiple RSS
  Fingerprint based Indoor Localization in RIS-Assisted 5G Wireless
  Communication System},'' \emph{The International Archives of the
  Photogrammetry, Remote Sensing and Spatial Information Sciences}, vol.
  XLVI-3/W1-2022, pp. 287--292, 2022.

\bibitem{9902991}
X.~Gan, C.~Huang, Z.~Yang, C.~Zhong, and Z.~Zhang, ``{Near-Field Localization
  for Holographic RIS Assisted mmWave Systems},'' \emph{IEEE Communications
  Letters}, vol.~27, no.~1, pp. 140--144, 2023.

\bibitem{10536135}
N.~González-Prelcic, M.~F. Keskin, O.~Kaltiokallio, M.~Valkama, D.~Dardari,
  X.~Shen, Y.~Shen, M.~Bayraktar, and H.~Wymeersch, ``The integrated sensing
  and communication revolution for 6g: Vision, techniques, and applications,''
  \emph{Proceedings of the IEEE}, pp. 1--0, 2024.

\bibitem{10188352}
K.~Muthineni, A.~Artemenko, J.~Vidal, and M.~Nájar, ``A survey of 5g-based
  positioning for industry 4.0: State of the art and enhanced techniques,'' in
  \emph{2023 Joint European Conference on Networks and Communications \& 6G
  Summit (EuCNC/6G Summit)}, 2023, pp. 120--125.

\bibitem{10541024}
B.~Rother, N.~Kalis, C.~Haubelt, and F.~Golatowski, ``Localization in 6g: A
  journey along existing wireless communication technologies,'' in \emph{2024
  IEEE 20th International Conference on Factory Communication Systems (WFCS)},
  2024, pp. 1--7.

\bibitem{Zhang2020TowardsUP}
H.~Zhang, H.~Zhang, B.~Di, K.~Bian, Z.~Han, and L.~Song, ``{Towards Ubiquitous
  Positioning by Leveraging Reconfigurable Intelligent Surface},'' \emph{IEEE
  Communications Letters}, vol.~25, pp. 284--288, 2020.

\bibitem{8466374}
C.~Liaskos, S.~Nie, A.~Tsioliaridou, A.~Pitsillides, S.~Ioannidis, and
  I.~Akyildiz, ``{A New Wireless Communication Paradigm through
  Software-Controlled Metasurfaces},'' \emph{IEEE Communications Magazine},
  vol.~56, no.~9, pp. 162--169, 2018.

\bibitem{8910627}
Q.~Wu and R.~Zhang, ``{Towards Smart and Reconfigurable Environment:
  Intelligent Reflecting Surface Aided Wireless Network},'' \emph{IEEE
  Communications Magazine}, vol.~58, no.~1, pp. 106--112, 2020.

\bibitem{9424177}
Y.~Liu, X.~Liu, X.~Mu, T.~Hou, J.~Xu, M.~Di~Renzo, and N.~Al-Dhahir,
  ``{Reconfigurable Intelligent Surfaces: Principles and Opportunities},''
  \emph{IEEE Communications Surveys \& Tutorials}, vol.~23, no.~3, pp.
  1546--1577, 2021.

\bibitem{9365009}
S.~Zeng, H.~Zhang, B.~Di, Y.~Tan, Z.~Han, H.~V. Poor, and L.~Song,
  ``{Reconfigurable Intelligent Surfaces in 6G: Reflective, Transmissive, or
  Both?}'' \emph{IEEE Communications Letters}, vol.~25, no.~6, pp. 2063--2067,
  2021.

\bibitem{ozturk2023ris}
C.~Ozturk, M.~F. Keskin, H.~Wymeersch, and S.~Gezici, ``{RIS-aided near-field
  localization under phase-dependent amplitude variations},'' \emph{IEEE
  Transactions on Wireless Communications}, vol.~22, no.~8, pp. 5550--5566,
  2023.

\bibitem{alexandropoulos2023hybrid}
G.~C. Alexandropoulos, N.~Shlezinger, I.~Alamzadeh, M.~F. Imani, H.~Zhang, and
  Y.~C. Eldar, ``Hybrid reconfigurable intelligent metasurfaces: Enabling
  simultaneous tunable reflections and sensing for 6g wireless
  communications,'' \emph{IEEE Vehicular Technology Magazine}, vol.~19, no.~1,
  pp. 75--84, 2024.

\bibitem{9437234}
J.~Xu, Y.~Liu, X.~Mu, and O.~A. Dobre, ``{STAR-RISs: Simultaneous Transmitting
  and Reflecting Reconfigurable Intelligent Surfaces},'' \emph{IEEE
  Communications Letters}, vol.~25, no.~9, pp. 3134--3138, 2021.

\bibitem{8811733}
Q.~Wu and R.~Zhang, ``{Intelligent Reflecting Surface Enhanced Wireless Network
  via Joint Active and Passive Beamforming},'' \emph{IEEE Transactions on
  Wireless Communications}, vol.~18, no.~11, pp. 5394--5409, 2019.

\bibitem{8930608}
Q.~Wu, ``{Beamforming Optimization for Wireless Network Aided by Intelligent
  Reflecting Surface With Discrete Phase Shifts},'' \emph{IEEE Transactions on
  Communications}, vol.~68, no.~3, pp. 1838--1851, 2020.

\bibitem{Liu2021STARST}
Y.~Liu, X.~Mu, J.~Xu, R.~Schober, Y.~Hao, H.~V. Poor, and L.~H. Hanzo, ``{STAR:
  Simultaneous Transmission and Reflection for 360° Coverage by Intelligent
  Surfaces},'' \emph{IEEE Wireless Communications}, vol.~28, pp. 102--109,
  2021.

\bibitem{9570143}
X.~Mu, Y.~Liu, L.~Guo, J.~Lin, and R.~Schober, ``{Simultaneously Transmitting
  and Reflecting (STAR) RIS Aided Wireless Communications},'' \emph{IEEE
  Transactions on Wireless Communications}, vol.~21, no.~5, pp. 3083--3098,
  2022.

\bibitem{news}
\BIBentryALTinterwordspacing
N.~DOCOMO, ``{DOCOMO conducts world’s first successful trial of transparent
  dynamic metasurface},'' 2020. [Online]. Available:
  \url{https://www.nttdocomo.co.jp/english/info/mediacenter/pr/2020/0117
  00.html}
\BIBentrySTDinterwordspacing

\bibitem{li2022reconfigurable}
Q.~Li, M.~El-Hajjar, I.~Hemadeh, A.~Shojaeifard, A.~A.~M. Mourad, B.~Clerckx,
  and L.~Hanzo, ``Reconfigurable intelligent surfaces relying on non-diagonal
  phase shift matrices,'' \emph{IEEE Transactions on Vehicular Technology},
  vol.~71, no.~6, pp. 6367--6383, 2022.

\bibitem{najafi2020physics}
M.~Najafi, V.~Jamali, R.~Schober, and H.~V. Poor, ``{Physics-based modeling and
  scalable optimization of large intelligent reflecting surfaces},'' \emph{IEEE
  Transactions on Communications}, vol.~69, no.~4, pp. 2673--2691, 2020.

\bibitem{9833921}
G.~Mylonopoulos, C.~D’Andrea, and S.~Buzzi, ``{Active Reconfigurable
  Intelligent Surfaces for User Localization in mmWave MIMO Systems},''
  \emph{2022 IEEE 23rd International Workshop on Signal Processing Advances in
  Wireless Communication (SPAWC)}, pp. 1--5, 2022.

\bibitem{9998527}
Z.~Zhang, L.~Dai, X.~Chen, C.~Liu, F.~Yang, R.~Schober, and H.~V. Poor,
  ``{Active RIS vs. Passive RIS: Which Will Prevail in 6G?}'' \emph{IEEE
  Transactions on Communications}, vol.~71, no.~3, pp. 1707--1725, 2023.

\bibitem{article}
K.~Zhi, C.~Pan, H.~Ren, M.~Chai, and M.~Elkashlan, ``{Active RIS Versus Passive
  RIS: Which is Superior With the Same Power Budget?}'' \emph{IEEE
  Communications Letters}, vol.~26, pp. 1150--1154, 05 2022.

\bibitem{9893187}
H.~Wymeersch and G.~Seco~Granados, ``{Radio Localization and Sensing—Part I:
  Fundamentals},'' \emph{IEEE Communications Letters}, vol.~26, no.~12, pp.
  2816--2820, 2022.

\bibitem{9893114}
H.~Wymeersch and G.~Seco-Granados, ``{Radio Localization and Sensing—Part II:
  State-of-the-Art and Challenges},'' \emph{IEEE Communications Letters},
  vol.~26, no.~12, pp. 2821--2825, 2022.

\bibitem{braun2014ofdm}
K.~M. Braun, ``Ofdm adar algorithms in mobile communication networks,'' Ph.D.
  dissertation, Karlsruher Institut f{\"u}r Technologie (KIT), 2014.

\bibitem{roemer2014analytical}
F.~Roemer, M.~Haardt, and G.~Del~Galdo, ``Analytical performance assessment of
  multi-dimensional matrix-and tensor-based esprit-type algorithms,''
  \emph{IEEE Transactions on Signal Processing}, vol.~62, no.~10, pp.
  2611--2625, 2014.

\bibitem{8240645}
A.~Shahmansoori, G.~E. Garcia, G.~Destino, G.~Seco-Granados, and H.~Wymeersch,
  ``{Position and Orientation Estimation Through Millimeter-Wave MIMO in 5G
  Systems},'' \emph{IEEE Transactions on Wireless Communications}, vol.~17,
  no.~3, pp. 1822--1835, 2018.

\bibitem{9247315}
I.~Yildirim, A.~Uyrus, and E.~Basar, ``{Modeling and Analysis of Reconfigurable
  Intelligent Surfaces for Indoor and Outdoor Applications in Future Wireless
  Networks},'' \emph{IEEE Transactions on Communications}, vol.~69, no.~2, pp.
  1290--1301, 2021.

\bibitem{8108330}
S.~Hu, F.~Rusek, and O.~Edfors, ``{The Potential of Using Large Antenna Arrays
  on Intelligent Surfaces},'' \emph{2017 IEEE 85th Vehicular Technology
  Conference (VTC Spring)}, pp. 1--6, 2017.

\bibitem{hexa2}
\BIBentryALTinterwordspacing
H.~W. et~al., ``{Localisation and sensing use cases and gap analysis},'' 2021.
  [Online]. Available:
  \url{https://hexa-x.eu/wp-content/uploads/2022/02/Hexa-X_D3.1_v1.4.pdf}
\BIBentrySTDinterwordspacing

\bibitem{7426565}
K.~Witrisal, P.~Meissner, E.~Leitinger, Y.~Shen, C.~Gustafson, F.~Tufvesson,
  K.~Haneda, D.~Dardari, A.~F. Molisch, A.~Conti, and M.~Z. Win,
  ``{High-Accuracy Localization for Assisted Living: 5G systems will turn
  multipath channels from foe to friend},'' \emph{IEEE Signal Processing
  Magazine}, vol.~33, no.~2, pp. 59--70, 2016.

\bibitem{8752016}
A.~Shahmansoori, B.~Uguen, G.~Destino, G.~Seco-Granados, and H.~Wymeersch,
  ``{Tracking Position and Orientation Through Millimeter Wave Lens MIMO in 5G
  Systems},'' \emph{IEEE Signal Processing Letters}, vol.~26, no.~8, pp.
  1222--1226, 2019.

\bibitem{8647451}
R.~Mendrzik, H.~Wymeersch, and G.~Bauch, ``{Joint Localization and Mapping
  Through Millimeter Wave MIMO in 5G Systems},'' \emph{2018 IEEE Global
  Communications Conference (GLOBECOM)}, pp. 1--6, 2018.

\bibitem{8515231}
R.~Mendrzik, H.~Wymeersch, G.~Bauch, and Z.~Abu-Shaban, ``{Harnessing NLOS
  Components for Position and Orientation Estimation in 5G Millimeter Wave
  MIMO},'' \emph{IEEE Transactions on Wireless Communications}, vol.~18, no.~1,
  pp. 93--107, 2019.

\bibitem{9140329}
M.~Di~Renzo, A.~Zappone, M.~Debbah, M.-S. Alouini, C.~Yuen, J.~de~Rosny, and
  S.~Tretyakov, ``{Smart Radio Environments Empowered by Reconfigurable
  Intelligent Surfaces: How It Works, State of Research, and The Road Ahead},''
  \emph{IEEE Journal on Selected Areas in Communications}, vol.~38, no.~11, pp.
  2450--2525, 2020.

\bibitem{9148610}
A.~U. Makarfi, K.~M. Rabie, O.~Kaiwartya, O.~S. Badarneh, X.~Li, and R.~Kharel,
  ``{Reconfigurable Intelligent Surface Enabled IoT Networks in Generalized
  Fading Channels},'' \emph{ICC 2020 - 2020 IEEE International Conference on
  Communications (ICC)}, pp. 1--6, 2020.

\bibitem{Cao2021SumRateMF}
Y.~Cao, T.~Lv, W.~Ni, and Z.~Lin, ``{Sum-Rate Maximization for
  Multi-Reconfigurable Intelligent Surface-Assisted Device-to-Device
  Communications},'' \emph{IEEE Transactions on Communications}, vol.~69, pp.
  7283--7296, 2021.

\bibitem{9369969}
X.~Mu, Y.~Liu, L.~Guo, J.~Lin, and R.~Schober, ``{Intelligent Reflecting
  Surface Enhanced Indoor Robot Path Planning: A Radio Map-Based Approach},''
  \emph{IEEE Transactions on Wireless Communications}, vol.~20, no.~7, pp.
  4732--4747, 2021.

\bibitem{9082625}
H.~Chen, T.~Ballal, A.~H. Muqaibel, X.~Zhang, and T.~Y. Al-Naffouri, ``{Air
  Writing via Receiver Array-Based Ultrasonic Source Localization},''
  \emph{IEEE Transactions on Instrumentation and Measurement}, vol.~69, no.~10,
  pp. 8088--8101, 2020.

\bibitem{9293395}
H.~Zhang, J.~Hu, H.~Zhang, B.~Di, K.~Bian, Z.~Han, and L.~Song, ``{MetaRadar:
  Indoor Localization by Reconfigurable Metamaterials},'' \emph{IEEE
  Transactions on Mobile Computing}, vol.~21, no.~8, pp. 2895--2908, 2022.

\bibitem{zheng2023jrcup}
P.~Zheng, H.~Chen, T.~Ballal, M.~Valkama, H.~Wymeersch, and T.~Y. Al-Naffouri,
  ``Jrcup: Joint ris calibration and user positioning for 6g wireless
  systems,'' \emph{IEEE Transactions on Wireless Communications}, vol.~23,
  no.~6, pp. 6683--6698, 2024.

\bibitem{10012935}
M.~Ammous and S.~Valaee, ``{Cooperative Positioning with the Aid of
  Reconfigurable Intelligent Surfaces and Zero Access Points},'' \emph{2022
  IEEE 96th Vehicular Technology Conference (VTC2022-Fall)}, pp. 1--5, 2022.

\bibitem{10100675}
F.~Jiang, A.~Abrardo, K.~Keykhosravi, H.~Wymeersch, D.~Dardari, and
  M.~Di~Renzo, ``Two-timescale transmission design and ris optimization for
  integrated localization and communications,'' \emph{IEEE Transactions on
  Wireless Communications}, vol.~22, no.~12, pp. 8587--8602, 2023.

\bibitem{9852985}
T.~V. Nguyen, D.~N. Nguyen, M.~D. Renzo, and R.~Zhang, ``Leveraging secondary
  reflections and mitigating interference in multi-irs/ris aided wireless
  networks,'' \emph{IEEE Transactions on Wireless Communications}, vol.~22,
  no.~1, pp. 502--517, 2023.

\bibitem{9771731}
T.~V. Nguyen and D.~N. Nguyen, ``Secondary reflections amongst multiple irss:
  Friends or foes?'' in \emph{2022 IEEE Wireless Communications and Networking
  Conference (WCNC)}, 2022, pp. 1347--1352.

\bibitem{9860999}
M.~Ammous and S.~Valaee, ``{Positioning and Tracking Using Reconfigurable
  Intelligent Surfaces and Extended Kalman Filter},'' \emph{2022 IEEE 95th
  Vehicular Technology Conference: (VTC2022-Spring)}, pp. 1--6, 2022.

\bibitem{882463}
E.~Wan and R.~Van Der~Merwe, ``{The unscented Kalman filter for nonlinear
  estimation},'' \emph{Proceedings of the IEEE 2000 Adaptive Systems for Signal
  Processing, Communications, and Control Symposium (Cat. No.00EX373)}, pp.
  153--158, 2000.

\bibitem{9081910}
T.~Koike-Akino, P.~Wang, M.~Pajovic, H.~Sun, and P.~V. Orlik,
  ``{Fingerprinting-Based Indoor Localization With Commercial MMWave WiFi: A
  Deep Learning Approach},'' \emph{IEEE Access}, vol.~8, pp. 84\,879--84\,892,
  2020.

\bibitem{8292280}
J.~Vieira, E.~Leitinger, M.~Sarajlic, X.~Li, and F.~Tufvesson, ``{Deep
  convolutional neural networks for massive MIMO fingerprint-based
  positioning},'' \emph{2017 IEEE 28th Annual International Symposium on
  Personal, Indoor, and Mobile Radio Communications (PIMRC)}, pp. 1--6, 2017.

\bibitem{kim2022risenabled}
H.~Kim, H.~Chen, M.~F. Keskin, Y.~Ge, K.~Keykhosravi, G.~C. Alexandropoulos,
  S.~Kim, and H.~Wymeersch, ``Ris-enabled and access-point-free simultaneous
  radio localization and mapping,'' \emph{IEEE Transactions on Wireless
  Communications}, 2023.

\bibitem{9685930}
Z.~Yang, H.~Zhang, B.~Di, H.~Zhang, K.~Bian, and L.~Song, ``{Wireless Indoor
  Simultaneous Localization and Mapping Using Reconfigurable Intelligent
  Surface},'' \emph{2021 IEEE Global Communications Conference (GLOBECOM)}, pp.
  1--6, 2021.

\bibitem{mitchell1997machine}
T.~M. Mitchell, ``{Machine learning},'' 1997.

\bibitem{s16050707}
A.~Alarifi, A.~Al-Salman, M.~Alsaleh, A.~Alnafessah, S.~Al-Hadhrami, M.~A.
  Al-Ammar, and H.~S. Al-Khalifa, ``{Ultra Wideband Indoor Positioning
  Technologies: Analysis and Recent Advances},'' \emph{Sensors}, vol.~16,
  no.~5, 2016.

\bibitem{8688470}
X.~Guo, N.~R. Elikplim, N.~Ansari, L.~Li, and L.~Wang, ``{Robust WiFi
  Localization by Fusing Derivative Fingerprints of RSS and Multiple
  Classifiers},'' \emph{IEEE Transactions on Industrial Informatics}, vol.~16,
  no.~5, pp. 3177--3186, 2020.

\bibitem{7949029}
X.~Tian, W.~Li, Y.~Yang, Z.~Zhang, and X.~Wang, ``{Optimization of Fingerprints
  Reporting Strategy for WLAN Indoor Localization},'' \emph{IEEE Transactions
  on Mobile Computing}, vol.~17, no.~2, pp. 390--403, 2018.

\bibitem{8827665}
K.-H. Lam, C.-C. Cheung, and W.-C. Lee, ``{RSSI-Based LoRa Localization Systems
  for Large-Scale Indoor and Outdoor Environments},'' \emph{IEEE Transactions
  on Vehicular Technology}, vol.~68, no.~12, pp. 11\,778--11\,791, 2019.

\bibitem{hexa}
\BIBentryALTinterwordspacing
K.~R. et~al., ``{Towards TBPS communication in 6G: Use cases and gap
  analysis},'' 2021. [Online]. Available: \url{https://hexa-x.eu/wpcontent/
  uploads/2021/06/Hexa-X_D2.1.pdf}
\BIBentrySTDinterwordspacing

\bibitem{8356190}
Z.~Abu-Shaban, X.~Zhou, T.~Abhayapala, G.~Seco-Granados, and H.~Wymeersch,
  ``{Error Bounds for Uplink and Downlink 3D Localization in 5G Millimeter Wave
  Systems},'' \emph{IEEE Transactions on Wireless Communications}, vol.~17,
  no.~8, pp. 4939--4954, 2018.

\bibitem{9997557}
N.~Kouzayha, M.~A. Kishk, H.~Sarieddeen, M.-S. Alouini, and T.~Y. Al-Naffouri,
  ``{Coexisting Terahertz and RF Finite Wireless Networks: Coverage and Rate
  Analysis},'' \emph{IEEE Transactions on Wireless Communications}, vol.~22,
  no.~7, pp. 4873--4889, 2023.

\bibitem{9724202}
X.~Shao, C.~You, W.~Ma, X.~Chen, and R.~Zhang, ``{Target Sensing With
  Intelligent Reflecting Surface: Architecture and Performance},'' \emph{IEEE
  Journal on Selected Areas in Communications}, vol.~40, no.~7, pp. 2070--2084,
  2022.

\bibitem{ghiasvand2022miso}
S.~Ghiasvand, A.~Nasri, A.~H.~A. Bafghi, and M.~Nasiri-Kenari, ``{MISO Wireless
  Localization in The Presence of Reconfigurable Intelligent Surface},''
  \emph{arXiv}, 2022.

\bibitem{10042425}
Z.~Xing, R.~Wang, and X.~Yuan, ``{Joint Active and Passive Beamforming Design
  for Reconfigurable Intelligent Surface Enabled Integrated Sensing and
  Communication},'' \emph{IEEE Transactions on Communications}, vol.~71, no.~4,
  pp. 2457--2474, 2023.

\bibitem{9750199}
A.~Nasri, A.~H.~A. Bafghi, and M.~Nasiri-Kenari, ``{Wireless Localization in
  the Presence of Intelligent Reflecting Surface},'' \emph{IEEE Wireless
  Communications Letters}, vol.~11, no.~7, pp. 1315--1319, 2022.

\bibitem{9434917}
H.~Zhang, H.~Zhang, B.~Di, K.~Bian, Z.~Han, C.~Xu, D.~Zhang, and L.~Song,
  ``{RSS Fingerprinting Based Multi-user Outdoor Localization Using
  Reconfigurable Intelligent Surfaces},'' \emph{2021 15th International
  Symposium on Medical Information and Communication Technology (ISMICT)}, pp.
  167--172, 2021.

\bibitem{9860384}
Z.~Wang, Z.~Liu, Y.~Shen, A.~Conti, and M.~Z. Win, ``{Wideband Localization
  with Reconfigurable Intelligent Surfaces},'' \emph{2022 IEEE 95th Vehicular
  Technology Conference: (VTC2022-Spring)}, pp. 1--6, 2022.

\bibitem{9860587}
M.~Luan, B.~Wang, Z.~Chang, T.~Hämäläinen, Z.~Ling, and F.~Hu, ``{Joint
  Subcarrier and Phase Shifts Optimization for RIS-aided
  Localization-Communication System},'' \emph{2022 IEEE 95th Vehicular
  Technology Conference: (VTC2022-Spring)}, pp. 1--5, 2022.

\bibitem{9900336}
C.~Gaudreauand and A.~Chaaban, ``{Localization by Modulated Reconfigurable
  Intelligent Surfaces},'' \emph{IEEE Communications Letters}, vol.~26, no.~12,
  pp. 2904--2908, 2022.

\bibitem{s23020984}
X.~Luo and N.~Meratnia, ``{A Codeword-Independent Localization Technique for
  Reconfigurable Intelligent Surface Enhanced Environments Using Adversarial
  Learning},'' \emph{Sensors}, vol.~23, no.~2, 2023.

\bibitem{hong2023risposition}
S.~Hong, M.~Li, C.~Pan, M.~D. Renzo, W.~Zhang, and L.~Hanzo, ``{RIS-Position
  and Orientation Estimation in MIMO-OFDM Systems with Practical Scatterers},''
  \emph{arXiv}, 2023.

\bibitem{9977919}
Y.~Liu, S.~Hong, C.~Pan, Y.~Wang, Y.~Pan, and M.~Chen, ``{Cramér-Rao Lower
  Bound Analysis of Multiple-RIS-Aided mmWave Positioning Systems},''
  \emph{2022 IEEE 33rd Annual International Symposium on Personal, Indoor and
  Mobile Radio Communications (PIMRC)}, pp. 1110--1115, 2022.

\bibitem{10113892}
D.-R. Emenonye, H.~S. Dhillon, and R.~Michael~Buehrer, ``{RIS-Aided
  Localization under Position and Orientation Offsets in the Near and Far
  Field},'' \emph{IEEE Transactions on Wireless Communications}, pp. 1--1,
  2023.

\bibitem{10285302}
J.~Hu, Z.~Chen, T.~Zheng, R.~Schober, and J.~Luo, ``Holofed:
  Environment-adaptive positioning via multi-band reconfigurable holographic
  surfaces and federated learning,'' \emph{IEEE Journal on Selected Areas in
  Communications}, vol.~41, no.~12, pp. 3736--3751, 2023.

\bibitem{10216343}
T.~Ma, Y.~Xiao, X.~Lei, and M.~Xiao, ``Integrated sensing and communication for
  wireless extended reality (xr) with reconfigurable intelligent surface,''
  \emph{IEEE Journal of Selected Topics in Signal Processing}, vol.~17, no.~5,
  pp. 980--994, 2023.

\bibitem{10325432}
Y.~Zhao, P.~Chen, M.~Sun, T.~Luo, and Z.~Cao, ``Ris-aided sensing system with
  localization function: Fundamental and practical design,'' \emph{IEEE Sensors
  Journal}, vol.~24, no.~1, pp. 506--514, 2024.

\bibitem{10478036}
S.~Sardellitti, P.~D. Lorenzo, and S.~Barbarossa, ``Ris-aided wireless
  fingerprinting localization based on multilayer graph representations,''
  \emph{IEEE Communications Letters}, vol.~28, no.~5, pp. 1043--1047, 2024.

\bibitem{10373864}
T.~Chao, C.~C. Fung, Z.-E. Ni, and M.~Servetnyk, ``Joint beamforming and aerial
  irs positioning design for irs-assisted miso system with multiple access
  points,'' \emph{IEEE Open Journal of the Communications Society}, vol.~5, pp.
  612--632, 2024.

\bibitem{10257295}
M.~S. Siraj, A.~B. Rahman, E.~E. Tsiropoulou, S.~Papavassiliou, and
  J.~Plusquellic, ``Symbiotic positioning, navigation, and timing,'' in
  \emph{2023 19th International Conference on Distributed Computing in Smart
  Systems and the Internet of Things (DCOSS-IoT)}, 2023, pp. 261--268.

\bibitem{9782100}
A.~Fascista, M.~F. Keskin, A.~Coluccia, H.~Wymeersch, and G.~Seco-Granados,
  ``{RIS-Aided Joint Localization and Synchronization With a Single-Antenna
  Receiver: Beamforming Design and Low-Complexity Estimation},'' \emph{IEEE
  Journal of Selected Topics in Signal Processing}, vol.~16, no.~5, pp.
  1141--1156, 2022.

\bibitem{He2022SimultaneousIA}
J.~He, A.~Fakhreddine, and G.~C. Alexandropoulos, ``{Simultaneous Indoor and
  Outdoor 3D Localization with STAR-RIS-Assisted Millimeter Wave Systems},''
  \emph{2022 IEEE 96th Vehicular Technology Conference (VTC2022-Fall)}, pp.
  1--6, 2022.

\bibitem{xanthos2022joint}
Y.~Xanthos, W.~Lyu, S.~Yang, C.~Assi, X.~Zou, and N.~Wei, ``{Joint Localization
  and Beamforming for Reconfigurable Intelligent Surface Aided 5G mmWave
  Communication Systems},'' \emph{arXiv preprint arXiv:2210.17530}, 2022.

\bibitem{Keykhosravi2022RISEnabledSL}
K.~Keykhosravi, G.~Seco-Granados, G.~C. Alexandropoulos, and H.~Wymeersch,
  ``{RIS-Enabled Self-Localization: Leveraging Controllable Reflections With
  Zero Access Points},'' \emph{ICC 2022 - IEEE International Conference on
  Communications}, pp. 2852--2857, 2022.

\bibitem{9540372}
Y.~Lin, S.~Jin, M.~Matthaiou, and X.~You, ``{Channel Estimation and User
  Localization for IRS-Assisted MIMO-OFDM Systems},'' \emph{IEEE Transactions
  on Wireless Communications}, vol.~21, no.~4, pp. 2320--2335, 2022.

\bibitem{10012776}
Y.~Lu, H.~Chen, J.~Talvitie, H.~Wymeersch, and M.~Valkama, ``{Joint RIS
  Calibration and Multi-User Positioning},'' \emph{2022 IEEE 96th Vehicular
  Technology Conference (VTC2022-Fall)}, pp. 1--6, 2022.

\bibitem{9827856}
B.~Teng, X.~Yuan, R.~Wang, and S.~Jin, ``{Bayesian User Tracking for
  Reconfigurable Intelligent Surface Aided mmWave MIMO System},'' \emph{2022
  IEEE 12th Sensor Array and Multichannel Signal Processing Workshop (SAM)},
  pp. 201--205, 2022.

\bibitem{9977483}
M.~Ammous and S.~Valaee, ``{Cooperative Positioning with the Aid of
  Reconfigurable Intelligent Surfaces and Device-to-Device Communications in
  mmWave},'' \emph{2022 IEEE 33rd Annual International Symposium on Personal,
  Indoor and Mobile Radio Communications (PIMRC)}, pp. 683--688, 2022.

\bibitem{10124713}
J.~He, A.~Fakhreddine, C.~Vanwynsberghe, H.~Wymeersch, and G.~C.
  Alexandropoulos, ``{3D Localization with a Single Partially-Connected
  Receiving RIS: Positioning Error Analysis and Algorithmic Design},''
  \emph{IEEE Transactions on Vehicular Technology}, pp. 1--13, 2023.

\bibitem{10096904}
P.~Zheng, H.~Chen, T.~Ballal, H.~Wymeersch, and T.~Y. Al-Naffouri,
  ``{Misspecified Cramér-Rao Bound of RIS-Aided Localization Under Geometry
  Mismatch},'' \emph{ICASSP 2023 - 2023 IEEE International Conference on
  Acoustics, Speech and Signal Processing (ICASSP)}, pp. 1--5, 2023.

\bibitem{he2023compressed}
J.~He, A.~Fakhreddine, H.~Wymeersch, and G.~C. Alexandropoulos,
  ``{Compressed-sensing-based 3D localization with distributed passive
  reconfigurable intelligent surfaces},'' \emph{ICASSP 2023-2023 IEEE
  International Conference on Acoustics, Speech and Signal Processing
  (ICASSP)}, pp. 1--5, 2023.

\bibitem{wang2023target}
P.~Wang, W.~Mei, J.~Fang, and R.~Zhang, ``Target-mounted intelligent reflecting
  surface for joint location and orientation estimation,'' \emph{IEEE Journal
  on Selected Areas in Communications}, 2023.

\bibitem{asif2023isac}
M.~Asif~Haider, Y.~D. Zhang, and E.~Aboutanios, ``{ISAC system assisted by RIS
  with sparse active elements},'' \emph{EURASIP Journal on Advances in Signal
  Processing}, vol. 2023, no.~1, pp. 1--22, 2023.

\bibitem{chen2023multirisenabled}
H.~Chen, P.~Zheng, M.~F. Keskin, T.~Al-Naffouri, and H.~Wymeersch,
  ``Multi-ris-enabled 3d sidelink positioning,'' \emph{IEEE Transactions on
  Wireless Communications}, 2024.

\bibitem{9528041}
K.~Keykhosravi, M.~F. Keskin, S.~Dwivedi, G.~Seco-Granados, and H.~Wymeersch,
  ``{Semi-Passive 3D Positioning of Multiple RIS-Enabled Users},'' \emph{IEEE
  Transactions on Vehicular Technology}, vol.~70, no.~10, pp. 11\,073--11\,077,
  2021.

\bibitem{9774917}
K.~Keykhosravi, M.~F. Keskin, G.~Seco-Granados, P.~Popovski, and H.~Wymeersch,
  ``{RIS-Enabled SISO Localization Under User Mobility and Spatial-Wideband
  Effects},'' \emph{IEEE Journal of Selected Topics in Signal Processing},
  vol.~16, no.~5, pp. 1125--1140, 2022.

\bibitem{10001508}
Z.~Shao, X.~Yuan, W.~Zhang, and M.~Di~Renzo, ``{Joint Localization and
  Information Transfer for RIS Aided Full-Duplex Systems},'' \emph{GLOBECOM
  2022 - 2022 IEEE Global Communications Conference}, pp. 3253--3258, 2022.

\bibitem{ghazalian2022joint}
R.~Ghazalian, H.~Chen, G.~C. Alexandropoulos, G.~Seco-Granados, H.~Wymeersch,
  and R.~J{\"a}ntti, ``Joint user localization and location calibration of a
  hybrid reconfigurable intelligent surface,'' \emph{IEEE Transactions on
  Vehicular Technology}, 2023.

\bibitem{9771873}
Q.~Cheng, L.~Li, M.-M. Zhao, and M.-J. Zhao, ``{Cooperative Localization for
  Reconfigurable Intelligent Surface-Aided mmWave Systems},'' \emph{2022 IEEE
  Wireless Communications and Networking Conference (WCNC)}, pp. 1051--1056,
  2022.

\bibitem{10070578}
Z.~Xing, R.~Wang, X.~Yuan, and J.~Wu, ``{Location Information Assisted
  Beamforming Design for Reconfigurable Intelligent Surface Aided Communication
  Systems},'' \emph{IEEE Transactions on Wireless Communications}, pp. 1--1,
  2023.

\bibitem{9771819}
P.~Gao, L.~Lian, and J.~Yu, ``{Optimal Passive Beamforming for Cooperative
  Localization with RIS-Assisted mmWave Systems},'' \emph{2022 IEEE Wireless
  Communications and Networking Conference (WCNC)}, pp. 222--227, 2022.

\bibitem{9794297}
P.~Gao and L.~Lian, ``{Wireless Area Positioning in RIS-Assisted mmWave
  Systems: Joint Passive and Active Beamforming Design},'' \emph{IEEE Signal
  Processing Letters}, vol.~29, pp. 1372--1376, 2022.

\bibitem{10052208}
J.~Feng, P.~Zhang, L.~Huang, and G.~Qian, ``{Reconfigurable Intelligent Surface
  Aided DFRC Vehicular Networks},'' \emph{2023 6th World Conference on
  Computing and Communication Technologies (WCCCT)}, pp. 1--6, 2023.

\bibitem{10130575}
Z.~Li, Z.~Gao, and T.~Li, ``{Sensing User's Channel and Location with Terahertz
  Extra-Large Reconfigurable Intelligent Surface under Hybrid-Field Beam Squint
  Effect},'' \emph{IEEE Journal of Selected Topics in Signal Processing}, pp.
  1--16, 2023.

\bibitem{10193850}
M.~Eskandari and A.~V. Savkin, ``Slaps: Simultaneous localization and phase
  shift for a ris-equipped uav in 5g/6g wireless communication networks,''
  \emph{IEEE Transactions on Intelligent Vehicles}, vol.~8, no.~12, pp.
  4722--4733, 2023.

\bibitem{10570712}
A.~Fazli, E.~Tohidi, Z.~Utkovski, and S.~Stańczak, ``User-centric monostatic
  sensing aided by reconfigurable intelligent surfaces,'' in \emph{2024 IEEE
  Wireless Communications and Networking Conference (WCNC)}, 2024, pp. 01--06.

\bibitem{10571090}
D.~Wang, A.~Bazzi, and M.~Chafii, ``Ris-enabled integrated sensing and
  communication for 6g systems,'' in \emph{2024 IEEE Wireless Communications
  and Networking Conference (WCNC)}, 2024, pp. 1--6.

\bibitem{10108969}
E.~Moeen~Taghavi, R.~Hashemi, N.~Rajatheva, and M.~Latva-Aho,
  ``Environment-aware joint active/passive beamforming for ris-aided
  communications leveraging channel knowledge map,'' \emph{IEEE Communications
  Letters}, vol.~27, no.~7, pp. 1824--1828, 2023.

\bibitem{10234214}
D.-R. Emenonye, H.~S. Dhillon, and R.~M. Buehrer, ``Fundamentals of ris-aided
  localization in the far-field,'' \emph{IEEE Transactions on Wireless
  Communications}, vol.~23, no.~4, pp. 3408--3424, 2024.

\bibitem{10233198}
H.~Kim, H.~Chen, M.~F. Keskin, Y.~Ge, K.~Keykhosravi, G.~C. Alexandropoulos,
  S.~Kim, and H.~Wymeersch, ``Ris-enabled and access-point-free simultaneous
  radio localization and mapping,'' \emph{IEEE Transactions on Wireless
  Communications}, vol.~23, no.~4, pp. 3344--3360, 2024.

\bibitem{10464638}
A.~Mondal and S.~Biswas, ``Ris-assisted environment aware positioning in 6g
  wireless networks,'' in \emph{2023 IEEE Globecom Workshops (GC Wkshps)},
  2023, pp. 1892--1897.

\bibitem{10464850}
J.~Li, W.~Wang, R.~Jiang, and S.~Huang, ``Csinet-former network for bilateral
  user 3d localization in star-ris-assisted miso systems,'' in \emph{2023 IEEE
  Globecom Workshops (GC Wkshps)}, 2023, pp. 1886--1891.

\bibitem{10542461}
A.~Fadakar, M.~Sabbaghian, and H.~Wymeersch, ``Multi-ris-assisted 3d
  localization and synchronization via deep learning,'' \emph{IEEE Open Journal
  of the Communications Society}, vol.~5, pp. 3299--3314, 2024.

\bibitem{10471980}
D.~Ma, Z.~Bai, J.~Zhao, H.~Xu, Z.~Liu, D.~Zhou, M.~Jiang, and K.~S. Kwak,
  ``Dual-ris assisted 3d positioning and beamforming design in isac system,''
  in \emph{2024 26th International Conference on Advanced Communications
  Technology (ICACT)}, 2024, pp. 187--192.

\bibitem{10225368}
R.~Ghazalian, H.~Chen, G.~C. Alexandropoulos, G.~Seco-Granados, H.~Wymeersch,
  and R.~Jäntti, ``Joint user localization and location calibration of a
  hybrid reconfigurable intelligent surface,'' \emph{IEEE Transactions on
  Vehicular Technology}, vol.~73, no.~1, pp. 1435--1440, 2024.

\bibitem{10571832}
W.~Jia, J.~Cao, M.~Li, and Z.~Yu, ``User sensing and localization with
  reconfigurable intelligent surface for terahertz massive mimo systems,''
  \emph{IEEE Access}, vol.~12, pp. 91\,089--91\,100, 2024.

\bibitem{10464904}
K.~Meng, Q.~Wu, W.~Chen, and D.~Li, ``Vehicle-mounted intelligent surface for
  cooperative localization in cellular networks,'' in \emph{2023 IEEE Globecom
  Workshops (GC Wkshps)}, 2023, pp. 1904--1909.

\bibitem{10188365}
Z.~Ye, F.~Junaid, R.~Nilsson, and J.~van~de Beek, ``Autonomous single antenna
  receiver localization and tracking with ris and ekf,'' in \emph{2023 Joint
  European Conference on Networks and Communications \& 6G Summit (EuCNC/6G
  Summit)}, 2023, pp. 216--221.

\bibitem{10543050}
X.~Gan, C.~Huang, Z.~Yang, C.~Zhong, X.~Chen, Z.~Zhang, Q.~Guo, C.~Yuen, and
  M.~Debbah, ``Bayesian learning for double-ris aided isac systems with
  superimposed pilots and data,'' \emph{IEEE Journal of Selected Topics in
  Signal Processing}, pp. 1--16, 2024.

\bibitem{10283493}
D.-R. Emenonye, H.~S. Dhillon, and R.~M. Buehrer, ``Limitations of ris-aided
  localization: Inspecting the relationships between channel parameters,'' in
  \emph{2023 IEEE International Conference on Communications Workshops (ICC
  Workshops)}, 2023, pp. 1890--1895.

\bibitem{10582646}
U.~Mutlu, M.~Bilim, and Y.~Kabalci, ``Joint channel estimation and localization
  in ris assisted ofdm-mimo system,'' in \emph{2024 6th Global Power, Energy
  and Communication Conference (GPECOM)}, 2024, pp. 687--692.

\bibitem{10416244}
J.~Du, Y.~Cheng, L.~Jin, S.~Li, and F.~Gao, ``Nested tensor-based integrated
  sensing and communication in ris-assisted thz mimo systems,'' \emph{IEEE
  Transactions on Signal Processing}, vol.~72, pp. 1141--1157, 2024.

\bibitem{10143971}
M.~Mizmizi, D.~Tagliaferri, D.~Badini, and U.~Spagnolini, ``Target-to-user
  association in isac systems with vehicle-lodged ris,'' \emph{IEEE Wireless
  Communications Letters}, vol.~12, no.~9, pp. 1558--1562, 2023.

\bibitem{10436573}
Q.~Wang, L.~Liu, S.~Zhang, B.~Di, and F.~C.~M. Lau, ``A heterogeneous 6g
  networked sensing architecture with active and passive anchors,'' \emph{IEEE
  Transactions on Wireless Communications}, pp. 1--1, 2024.

\bibitem{10565756}
J.~Hu, D.~Niyato, and J.~Luo, ``Cross-domain learning framework for tracking
  users in ris-aided multi-band isac systems with sparse labeled data,''
  \emph{IEEE Journal on Selected Areas in Communications}, pp. 1--1, 2024.

\bibitem{10552748}
J.~Du, M.~He, J.~He, J.~Liu, L.~Jin, and Y.~Guan, ``A tensor-based signal
  processing for isac using c-drcnn in ris-assisted mmwave mimo-ofdm systems,''
  \emph{IEEE Internet of Things Journal}, pp. 1--1, 2024.

\bibitem{chen2023RISs}
H.~Chen, H.~Kim, M.~Ammous, G.~Seco-Granados, G.~C. Alexandropoulos, S.~Valaee,
  and H.~Wymeersch, ``Riss and sidelink communications in smart cities: The key
  to seamless localization and sensing,'' \emph{IEEE Communications Magazine},
  vol.~61, no.~8, pp. 140--146, 2023.

\bibitem{9548046}
C.~L. Nguyen, O.~Georgiou, G.~Gradoni, and M.~Di~Renzo, ``{Wireless
  Fingerprinting Localization in Smart Environments Using Reconfigurable
  Intelligent Surfaces},'' \emph{IEEE Access}, vol.~9, pp. 135\,526--135\,541,
  2021.

\bibitem{10118610}
G.~Zhang, D.~Zhang, Y.~He, J.~Chen, F.~Zhou, and Y.~Chen, ``{Passive Human
  Localization with the Aid of Reconfigurable Intelligent Surface},''
  \emph{2023 IEEE Wireless Communications and Networking Conference (WCNC)},
  pp. 1--6, 2023.

\bibitem{zhang2023multi}
G.~Zhang, D.~Zhang, Y.~He, J.~Chen, and F.~Zhou, ``{Multi-person passive wifi
  indoor localization with intelligent reflecting surface},'' \emph{IEEE
  Transactions on Wireless Communications}, 2023.

\bibitem{vaca2023radio}
C.~J. Vaca-Rubio, P.~Ramirez-Espinosa, K.~Kansanen, Z.-H. Tan, and
  E.~de~Carvalho, ``{Radio sensing with large intelligent surface for 6G},''
  \emph{ICASSP 2023-2023 IEEE International Conference on Acoustics, Speech and
  Signal Processing (ICASSP)}, pp. 1--5, 2023.

\bibitem{9827014}
N.~Afzali, M.~J. Omidi, K.~Navaie, and N.~S. Moayedian, ``{Low Complexity
  Multi-User Indoor Localization Using Reconfigurable Intelligent Surface},''
  \emph{2022 30th International Conference on Electrical Engineering (ICEE)},
  pp. 731--736, 2022.

\bibitem{10124023}
Y.~Wang, I.~W.-H. Ho, S.~Zhang, and Y.~Wang, ``{Intelligent Reflecting Surface
  enabled Fingerprinting-based Localization with Deep Reinforcement
  Learning},'' \emph{IEEE Transactions on Vehicular Technology}, pp. 1--11,
  2023.

\bibitem{9836317}
T.~Ma, Y.~Xiao, X.~Lei, W.~Xiong, and M.~Xiao, ``{Distributed Reconfigurable
  Intelligent Surfaces Assisted Indoor Positioning},'' \emph{IEEE Transactions
  on Wireless Communications}, vol.~22, no.~1, pp. 47--58, 2023.

\bibitem{9726785}
G.~C. Alexandropoulos, I.~Vinieratou, and H.~Wymeersch, ``{Localization via
  Multiple Reconfigurable Intelligent Surfaces Equipped With Single Receive RF
  Chains},'' \emph{IEEE Wireless Communications Letters}, vol.~11, no.~5, pp.
  1072--1076, 2022.

\bibitem{9838395}
B.~Luo, M.~Dong, H.~Wu, Y.~Li, L.~Yang, X.~Chen, and B.~Bai, ``{Reconfigurable
  Intelligent Surface Assisted Millimeter Wave Indoor Localization Systems},''
  \emph{ICC 2022 - IEEE International Conference on Communications}, pp.
  4535--4540, 2022.

\bibitem{10123105}
T.~Wu, C.~Pan, Y.~Pan, H.~Ren, M.~Elkashlan, and C.-X. Wang, ``{Fingerprint
  Based mmWave Positioning System Aided by Reconfigurable Intelligent
  Surface},'' \emph{IEEE Wireless Communications Letters}, pp. 1--1, 2023.

\bibitem{9903366}
G.~Stratidakis, S.~Droulias, and A.~Alexiou, ``{Optimal Position and
  Orientation Study of Reconfigurable Intelligent Surfaces in a Mobile User
  Environment},'' \emph{IEEE Transactions on Antennas and Propagation},
  vol.~70, no.~10, pp. 8863--8871, 2022.

\bibitem{10333393}
F.~Wang, X.~Houv, X.~Wang, X.~Li, L.~Chen, and T.~Asai, ``Reconfigurable
  intelligent surface aided joint communication and positioning,'' in
  \emph{2023 IEEE 98th Vehicular Technology Conference (VTC2023-Fall)}, 2023,
  pp. 1--6.

\bibitem{umer2024}
A.~Umer, I.~Müürsepp, and M.~M. Alam, ``Reconfigurable intelligent surfaces
  in dynamic rich scattering environments: Bilstm-based optimization for
  accurate user localization,'' pp. 122--126, 2024.

\bibitem{10188251}
S.~Bazin and K.~Navaie, ``Improving indoor positioning accuracy using ris-based
  rss optimization,'' in \emph{2023 Joint European Conference on Networks and
  Communications \& 6G Summit (EuCNC/6G Summit)}, 2023, pp. 663--668.

\bibitem{10440181}
Z.~Ye, F.~Junaid, E.~Ibrahim, R.~Nilsson, and J.~Van De~Beek, ``Monostatic
  sensing for passive ris localization and tracking,'' \emph{IEEE Wireless
  Communications Letters}, vol.~13, no.~5, pp. 1260--1264, 2024.

\bibitem{9674914}
S.~Huang, B.~Wang, Y.~Zhao, and M.~Luan, ``{Near-Field RSS-Based Localization
  Algorithms Using Reconfigurable Intelligent Surface},'' \emph{IEEE Sensors
  Journal}, vol.~22, no.~4, pp. 3493--3505, 2022.

\bibitem{9625826}
D.~Dardari, N.~Decarli, A.~Guerra, and F.~Guidi, ``{LOS/NLOS Near-Field
  Localization With a Large Reconfigurable Intelligent Surface},'' \emph{IEEE
  Transactions on Wireless Communications}, vol.~21, no.~6, pp. 4282--4294,
  2022.

\bibitem{9650561}
M.~Luan, B.~Wang, Y.~Zhao, Z.~Feng, and F.~Hu, ``{Phase Design and Near-Field
  Target Localization for RIS-Assisted Regional Localization System},''
  \emph{IEEE Transactions on Vehicular Technology}, vol.~71, no.~2, pp.
  1766--1777, 2022.

\bibitem{9838889}
C.~Öztürk, M.~F. Keskin, H.~Wymeersch, and S.~Gezici, ``{On the Impact of
  Hardware Impairments on RIS-aided Localization},'' \emph{ICC 2022 - IEEE
  International Conference on Communications}, pp. 2846--2851, 2022.

\bibitem{9860413}
M.~Rahal, B.~Denis, K.~Keykhosravi, M.~F. Keskin, B.~Uguen, and H.~Wymeersch,
  ``{Constrained RIS Phase Profile Optimization and Time Sharing for Near-field
  Localization},'' \emph{2022 IEEE 95th Vehicular Technology Conference:
  (VTC2022-Spring)}, pp. 1--6, 2022.

\bibitem{10000689}
O.~Rinchi, A.~Elzanaty, and A.~Alsharoa, ``{Single-Snapshot Localization for
  Near-Field RIS Model Using Atomic Norm Minimization},'' \emph{GLOBECOM 2022 -
  2022 IEEE Global Communications Conference}, pp. 2432--2437, 2022.

\bibitem{9940381}
F.~Zhang, M.-M. Zhao, M.~Lei, and M.~Zhao, ``{Joint Power Allocation and
  Phase-Shift Design for RIS-Aided Cooperative Near-Field Localization},''
  \emph{2022 International Symposium on Wireless Communication Systems
  (ISWCS)}, pp. 1--6, 2022.

\bibitem{rahal2023performance}
M.~Rahal, B.~Denis, K.~Keykhosravi, M.~F. Keskin, B.~Uguen, G.~C.
  Alexandropoulos, and H.~Wymeersch, ``Performance of ris-aided near-field
  localization under beams approximation from real hardware characterization,''
  \emph{EURASIP Journal on Wireless Communications and Networking}, vol. 2023,
  no.~1, p.~86, 2023.

\bibitem{ozturk2023risaided}
C.~Ozturk, M.~F. Keskin, V.~Sciancalepore, H.~Wymeersch, and S.~Gezici,
  ``Ris-aided localization under pixel failures,'' \emph{IEEE Transactions on
  Wireless Communications}, 2024.

\bibitem{9921216}
X.~Zhang and H.~Zhang, ``{Hybrid Reconfigurable Intelligent Surfaces-Assisted
  Near-Field Localization},'' \emph{IEEE Communications Letters}, vol.~27,
  no.~1, pp. 135--139, 2023.

\bibitem{9941256}
Y.~Jiang, F.~Gao, M.~Jian, S.~Zhang, and W.~Zhang, ``{Reconfigurable
  Intelligent Surface for Near Field Communications: Beamforming and
  Sensing},'' \emph{IEEE Transactions on Wireless Communications}, vol.~22,
  no.~5, pp. 3447--3459, 2023.

\bibitem{10001209}
R.~Ghazalian, K.~Keykhosravi, H.~Chen, H.~Wymeersch, and R.~Jäntti,
  ``{Bi-Static Sensing for Near-Field RIS Localization},'' \emph{GLOBECOM 2022
  - 2022 IEEE Global Communications Conference}, pp. 6457--6462, 2022.

\bibitem{10001107}
Y.~Pan, C.~Pan, S.~Jin, and J.~Wang, ``{Localization in the Near Field of a
  RIS-Assisted mmWave/subTHz System},'' \emph{GLOBECOM 2022 - 2022 IEEE Global
  Communications Conference}, pp. 3905--3910, 2022.

\bibitem{9940428}
Z.~Li, Z.~Wan, K.~Ying, Y.~Mei, M.~Ke, and Z.~Gao, ``{Reconfigurable
  Intelligent Surface Assisted Localization Over Near-Field Beam Squint
  Effect},'' \emph{2022 International Symposium on Wireless Communication
  Systems (ISWCS)}, pp. 1--6, 2022.

\bibitem{10149471}
Y.~Pan, C.~Pan, S.~Jin, and J.~Wang, ``{RIS-Aided Near-Field Localization and
  Channel Estimation for the Terahertz System},'' \emph{IEEE Journal of
  Selected Topics in Signal Processing}, pp. 1--14, 2023.

\bibitem{10436860}
P.~Zheng, H.~Chen, H.~Wymeersch, and T.~Y. Al-Naffouri, ``Near field sidelink
  positioning through a single active ris,'' in \emph{GLOBECOM 2023 - 2023 IEEE
  Global Communications Conference}, 2023, pp. 2511--2516.

\bibitem{10140033}
D.-R. Emenonye, A.~Pradhan, H.~S. Dhillon, and R.~M. Buehrer, ``Otfs enabled
  ris- aided localization: Fundamental limits and potential drawbacks,'' in
  \emph{2023 IEEE/ION Position, Location and Navigation Symposium (PLANS)},
  2023, pp. 344--353.

\bibitem{10557490}
M.~Rahal, B.~Denis, M.~F. Keskin, B.~Uguen, and H.~Wymeersch, ``Ris-enabled
  nlos near-field joint position and velocity estimation under user mobility,''
  \emph{IEEE Journal of Selected Topics in Signal Processing}, pp. 1--12, 2024.

\bibitem{10529957}
H.~Chen, M.~F. Keskin, A.~Sakhnini, N.~Decarli, S.~Pollin, D.~Dardari, and
  H.~Wymeersch, ``6g localization and sensing in the near field: Features,
  opportunities, and challenges,'' \emph{IEEE Wireless Communications}, pp.
  1--8, 2024.

\bibitem{7762095}
A.~Yassin, Y.~Nasser, M.~Awad, A.~Al-Dubai, R.~Liu, C.~Yuen, R.~Raulefs, and
  E.~Aboutanios, ``{Recent Advances in Indoor Localization: A Survey on
  Theoretical Approaches and Applications},'' \emph{IEEE Communications Surveys
  \& Tutorials}, vol.~19, no.~2, pp. 1327--1346, 2017.

\bibitem{9625032}
M.~A. Uusitalo, P.~Rugeland, M.~R. Boldi, E.~C. Strinati, P.~Demestichas,
  M.~Ericson, G.~P. Fettweis, M.~C. Filippou, A.~Gati, M.-H. Hamon,
  M.~Hoffmann, M.~Latva-Aho, A.~Pärssinen, B.~Richerzhagen, H.~Schotten,
  T.~Svensson, G.~Wikström, H.~Wymeersch, V.~Ziegler, and Y.~Zou, ``{6G
  Vision, Value, Use Cases and Technologies From European 6G Flagship Project
  Hexa-X},'' \emph{IEEE Access}, vol.~9, pp. 160\,004--160\,020, 2021.

\bibitem{9569364}
H.~Wymeersch, D.~Shrestha, C.~M. de~Lima, V.~Yajnanarayana, B.~Richerzhagen,
  M.~F. Keskin, K.~Schindhelm, A.~Ramirez, A.~Wolfgang, M.~F. de~Guzman,
  K.~Haneda, T.~Svensson, R.~Baldemair, and S.~Parkvall, ``{Integration of
  Communication and Sensing in 6G: a Joint Industrial and Academic
  Perspective},'' \emph{2021 IEEE 32nd Annual International Symposium on
  Personal, Indoor and Mobile Radio Communications (PIMRC)}, pp. 1--7, 2021.

\bibitem{9679804}
R.~Liu, Q.~Wu, M.~Di~Renzo, and Y.~Yuan, ``{A Path to Smart Radio Environments:
  An Industrial Viewpoint on Reconfigurable Intelligent Surfaces},'' \emph{IEEE
  Wireless Communications}, vol.~29, no.~1, pp. 202--208, 2022.

\bibitem{Tang2019MIMOTT}
W.~Tang, J.~Y. Dai, M.~Z. Chen, K.~Wong, X.~Li, X.~Zhao, S.~Jin, Q.~Cheng, and
  T.~jun Cui, ``{MIMO Transmission Through Reconfigurable Intelligent Surface:
  System Design, Analysis, and Implementation},'' \emph{IEEE Journal on
  Selected Areas in Communications}, vol.~38, pp. 2683--2699, 2019.

\bibitem{9020088}
L.~Dai, B.~Wang, M.~Wang, X.~Yang, J.~Tan, S.~Bi, S.~Xu, F.~Yang, Z.~Chen,
  M.~D. Renzo, C.-B. Chae, and L.~Hanzo, ``{Reconfigurable Intelligent
  Surface-Based Wireless Communications: Antenna Design, Prototyping, and
  Experimental Results},'' \emph{IEEE Access}, vol.~8, pp. 45\,913--45\,923,
  2020.

\bibitem{Pei2021RISAidedWC}
X.~Pei, H.~Yin, L.~Tan, L.~Cao, Z.~Li, K.~Wang, K.~Zhang, and E.~Bj{\"o}rnson,
  ``{RIS-Aided Wireless Communications: Prototyping, Adaptive Beamforming, and
  Indoor/Outdoor Field Trials},'' \emph{IEEE Transactions on Communications},
  vol.~69, pp. 8627--8640, 2021.

\bibitem{9473842}
R.~Fara, D.-T. Phan-Huy, P.~Ratajczak, A.~Ourir, M.~Di~Renzo, and J.~De~Rosny,
  ``{Reconfigurable Intelligent Surface-Assisted Ambient Backscatter
  Communications – Experimental Assessment},'' \emph{2021 IEEE International
  Conference on Communications Workshops (ICC Workshops)}, pp. 1--7, 2021.

\bibitem{alexandropoulos2023risenabled}
G.~C. Alexandropoulos, D.-T. Phan-Huy, K.~D. Katsanos, M.~Crozzoli,
  H.~Wymeersch, P.~Popovski, P.~Ratajczak, Y.~B{\'e}n{\'e}dic, M.-H. Hamon,
  S.~H. Gonzalez \emph{et~al.}, ``Ris-enabled smart wireless environments:
  Deployment scenarios, network architecture, bandwidth and area of
  influence,'' \emph{EURASIP Journal on Wireless Communications and
  Networking}, vol. 2023, no.~1, p. 103, 2023.

\bibitem{10032155}
E.~J. Khatib, C.~S. Álvarez Merino, H.~Q. Luo-Chen, and R.~B. Moreno,
  ``Designing a 6g testbed for location: Use cases, challenges, enablers and
  requirements,'' \emph{IEEE Access}, vol.~11, pp. 10\,053--10\,091, 2023.

\bibitem{10525242}
S.~Kerboeuf, P.~Porambage, A.~Jain, P.~Rugeland, G.~Wikström, M.~Ericson,
  D.~Thai~Bui, A.~Outtagarts, H.~Karvonen, P.~Alemany, R.~Muñoz, R.~Vilalta,
  P.~Botsinis, A.~Ramos, J.~Castaneda~Cisneros, M.~Karaca, C.~Karousatou,
  S.~Barmpounakis, P.~Demestichas, A.~Zafeiropoulos, I.~Tzanettis,
  S.~Papavassiliou, P.~G. Giardina, G.~Landi, B.~Han, A.~Nimr, and M.~A.
  Uusitalo, ``Design methodology for 6g end-to-end system: Hexa-x-ii
  perspective,'' \emph{IEEE Open Journal of the Communications Society},
  vol.~5, pp. 3368--3394, 2024.

\bibitem{strinati2021reconfigurable}
E.~C. Strinati, G.~C. Alexandropoulos, H.~Wymeersch, B.~Denis,
  V.~Sciancalepore, R.~D'Errico, A.~Clemente, D.-T. Phan-Huy, E.~De~Carvalho,
  and P.~Popovski, ``{Reconfigurable, intelligent, and sustainable wireless
  environments for 6G smart connectivity},'' \emph{IEEE Communications
  Magazine}, vol.~59, no.~10, pp. 99--105, 2021.

\bibitem{9360709}
X.~Yuan, Y.-J.~A. Zhang, Y.~Shi, W.~Yan, and H.~Liu,
  ``{Reconfigurable-Intelligent-Surface Empowered Wireless Communications:
  Challenges and Opportunities},'' \emph{IEEE Wireless Communications},
  vol.~28, no.~2, pp. 136--143, 2021.

\bibitem{10158690}
J.~An, C.~Xu, D.~W.~K. Ng, G.~C. Alexandropoulos, C.~Huang, C.~Yuen, and
  L.~Hanzo, ``Stacked intelligent metasurfaces for efficient holographic mimo
  communications in 6g,'' \emph{IEEE Journal on Selected Areas in
  Communications}, vol.~41, no.~8, pp. 2380--2396, 2023.

\bibitem{10557708}
J.~An, C.~Yuen, Y.~L. Guan, M.~D. Renzo, M.~Debbah, H.~V. Poor, and L.~Hanzo,
  ``Two-dimensional direction-of-arrival estimation using stacked intelligent
  metasurfaces,'' \emph{IEEE Journal on Selected Areas in Communications},
  vol.~42, no.~10, pp. 2786--2802, 2024.

\end{thebibliography}
%\printbibliography

\vspace{-1cm}
%\vskip -2\baselineskip plus -1fil
\begin{IEEEbiography}[{\includegraphics[width=1in,height=1.25in,clip,keepaspectratio]{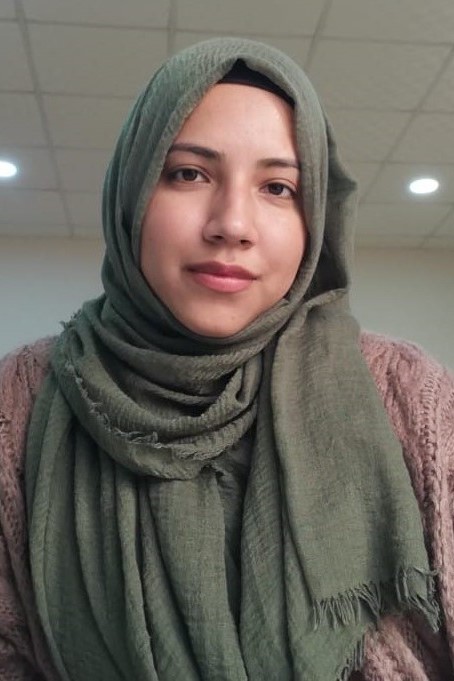}}]{Anum Umer} 
received the B.E. degree in electrical (telecommunication) engineering and the M.S. degree in electrical engineering from the National University of Science and Technology (NUST), Pakistan, in 2015 and 2017, respectively, where she was a Research Associate with the System Analysis and Verification Lab, School of Electrical Engineering and Computer Science in 2018.  During 2019-2022, she was a Research Engineer with the Research and Development Wing, NUST. She is currently pursuing her Ph.D. degree in Information and Communication Technology from the Thomas Johann Seebeck Department of Electronics, Tallinn University of Technology, Estonia. Her areas of research include wireless communication, localization, and sensing.
\end{IEEEbiography}
%\vskip -2\baselineskip plus -1fil
\vspace{-1cm}
\begin{IEEEbiography}[{\includegraphics[width=1in,height=1in,clip,keepaspectratio]{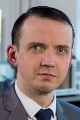}}]{Ivo M\"{u}\"{u}rsepp} 
was born in Tallinn, Estonia in 1980. He received bachelors and MSc degrees in telecommunication from the Tallinn University of Technology (TUT), Estonia in 2002 and 2004 respectively. PhD degree in telecommunication from Tallinn University of Technology was received in 2013. In 2002 he joined department of Radio- and Communication Technology of TUT as teaching assistant. Currently he works in Thomas Johann Seebeck department of Electronics as senior lecturer. Has also been employed for 5 years by Cyber Command of Estonian Defence Forces. His research interest includes signal processing, RF engineering, radio communication and mobile positioning.
\end{IEEEbiography}
\vspace{-1cm}
%\newpage
%\vskip 1\baselineskip plus -1fil
\begin{IEEEbiography}[{\includegraphics[width=1in,height=1.25in,clip,keepaspectratio]{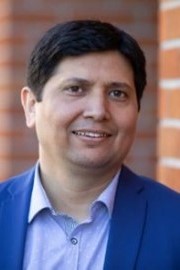}}]{Muhammad Mahtab Alam} 
(Senior Member, IEEE) received the M.Sc. degree in electrical engineering from Aalborg University, Denmark, in 2007, and the Ph.D. degree in signal processing and telecommunication from the INRIA Research Center, University of Rennes 1, France, in 2013. From 2014 to 2016, he was Post-Doctoral Research at the Qatar Mobility Innovation Center, Qatar. In 2016, he joined as the European Research Area Chair and as an Associate Professor with the Thomas Johann Seebeck Department of Electronics, Tallinn University of Technology, where he was elected as a Professor in 2018 and Tenured Full Professor in 2021. Since 2019, he has been the Communication Systems Research Group Leader. He has over 15 years of combined academic and industrial multinational experiences while working in Denmark, Belgium, France, Qatar, and Estonia. He has several leading roles as PI in multimillion Euros international projects funded by European Commission (Horizon Europe LATEST-5GS, 5G-TIMBER, H2020 5GROUTES, NATOSPS (G5482), Estonian Research Council (PRG424), Telia Industrial Grant etc. He is an author and co-author of more than 100 research publications. He is actively supervising a number of Ph.D. and Postdoc Researchers. He is also a contributor in two standardization bodies (ETSI SmartBAN, IEEE-GeeenICT-EECH), including ‘‘Rapporteur’’ of work item: DTR/ SmartBAN-0014. His research focuses on the fields of wireless communications–connectivity, mobile positioning, 5G/6G services and applications.
\end{IEEEbiography}
\newpage
%\vskip 0pt plus -1fil
%\vskip -2\baselineskip plus -1fil
%\vspace{-1cm}
\begin{IEEEbiography}[{\includegraphics[width=1in,height=1.25in,clip,keepaspectratio]{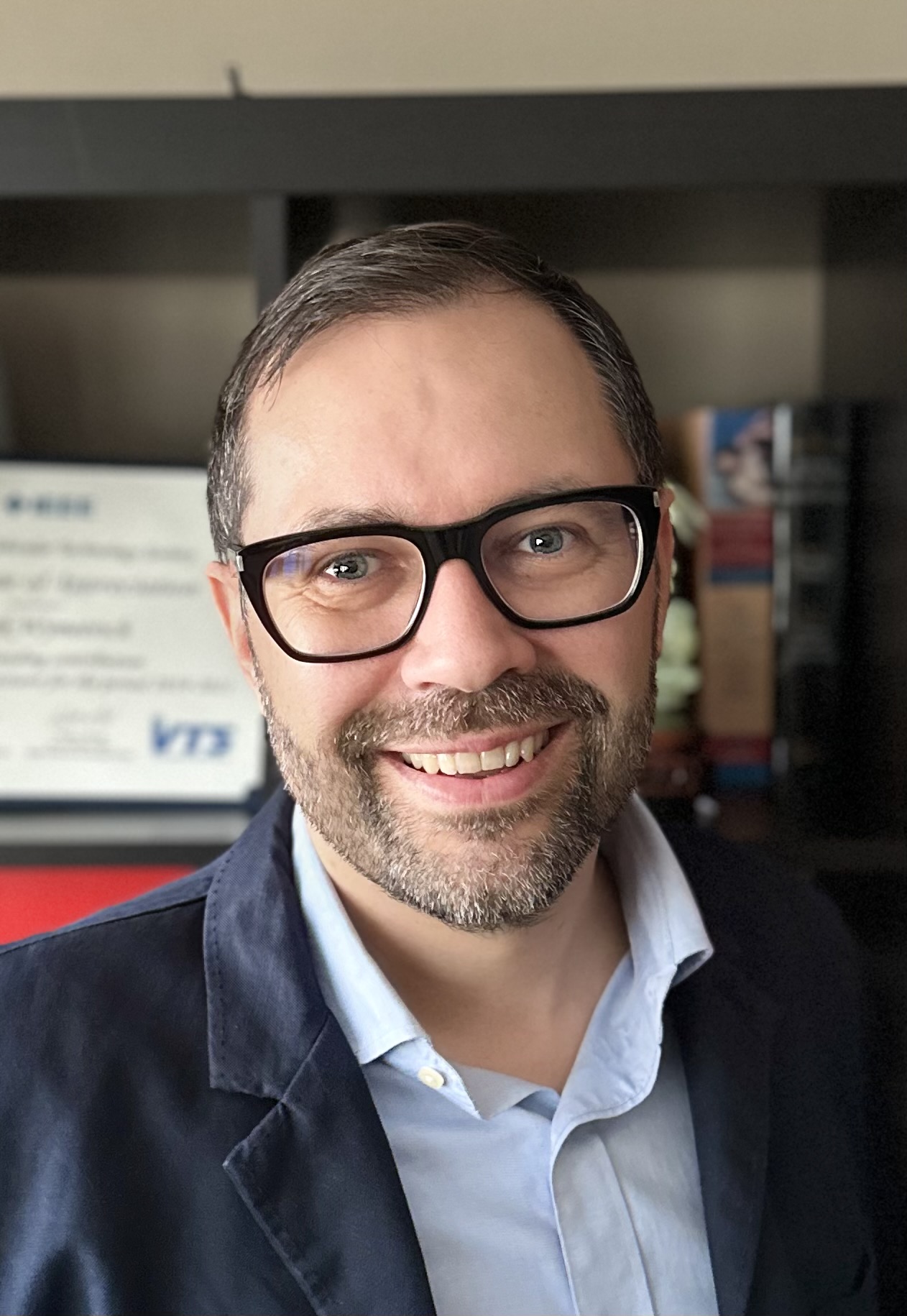}}]{Henk~Wymeersch}
 	(S'01, M'05, SM'19, F'24) obtained the Ph.D. degree in Electrical Engineering/Applied Sciences in 2005 from Ghent University, Belgium. He is currently a Professor of Communication Systems with the Department of Electrical Engineering at Chalmers University of Technology, Sweden. He is also a Distinguished Research Associate with Eindhoven University of Technology. Prior to joining Chalmers, he was a postdoctoral researcher from 2005 until 2009 with the Laboratory for Information and Decision Systems at the Massachusetts Institute of Technology. Prof. Wymeersch served as Associate Editor for IEEE Communication Letters (2009-2013), IEEE Transactions on Wireless Communications (since 2013), and IEEE Transactions on Communications (2016-2018) and is currently Senior Member of the IEEE Signal Processing Magazine Editorial Board.  During 2019-2021, he was an IEEE Distinguished Lecturer with the Vehicular Technology Society.  His current research interests include the convergence of communication and sensing, in a 5G and Beyond 5G context. 
\end{IEEEbiography}

%\newpage
%\includepdf[pages={{},{},-}]{IEEE_COMST_response_letter.pdf}
\end{document}